\documentclass[
 reprint,
 nofootinbib,
 showkeys,
 amsmath,
 amssymb,
 aps,
 prb,
 floatfix,
 superscriptaddress,
 longbibliography
 ]{revtex4-2}

\usepackage{graphicx}
\usepackage{multirow}
\usepackage[dvipsnames]{xcolor}
\usepackage{dcolumn}
\usepackage{bm}
\usepackage{dsfont}
\usepackage{physics}
\usepackage{hyperref}

\usepackage{ulem}

\newcommand{\Ay}[0]{\mathcal{A}}
\newcommand{\bAy}[0]{\boldsymbol{\mathcal{A}}}

\DeclareMathOperator{\Log}{Log}
\DeclareMathOperator{\arctanh}{arctanh}
\DeclareMathOperator{\arcsinh}{arcsinh}
\DeclareMathOperator{\sgn}{sgn}
\DeclareMathOperator{\diag}{diag}

\begin{document}

\author{B.~A.~Levitan}
\email{Contact author: benjamin.levitan@usherbrooke.ca}
\affiliation{Department of Condensed Matter Physics, Weizmann Institute of Science, Rehovot 7610001, Israel}
\author{Y.~Oreg}
\affiliation{Department of Condensed Matter Physics, Weizmann Institute of Science, Rehovot 7610001, Israel}
\author{E.~Berg}
\affiliation{Department of Condensed Matter Physics, Weizmann Institute of Science, Rehovot 7610001, Israel}
\author{M.~S.~Rudner}
\affiliation{Department of Physics, University of Washington, Seattle, WA 98195-1560, USA}
\author{I.~Iorsh}
\affiliation{Department of Condensed Matter Physics, Weizmann Institute of Science, Rehovot 7610001, Israel}
\affiliation{Faculty of Physics, ITMO University, St.~Petersburg 197101, Russia}
\affiliation{Department of Physics, Engineering Physics and Astronomy, Queen’s University, Kingston, Ontario K7L 3N6, Canada}

\title{Linear spectroscopy of collective modes and the gap structure \\ in two-dimensional superconductors}

\date{\today}

\begin{abstract}
We consider optical response in multiband, multilayer two-dimensional superconductors. Within a simple model, we show that linear response to AC gating can detect collective modes of the condensate, such as Leggett and clapping modes. We show how trigonal warping of the superconducting order parameter can help facilitate detection of clapping modes. Taking rhombohedral trilayer graphene as an example, we consider several possible pairing mechanisms and show that all-electronic mechanisms may produce in-gap clapping modes. These modes, if present, should be detectable in the absorption of microwaves applied via the gate electrodes, which are necessary to enable superconductivity in this and many other settings; 
their detection would constitute strong evidence for unconventional pairing. Last, we show that absorption at frequencies above the superconducting gap $2 |\Delta|$ also contains a wealth of information about the gap structure. Our results suggest that linear spectroscopy can be a powerful tool for the characterization of  unconventional two-dimensional superconductors.
\end{abstract}

\maketitle

\section{Introduction}
In superconductors, a minimum of two collective modes arise: the Anderson-Bogoliubov-Goldstone (ABG) mode~\cite{anderson1958rpa, anderson1958newmethod, bogoljubov1958new, rickayzen1959collective}, corresponding to phase fluctuations of the complex order parameter, and the Higgs mode~\cite{littlewoodvarma1982higgs, shimano2020higgs}, corresponding to amplitude fluctuations. When the superconducting order parameter has more structure, its enlarged configuration space can produce a richer diversity of collective modes~\cite{bardasisschrieffer1961, leggett1966number, wolfle1976collective, hirashima1988collective,hirschfeld1989absorptionanisotropic,hirschfeld1992collectiveabsorption,yip1992circular,kee2000collective,higashitani2000order,higashitani2000response,balatsky2000collective,roy2008chiral,ota2011collective,carlstrom2011threeband,lin2012massless,stanev2012model,puetter2012identifying,kobayashi2013massless,marciani2013leggettTRSB,lin2014ground,maiti2015competing,cea2016leggettRaman,sun2020terahertz,poniatowski2022spectroscopic,gabriele2022plasma,sellati2023josephson}. Measurements of the collective mode spectrum (e.g., by electromagnetic absorption) can therefore provide a valuable tool for diagnosing the nature of the order parameter, and even the underlying pairing mechanism~\cite{hirschfeld1989absorptionanisotropic,hirschfeld1992collectiveabsorption,yip1992circular,ponomarev2004evidence,chubukov2009ramanpnictides,bohm2014balancing,maiti2016ramanBaSh,bohm2018microscopic,huang2018identifying,poniatowski2022spectroscopic,sarkar2024TRSBraman,sellati2024tilted}. 
In this work, we show how to excite collective modes in gated two-dimensional (2D) superconductors using the gate electrodes themselves, providing a natural way to probe the nature of superconductivity in a wide range of emerging new materials.

As a prototypical example, consider the Leggett mode in a multiband superconductor~\cite{leggett1966number}. One can think of this mode as a momentum-space analog of the Josephson effect, consisting of relative number and phase fluctuations between the condensates formed from different bands. Assuming that intraband interactions are attractive, when the Josephson coupling (i.e.,~interband pair swapping interaction) is not too strong, the Leggett mode can arise inside the superconducting gap, and it can therefore be underdamped. However, even when this mode is a well-defined excitation, crystal symmetries often render it invisible to linear electromagnetic response at $\vb{q} \rightarrow 0$~\cite{kamatani2022optical, nagashima2024classification}.
As we show below, in a gated 2D superconductor, static gating can break $M_z$ mirror symmetry, permitting linear access to the Leggett mode at $\vb{q} \rightarrow 0$. The same gate electrodes can then be used to apply an AC probe field, thereby exciting the mode. This approach, which is uniquely well-suited to gated superconductors, represents an alternative to Raman \cite{blumberg2007MgB2leggett,chubukov2009ramanpnictides,klein2010leggettraman,lin2012massless,bohm2014balancing,cea2016leggettRaman,maiti2016ramanBaSh,maiti2017raman,bohm2018microscopic,huang2018identifying,giorgianni2019leggett,zhao2020softleggett,beneklins2024selection,sarkar2024TRSBraman} and tunneling \cite{ponomarev2004evidence,lee2023tunneling} spectroscopy.
\begin{figure*}
\centering
\includegraphics[width=\textwidth]{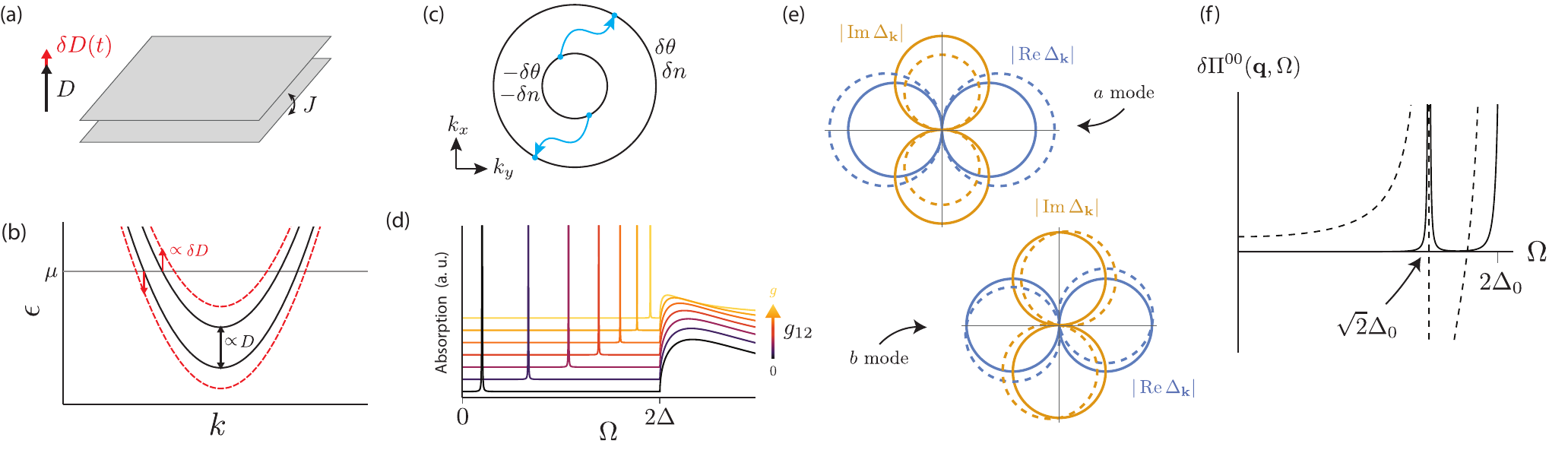}
\caption{\label{fig:toy_model} (a) A two-layer superconductor under a large constant electric displacement field $D$, driven by a perturbing AC field $\delta D(t)$. (b) Single-electron dispersion in the toy model (black), showing the effect of a static perturbing field (red). (c) Schematic electron-electron scattering between Fermi sheets, indicating the relative phase ($\delta \theta$) and number ($\delta n$) fluctuations of the Leggett mode. (d) Absorption spectrum given $\nu_1 = \nu_2$, $g_{11}^{(s)} = g_{22}^{(s)}$, and $\Delta_1 = \Delta_2 = \Delta$, for different values of interband coupling $g^{(s)}_{12}$ and $g^{(p)} = 0$. (e, f) The clapping modes in a chiral $p$-wave superconductor, using the toy model with $g_{\alpha \alpha'}^{(s)} = 0 > g^{(p)}$ and $|V| \gg \mu$. (e) Assuming a $p_x + i p_y$ ground state, the clapping modes at $\vb{q} \rightarrow 0$ correspond to fluctuations of the form $\Delta_{\vb{k}} (t) = \Delta_0 e^{i \varphi_{\vb{k}}} + \left( a(t) + i b(t) \right) e^{-i \varphi_{\vb{k}}}$. Solid lines show a polar plot of the equilibrium order parameter, dashed lines show the distortion associated with each mode. (f)~Real (dashed) and imaginary (solid) parts of the clapping mode contribution to the electronic compressibility for small but nonzero $\vb{q}$ in the circularly symmetric model, Eq.~\eqref{eq:circular_clapping_compressibility}; see text. We give $\Omega$ a small imaginary part by hand to give the pole a finite width.}
\end{figure*}

Besides the Leggett mode, another family of exotic collective modes arise in superconductors with spontaneously broken time-reversal symmetry. Known as ``generalized clapping modes''~\cite{wolfle1976collective, hirashima1988collective,hirschfeld1989absorptionanisotropic,yip1992circular,hirschfeld1992collectiveabsorption,kee2000collective, higashitani2000order, higashitani2000response, balatsky2000collective,poniatowski2022spectroscopic}, these amount to fluctuations in the uncondensed time-reversed partner of the condensed ground-state channel. For example, in a $p_x + i p_y$ chiral superconductor, clapping modes arise as fluctuations in the $p_x -i p_y$ channel, as shown in Fig.~\ref{fig:toy_model}(e). Linear coupling of light to clapping modes at $\vb{q} \rightarrow 0$ generically requires breaking of rotational symmetry due to the nonvanishing angular momentum of these modes. Direct observation of clapping modes in linear electromagnetic response is therefore typically not possible at $\vb{q} \rightarrow 0$ in 2D materials with $C_n$ rotational symmetry, where $n>2$. Working at finite $\vb{q}$, we focus on the experimentally relevant case where (discrete) rotational symmetry is preserved, and show that the wavevector required for an appreciable signal corresponds to length scales comparable to device size in the case of superconducting rhombohedral trilayer graphene (RTG)~\cite{zhou2021superconductivity}, which may host a chiral $p$-wave order parameter~\cite{ghazaryan2021unconventional,Qin2023,chatterjee2022inter,dong2023signatures}. Our results therefore point toward a particularly convenient method for characterizing pairing in gated 2D superconductors.

\section{The toy model}
To elucidate the basic ideas at work, we begin with a toy model which captures both the Leggett and clapping modes. The Hamiltonian is $H = H_0 + H_{\text{int}}$, with
\begin{multline}    \label{eq:toy_H_0}
    H_0
    =
    \sum_{\vb{k} \sigma}
    \begin{pmatrix}
        f^{\dag}_{\vb{k} t \sigma} & f^{\dag}_{\vb{k} b \sigma}
    \end{pmatrix}
    \begin{pmatrix}
        \xi_{\vb{k} 0} + V &   J                   \\
        J                   &   \xi_{\vb{k} 0} - V
    \end{pmatrix}
    \begin{pmatrix}
        f_{\vb{k} t \sigma}   \\
        f_{\vb{k} b \sigma}
    \end{pmatrix}
    \\
    =
    \sum_{\vb{k} \sigma}
    \begin{pmatrix}
        c^{\dag}_{\vb{k} 1 \sigma} & c^{\dag}_{\vb{k} 2 \sigma}
    \end{pmatrix}
    \begin{pmatrix}
        \xi_{\vb{k} 1}     &   0                   \\
        0                   &   \xi_{\vb{k} 2}
    \end{pmatrix}
    \begin{pmatrix}
        c_{\vb{k} 1 \sigma}   \\
        c_{\vb{k} 2 \sigma}
    \end{pmatrix}
\end{multline}
describing spin-1/2 fermions ($\sigma \in \lbrace \uparrow, \downarrow \rbrace$) in two layers ($l \in \lbrace t, b \rbrace$), each with bare dispersion $\xi_{\vb{k} 0} = \epsilon_{\vb{k} 0} - \mu$ relative to the Fermi level. We assume isotropic $\xi_{\vb{k} 0}$ for simplicity. Single-particle tunneling ($J$) couples the layers, and the electric potential drops by $2 V$ from top to bottom (we set the electron charge to $1$). Including interlayer tunneling, the band energies are $\xi_{\vb{k}, \alpha \in \lbrace 1, 2 \rbrace} = \xi_{\vb{k} 0} \pm \sqrt{J^2 + V^2}$. Notice that starting from $V = 0$, perturbing with a small $V$ produces energy shifts at second order in $V$. This reflects a more general principle: barring accidental degeneracies, any system which respects the mirror symmetry $M_z$ (which here exchanges the two layers) must have a spectrum which is even in the $M_z$-breaking out-of-plane displacement field, preventing linear access to the Leggett mode. In Fig.~\ref{fig:toy_model}(a), we assume a strong field ($|V| \gg |J|$), strongly breaking $M_z$, and rendering the band splitting essentially linear in $|V|$. This assumption is not essential to our calculations; any $V \ne 0$ will enable linear response to a subsequent perturbation $\delta V$.

We model interactions using
\begin{multline} \label{eq:toy_model_interaction_H}
    H_{\text{int}}
    =
    \frac{1}{L^2}
    \sum_{\substack{\vb{k k' q} \\ \alpha \alpha'}}
    \mathcal{V}_{\vb{k k'}}^{\alpha \alpha'}
    c^{\dag}_{\vb{k} \alpha \uparrow}
    c^{\dag}_{-\vb{k} + \vb{q}, \alpha \downarrow}
    c_{-\vb{k'} + \vb{q}, \alpha' \downarrow}
    c_{\vb{k'} \alpha' \uparrow},
\end{multline}
focusing on small $\vb{q}$ (i.e.,~the Cooper channel). $L$ is the linear dimension of the sample. For concreteness, we take
\begin{equation}
    \mathcal{V}_{\vb{k k'}}^{\alpha \alpha'}
    =
    g^{(s)}_{\alpha \alpha'}
    +
    2 g^{(p)} \cos(\varphi_{\vb{k}} - \varphi_{\vb{k'}}) \delta_{\alpha \alpha'},
\end{equation}
where $g^{(s)}_{\alpha \alpha} < 0$ binds $s$-wave pairs in band $\alpha$, $g^{(s)}_{1 2} = \left( g^{(s)}_{2 1} \right)^*$ Josephson couples the two bands, and $g^{(p)} < 0$ binds $p$-wave pairs. Here $\varphi_{\vb{k}}$ is the angle of $\vb{k}$ relative to the $k_x$ axis, i.e.,~$\vb{k} = |\vb{k}| (\cos \varphi_{\vb{k}}, \sin \varphi_{\vb{k}})$. Note that with the spin indices as written in Eq.~\eqref{eq:toy_model_interaction_H}, when considering $p$-wave pairing, we implicitly consider the $m_z = 0$ triplet component only; the extension to the more general case is straightforward. 

We base our analysis on the imaginary-time path integral. To identify the collective modes, we perform a Hubbard-Stratonovich decoupling in the Cooper channel, and expand the effective action to quadratic order in fluctuations $\phi$ around the mean-field saddle point~\cite{depalo1999effectiveaction,paramekanti2000action,sharapov2002effective,marciani2013leggettTRSB,cea2016leggettRaman,poniatowski2022spectroscopic}. Analytically continuing the Gaussian effective action to real frequency ($i \Omega_m \rightarrow \Omega + i 0^+$), the collective modes appear  as poles of the bosonic propagator for pair fluctuations. To evaluate the linear response to applied electric fields, we also compute the couplings between the collective modes and such fields. See Appendices \ref{appendix:toy_leggett}, \ref{appendix:leggett_fl}, and \ref{appendix:toy_clapping} for the details of our toy-model calculations.

First we focus on the Leggett mode, by setting $g^{(p)} = 0$ and $g^{(s)}_{\alpha \alpha'} \ne 0$. We assume $g^{(s)}_{\alpha \alpha} < 0$. The Leggett mode requires $\mu > \sqrt{J^2 + V^2}$ such that both bands are occupied; the two Fermi surfaces then develop superconducting gaps $\Delta_{1, 2}$. Leggett's original paper analyzed essentially this model---at $\vb{q} = 0$ and with $g^{(p)} = 0$---and found its relative-phase mode~\cite{leggett1966number}. Taking $|\Delta_1| \le |\Delta_2|$ without further loss of generality, we find that an in-gap Leggett mode requires $0 \le |g^{(s)}_{12}| / (g^{(s)}_{11} g^{(s)}_{22} - |g^{(s)}_{12}|^2) < \nu_2 \arcsin(|\Delta_1/\Delta_2|) / \sqrt{1 - (\Delta_1 / \Delta_2)^2}$, where $\nu_{\alpha}$ is the density of states per unit area per spin on the Fermi surface of band $\alpha$ in the normal state. (When $|\Delta_1| = |\Delta_2|$, the second inequality is irrelevant). Fermi-liquid interactions can modify this condition~\cite{maiti2017raman}; see Appendix \ref{appendix:leggett_fl} for details.

We now show how an AC modulation of the displacement field can probe the Leggett mode. The drive field corresponds to a small time-dependent perturbation around the static background, $V \rightarrow V_{\rm DC} + \delta V(t)$. The physical idea is that in the presence of the static $V_{\text{DC}} \ne 0$ (which breaks $M_z$), the perturbation $\delta V$ produces linear energy shifts in opposite directions for the two bands, as illustrated in Fig.~\ref{fig:toy_model}(a), directly driving a phase difference between the two condensates and hence exciting the Leggett mode. Keeping only terms up to linear order in $\delta V$ and working in the band basis, driving corresponds to a Hamiltonian term $H_{\delta V} = \sum_{\vb{k} \sigma} \delta V (t) (V_{\rm DC} / \sqrt{J^2 + V_{\rm DC}^2}) (c^{\dag}_{\vb{k} 1 \sigma} c_{\vb{k} 1 \sigma} - c^{\dag}_{\vb{k} 2 \sigma} c_{\vb{k} 2 \sigma})$. Integrating out the fermions in the presence of the drive yields a quadratic term $\sim \delta V^2$ and source terms $\sim \phi \, \delta V$ in the effective action, on top of the Gaussian action for pair fluctuations $\sim - \phi^{\dag} \mathcal{D}^{-1} \phi$. Integrating out the fluctuations $\phi$ then yields the full quadratic free-energy functional $\mathcal{F} [\delta V$], describing a capacitance (dependent on frequency $\Omega$):
\begin{equation}
    \frac{1}{L^2} \mathcal{F} [\delta V(\Omega)]
    =
    \frac{1}{2} \mathcal{C} (\Omega) \delta V (-\Omega) \delta V (\Omega).
\end{equation}

We provide an explicit calculation of the capacitance per unit area $\mathcal{C} (\Omega)$ in Appendix \ref{appendixsec:legget_driving}. In the symmetric case where $\Delta_1 = \Delta_2 = \Delta > 0$ and $\nu_1 = \nu_2 = \nu$, the collective-mode contribution simplifies to
\begin{equation}
    \mathcal{C}_{\phi} (\Omega)
    =
    \frac{4 V_{\text{DC}}^2}{V_{\text{DC}}^2 + J^2}
    \frac{(\nu \gamma \sec \gamma)^2}{
        \nu \gamma \tan \gamma
        - 2 \tilde{g}^{-1}
    },
\end{equation}
where
\begin{equation}
    \tilde{g}^{-1}
    =
    \big| g_{12}^{(s)} \big|
    /
    \Big(
        g_{11}^{(s)} g_{22}^{(s)} - \big| g_{12}^{(s)} \big|^2
    \Big)
\end{equation}
and
\begin{equation}
    \sin \gamma
    =
    (\Omega + i 0^+) / (2 |\Delta|).
\end{equation}
The Leggett mode frequency $\Omega_{\text{L}}$ solves $\nu \gamma_{\text{L}} \tan \gamma_{\text{L}} = 2 \tilde{g}^{-1}$, with $\sin \gamma_{\text{L}} = \Omega_{\text{L}} / (2 \Delta)$.
Near the positive-frequency pole, we can approximate
\begin{equation}
    \mathcal{C}_{\phi} (\Omega)
    \approx
    \frac{4 V_{\text{DC}}^2}{V_{\text{DC}}^2 + J^2}
    \frac{\nu \gamma_{\text{L}}^2}{
        \sin \gamma_{\text{L}}
        + \gamma_{\text{L}} \sec \gamma_{\text{L}}
    }
    \frac{2 \Delta}{\Omega - \Omega_{\text{L}} + i 0^{+}}.
\end{equation}

In the general (asymmetric) case, the key results are that: (i) The amplitude modes and ABG mode (at $\vb{q} \rightarrow 0$) \textit{do not} contribute to $\mathcal{C} (\Omega)$, since the amplitude modes are neutral scalars, and the ABG mode at $\vb{q} = 0$ corresponds to a fluctuation of the (conserved) total charge; and (ii) The Leggett mode \textit{does} contribute an in-gap pole to $\mathcal{C} (\Omega)$, provided that the background field $V_{\rm DC} \ne 0$ (and that the mode is in-gap). We make no attempt to explicitly compute the width of the pole, which is controlled by higher-order processes, and is model-dependent. However, we expect the resonance to be narrow, since no quasiparticle excitations are available to facilitate decay through low-order processes.

Figure \ref{fig:toy_model}(d) shows the total absorption ($-\Im \mathcal{C}$, including quasiparticle and collective-mode contributions) for the symmetric toy model. In this case, when the interactions become uniform ($g^{(s)}_{12} \rightarrow g^{(s)}_{11} = g^{(s)}_{22}$), the absorption peak corresponding to the Leggett mode approaches the gap edge. For illustrative purposes, in this figure we add a small imaginary part to $\Omega$, tuned by hand to give the pole an approximately equal width in each of the curves.

We now turn our attention to the clapping modes, setting $g^{(s)}_{\alpha \alpha'} = 0 > g^{(p)}$ and $|V| \gg \mu$. We project into the occupied (lower) band, and hence drop the band index $\alpha$. The gap equation has two degenerate solutions, corresponding to $p_{x} \pm i p_y$ pairing. We assume the $p_x + i p_y$ channel condenses, i.e.,~$\Delta_{\vb{k}} = \Delta_0 e^{i \varphi_{\vb{k}}}$ with $\Delta_0 > 0$. As detailed in Appendix \ref{appendix:toy_clapping}, fluctuations in the uncondensed $p_x - i p_y$ channel give rise to two real bosons $a$ and $b$; these are the clapping modes, which in this symmetric minimal model are degenerate at $\Omega = \sqrt{2} \Delta_0$.

To show how an AC electric field can probe the clapping modes, as in the Leggett case, we perturb around the static background field, $V \rightarrow V_{\text{DC}} + \delta V (t)$. Since we projected out the upper band, the only effect of the modulation is to shift the energy of the lower band; when the displacement field-induced interlayer potential is very large compared with the interlayer hopping matrix element, the electrons only live on one layer, so they only feel the perturbing field through its local potential. Therefore, we calculate the clapping mode contribution to the electronic compressibility $\Pi^{00}$. Unlike the Leggett mode, the clapping modes carry angular momentum $2$ relative to the ground state; this means that with circular symmetry, their couplings to the scalar potential must vanish at $\vb{q} = 0$. Hence, we calculate their compressibility contributions to leading nonvanishing order in $|\vb{q}|$.

Our calculation, detailed in Appendix \ref{appendix:toy_clapping}, mirrors that of Ref.~\cite{poniatowski2022spectroscopic}, generalized to nonzero momentum. After Hubbard-Stratonovich decoupling, the fermions encounter the total (fluctuating) pairing field
\begin{equation}
    \Delta_{\vb{k}, q}
    =
    \Delta^{(+)}_{q} e^{i \varphi_{\vb{k}}}
    +
    \Delta^{(-)}_{q} e^{- i \varphi_{\vb{k}}}.
\end{equation}
Here $q = (i \Omega_m, \vb{q})$ contains both the bosonic Matsubara frequency $\Omega_m = 2 \pi m T$ and momentum $\vb{q}$. In coordinate space, writing $x = (\tau, \vb{r})$, the Hubbard-Stratonovich bosons corresponding to the two pairing channels are
\begin{equation}
    \Delta^{(+)} (x)
    =
    e^{i \theta(x)} [\Delta_0 + h(x)]
\end{equation}
and
\begin{equation}
    \Delta^{(-)} (x)
    =
    e^{i \theta (x)} [a(x) + i b(x)].
\end{equation}
Here $\theta$ is the global phase (ABG) mode, $h$ the Higgs mode, and $a$ and $b$ the clapping modes. After integrating out the fermions, and neglecting the (massive) Higgs mode for brevity, the long-wavelength action at quadratic order is
\begin{multline}
    S
    =
    \sum_{q} \Bigg(
        - \mathcal{D}^{-1}_{a, q} a_{-q} a_{q}
        - \mathcal{D}^{-1}_{b, q} b_{-q} b_{q}
        \\ 
        +
        \Pi^{\mu \nu}_{q} \Ay^{\mu}_{-q} \Ay^{\nu}_{q}
        +
        \Pi^{\mu a}_{q} \Ay^{\mu}_{-q} a_{q}
        + 
        \Pi^{\mu b}_{q} \Ay^{\mu}_{-q} b_{q}
    \Bigg),
\end{multline}
where $\Ay = (\Ay^0, \bAy) = (A^0 + \partial_{\tau} \theta, \vb{A} - \grad \theta)$ is the gauge-invariant combination of the electromagnetic gauge potential $A$ and the spacetime gradient of the global phase $\theta$, and summation over $\mu, \nu \in \lbrace 0, x, y \rbrace$ is implied. Since we are only interested in the impact of the clapping modes on the compressibility, we drop all terms involving the spatial component $\bAy$ for simplicity. Physically, we expect that this approximation should not drastically change the result, since the ABG mode associated with $\grad \theta$ is either at a much lower energy (in a neutral superfluid) or much higher energy (after the Anderson-Higgs mechanism in a superconductor) than the clapping mode, for small but nonzero $\vb{q}$. We provide explicit expressions for the propagators $\mathcal{D}_{X q}$ and couplings $\Pi^{0X}_{q}$ ($X \in \lbrace a, b \rbrace$), to leading nonvanishing order in $|\vb{q}|$, in Eqs.~\eqref{eq:toy_clapping_couplings} and \eqref{eq:toy_clapping_propagator}, respectively.
Note that with circular symmetry, $\Pi^{0X}_{q}$ vanishes at $\vb{q} = 0$ and first shows up at $O(|\vb{q}|^2)$. The product $\Pi^{0X}_{q} \Ay^0_{-q} X_{q}$ must be scalar, so two factors of momentum are required to balance the angular momentum 2 of the clapping modes.

\begin{figure*}[ht!]
\centering
\includegraphics[width=\textwidth]{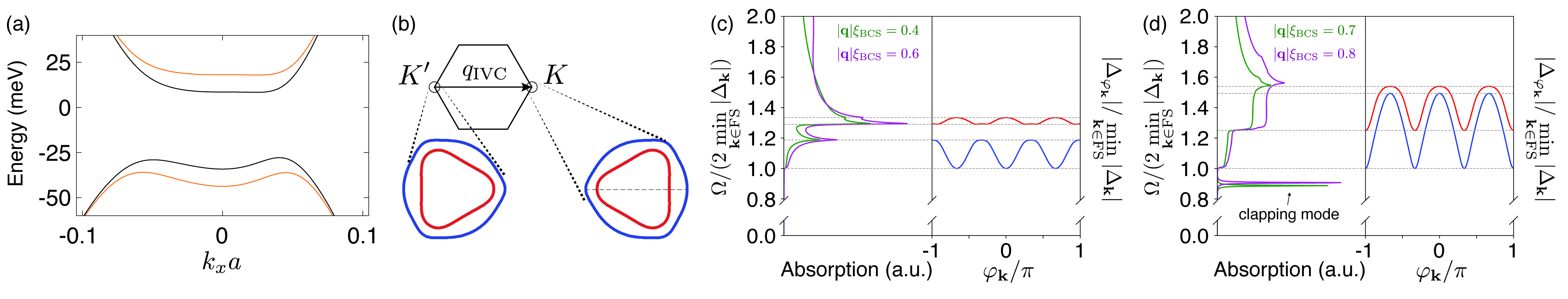}
\caption{\label{fig:RTG} (a) Electron dispersion of RTG under a static perpendicular electric field corresponding to a potential drop of $20$ meV (black) or $30$ meV (orange) between the top and bottom layers. Here, $k_x$ is measured relative to the $K$ point. (b) Fermi surfaces of RTG in each of its two valleys, for a potential drop of $20$ meV and a Fermi energy of $-29.3$ meV. The line cut through the dispersion relation [panel (a)] is along the dotted line schematically indicated by the dashed line in panel (b). (c-d) Absorption spectrum and gap structure for (c) phonon-mediated $s$-wave pairing, and (d) intervalley coherence fluctuation-mediated chiral $p$-wave pairing. Note the in-gap clapping mode in the $p$-wave case. The gap curves are colored according to the Fermi surface component on which they reside, matching panel (b).}
\end{figure*}

We obtain the clapping mode contribution to the compressibility $\delta \Pi^{00}_{q}$ by integrating out those modes, yielding
\begin{equation}
    \delta \Pi^{00}_{q}
    =
    \frac{1}{4} \Pi^{0 a}_{-q} \mathcal{D}_{a, q} \Pi^{0 a}_{q}
    + \frac{1}{4} \Pi^{0 b}_{-q} \mathcal{D}_{b, q} \Pi^{0 b}_{q}.
\end{equation}
Analytically continuing to real time ($i \Omega_m \rightarrow \Omega + i 0^+$), we find to leading order in $\vb{q}$ and for large $\mu$:
\begin{equation}    \label{eq:circular_clapping_compressibility}
    \delta \Pi^{00}_{q}
    =
    2 \nu \left(
        \frac{\pi}{4} \xi_{\text{BCS}} |\vb{q}|
    \right)^4
    \frac{1 - \gamma^2 (\cot \gamma - \tan \gamma)^2}{
        \gamma \cos^2 \gamma \sin 4 \gamma
    }.
\end{equation}

Here $\xi_{\text{BCS}} = v_{\text{F}} / (\pi \Delta_0)$ is the BCS coherence length, and $v_{\text{F}} = \sqrt{2 \mu / m}$ is the Fermi velocity. This result neglects the dispersion of the clapping modes. Note the pole at $\gamma = \pi / 4$, corresponding to $\Omega = \pm \sqrt{2} \Delta_0$: the clapping modes are visible in the compressibility, and hence in the capacitance. In the vicinity of the positive-frequency pole,
\begin{equation}
    \delta \Pi^{00}_{q}
    \approx
    \nu
    \left( \frac{\pi}{4} \right)^3
    \left( \xi_{\text{BCS}} |\vb{q}| \right)^4
    \frac{- \sqrt{2} \Delta_0}{
        \Omega - \sqrt{2} \Delta_0 + i 0^+
    }.
\end{equation}

As explained above, the factor of $|\vb{q}|^4$ in Eq.~\eqref{eq:circular_clapping_compressibility} is a consequence of circular symmetry. To obtain a signal at lower order in $|\vb{q}|$, something must break that symmetry to relax the angular momentum constraint. If the symmetry is broken completely, for example by a $\cos (\zeta) p_x + i \sin(\zeta) p_y$ order parameter with $\zeta$ away from high-symmetry values (as in Ref.~\cite{poniatowski2022spectroscopic}), then a signal can even arise at $\vb{q} \rightarrow 0$. A less drastic scenario is that rotation is explicitly broken down to a discrete subgroup dictated by the lattice structure. With $C_{3v}$ symmetry (possessed by RTG, for example), the couplings $\Pi^{0X}_{q}$ need only contain a single factor of $\vb{q}$, since $2 = -1 \pmod 3$. In Appendix \ref{appendixsec:trigonal_warping} we find that trigonal warping indeed yields a nonvanishing result at $O(|\vb{q}|^2)$, of similar qualitative form to Eq.~\eqref{eq:circular_clapping_compressibility}, but suppressed by the small factor $1 / (\xi_{\text{BCS}} k_{\text{F}})^2 \sim (\Delta_0 / \mu)^2$. 

\section{A more realistic model: rhombohedral trilayer graphene}

To go beyond our simple toy model, we consider superconducting RTG as an illustrative case study. We model the single-particle bandstructure of RTG using a $6\times6$ tight-binding Hamiltonian in its continuum limit, written in the basis of two atoms per unit cell in each of the three layers~\cite{zhang2010band,zhou2021half}; for convenience, we reproduce this Hamiltonian in Eq.~\eqref{eq:rtg_hamiltonian}. We limit ourselves to the range of parameters (static perpendicular displacement field and doping level) where superconductivity was observed~\cite{zhou2021superconductivity}, focusing on the SC2 phase. In this regime, prior to the onset of superconductivity, RTG contains one annular Fermi sea in each valley (two in total), whose Fermi surfaces show substantial trigonal warping [see Fig.~\ref{fig:RTG}(b)]. The specific pairing mechanism responsible for superconductivity in RTG is yet to be determined, so we consider conventional ($s$ wave, mediated by phonons) and unconventional (chiral $p$ wave, mediated by fluctuations of either charge density, as in the Kohn-Luttinger mechanism~\cite{ghazaryan2021unconventional, cea2022KL, Qin2023}, or intervalley coherence~\cite{chatterjee2022inter,dong2023signatures}) scenarios. See Appendix \ref{appendix:rtg_potentials} for our specific model interaction potentials.

To approach the RTG calculation, we lay out a general formalism for calculating superconducting collective modes and their response functions within the Gaussian pair-fluctuation paradigm. We provide a detailed derivation in Appendix \ref{appendix:general_case}, and summarize here only the key ideas. Our approach begins with a spectral decomposition of the interaction matrix $\mathcal{V}$ into its eigenbasis, $\mathcal{V}_{\vb{k k'}} = \sum_{\ell} g_{\ell} \chi^{(\ell)}_{\vb{k}} \chi^{(\ell)}_{\vb{k'}}$. We then decouple the interactions in the Cooper channel, using one complex Hubbard-Stratonovich boson $\Delta^{(\ell)}_{q}$ for each nonnull eigenvector $\chi^{(\ell)}$. These bosons combine to the total, fluctuating, pairing field
\begin{equation}
    \Delta_{\vb{k}, q} 
    =
    \sum_{\ell} i^{\rho_{\ell}} \Delta^{(\ell)}_{q} \chi^{(\ell)}_{\vb{k}}
\end{equation}
with $i^{\rho_{\ell}} = 1$ if $g_{\ell} < 0$ (attraction) and $i^{\rho_{\ell}} = i$ if $g_{\ell} > 0$ (repulsion) \cite{marciani2013leggettTRSB, dalal2023repulsion}. After integrating out the fermions, the spacetime-independent saddle points of the bosonic action $S [\Delta^{(\ell)}]$ correspond to solutions of the usual gap equation
\begin{multline}
    \Delta_{\vb{k}}
    =
    \frac{T}{L^2} \sum_{k'}
    \mathcal{V}_{\vb{k k'}} 
    \frac{\Delta_{\vb{k'}}}{(i \omega_{n'})^2 - E_{\vb{k'}}^2}
    \\
    =
    -\int \frac{\dd^2 \vb{k}}{(2 \pi)^2}
    \mathcal{V}_{\vb{k k'}}
    \tanh \left(
        \frac{E_{\vb{k'}}}{2 T}
    \right)
    \frac{\Delta_{\vb{k'}}}{2 E_{\vb{k'}}},
\end{multline}
where $\Delta_{\vb{k}}$ without a $q$ index denotes $\Delta_{\vb{k}, q=0}$ evaluated at the saddle point.

We then re-express the pairing bosons $\Delta^{(\ell)} (x)$ as real and imaginary fluctuations $\mathcal{R}^{(\ell)} (x)$ and $\mathcal{I}^{(\ell)} (x)$ around their saddle-point values, $\Delta^{(\ell)} (x) = \Delta^{(\ell)} + \mathcal{R}^{(\ell)} (x) + i \mathcal{I}^{(\ell)} (x)$. Going back to the Fourier domain, we expand the action to second order in the fluctuations $X, Y \in \lbrace \mathcal{R, I} \rbrace$:
\begin{equation}
    S_{\text{eff}}
    =
    - \sum_{\substack{q \ell \ell'  \\ X Y}}
    X^{(\ell)}_{-q} (\mathcal{D}^{-1}_{q})^{X Y}_{\ell \ell'} Y^{(\ell')}_{q},
\end{equation}
where the inverse propagator for pair fluctuations is
\begin{equation}
    (\mathcal{D}^{-1}_{q})_{\ell \ell'}^{X Y}
    =
    -\frac{1}{|g_{\ell}|} \delta_{X Y} \delta_{\ell \ell'}
    - \Pi^{X Y}_{q; \ell \ell'}.
\end{equation}
We provide expressions for the polarization components $\Pi^{XY}_{q; \ell \ell'}$ in terms of Fermi-surface integrals in Eqs.~\eqref{eq:polarizations}--\eqref{eq:qp_resp}. Note that $\mathcal{D}^{-1}_{q}$ is a matrix indexed by pairs $(X \ell; Y \ell')$, and the propagator $\mathcal{D}_{q}$ is its matrix inverse.

In the presence of driving by a perturbing field $d$, the fermionic action also contains an additional part $S_{\text{drive}}$, which is bilinear in fermion operators and proportional to $d$. Integrating out the fermions in the presence of driving then yields terms
\begin{equation}
    S_{\text{eff, drive}}
    =
    \sum_{q} \left[
        \Pi^{d d}_{q} d_{-q} d_{q}
        +
        \sum_{X \ell} \Pi^{d X}_{q; \ell} d_{-q} X^{(\ell)}_{q}
    \right].
\end{equation}
$\Pi^{dd}_{q}$ gives the quasiparticle response to the drive field (after analytic continuation to real frequency), and the collective mode response is obtained by integrating out the pair fluctuations: $\Pi^{dd}_{q} \rightarrow \Pi^{dd}_{q} + \delta \Pi^{dd}_{q}$, with
\begin{equation}
    \delta \Pi^{dd}_{q}
    =
    \frac{1}{4} \sum_{X Y \ell \ell'} 
    \Pi^{d X}_{-q; \ell}
    \left( \mathcal{D}_{q} \right)^{X Y}_{\ell \ell'}
    \Pi^{d Y}_{q; \ell'}.
\end{equation}
For our RTG calculation, where $d$ corresponds to a vertical electric field, the total capacitance is then proportional to the renormalized response $\Pi^{dd}_{q} + \delta \Pi^{dd}_{q}$. We also implement the Anderson-Higgs mechanism via self-consistent electrodynamic response; see Appendix \ref{appendix:anderson-higgs} for details.

Figures \ref{fig:RTG}(c) and \ref{fig:RTG}(d) show our numerical results for the absorption spectrum of superconducting RTG in the SC2 phase, accounting for the Anderson-Higgs mechanism, and working to leading order in $|\Delta| / \mu$ (hence neglecting the weak $O(|\vb{q}|^2)$ contribution discussed above, which is permitted by the substantial trigonal warping present in RTG). As in the toy model discussion, the absorption corresponds to the imaginary part of the capacitance. Figure \ref{fig:RTG}(c) shows the results for the $s$-wave case. Figure \ref{fig:RTG}(d) shows the results for the IVC-mediated $p$-wave case; the Kohn-Luttinger mechanism gives qualitatively similar results. Most importantly, the all-electronic mechanisms both show an in-gap clapping mode. While the clapping mode is visible only away from $\vb{q} = 0$, since superconducting RTG devices are comparable in size to the superconducting coherence length $\xi_{\text{BCS}}$, we expect finite size effects to render the clapping mode visible even without any special patterning of the gate electrodes.

Figures \ref{fig:RTG}(c) and \ref{fig:RTG}(d) also show that even without the collective modes, the absorption spectrum contains rich information about the gap structure. Extrema of the superconducting gap reveal themselves as sharp features in the absorption.

\section{Conclusion}
We have shown that simple linear spectroscopy via gate electrodes can be an invaluable tool for studying the pairing mechanisms in low-dimensional superconductors. Using a simple but powerful toy model, we showed how various exotic collective modes should be detectable in AC capacitance measurements. By considering the example of superconducting rhombohedral trilayer graphene (RTG) under a perpendicular displacement field, we further showed how the observation of an in-gap clapping mode could yield compelling evidence for unconventional, all-electronic pairing mechanisms. We also showed how linear spectroscopy above the quasiparticle excitation gap reveals information on the gap structure, in the form of sharp features at frequencies corresponding to gap extrema over the Fermi surface. 

The physics we have described should be accessible in  experiments. The superconducting phases of RTG~\cite{zhou2021superconductivity} require dual gating to be realized, so RTG is an ideal candidate material for the gate-based detection scheme we have proposed. These phases have transition temperatures between $\sim 40$ and $\sim 100$ mK, so one expects the gap and possible in-gap clapping modes to reside at the scale of a few GHz to 10s of GHz, easily within the range of microwave function generators. Further, we expect that similar phenomenology may occur in other quasi-two-dimensional superconductors which are sensitive to a displacement field, such as Bernal bilayer \cite{zhou2022isospin, zhang2023enhanced} and twisted trilayer \cite{chen2019signatures, park2021tunable, hao2021electric, kim2022evidence} graphene, as well as twisted bilayer WSe$_2$~\cite{guo2024superconductivity,xia2024unconventional}. Future work should study the collective mode spectroscopy of these systems.

\acknowledgments{
We thank Peter Hirschfeld, Saurabh Maiti, and Surajit Sarkar for their helpful comments on this manuscript. 
B.A.L.~was hosted at the Institute for Theoretical Physics at the University of Cologne during a considerable portion of the preparation of this manuscript, and gratefully acknowledges their generous support, as well as the financial support of the Zuckerman STEM Leadership Program. This work was supported by NSF-BSF Award No.~DMR-2310312, by the European Union’s Horizon 2020 research and innovation programme (Grant Agreements LEGOTOP No.~788715 and HQMAT No.~817799), and the DFG CRC Grant No.~SFB/TRR183. M.R.~gratefully acknowledges the Brown Institute for Basic Sciences, administered by Caltech and established by Ross M.~Brown, the University of Washington College of Arts and Sciences, and the Kenneth K.~Young Memorial Professorship for support. I.I.~acknowledges the support of the Natural Sciences and Engineering Research Council of Canada (NSERC) [Discovery Grant No.~2024-05599].
}

\appendix
\section{Toy model for the Leggett mode} \label{appendix:toy_leggett}
In this Appendix, we consider a minimal model containing a Leggett mode,
\begin{multline}
    S[\overline{c}, c]
    =
    S_0 + S_{\rm int}
    \\
    =
    \sum_{k \alpha \sigma}
    \overline{c}_{k \alpha \sigma}
    \left[ -i \omega_n + \xi_{\vb{k} \alpha} \right] 
    c_{k \alpha \sigma}
    \\
    +
    \frac{T}{L^2} \sum_{\substack{k k' q \\ \alpha \alpha'}}
    g_{\alpha \alpha'}
    \overline{c}_{k \alpha \uparrow}
    \overline{c}_{-k+q, \alpha \downarrow}
    c_{-k'+q, \alpha' \downarrow}
    c_{k' \alpha' \uparrow}.
\end{multline}
This action corresponds to the toy model given in the main text, with $g^{(p)} = 0$. $L$ is the linear size of the periodic system. $k$ denotes the three-vector $(i \omega_n, \vb{k})$, with $\omega_n = (2 n + 1) \pi T$ a fermionic Matsubara frequency at temperature $T$ (analogously for $k'$). $q$ is a similar three-vector, but with a bosonic frequency component $\Omega_m = 2 \pi m T$. $\alpha \in \lbrace 1, 2 \rbrace$ labels the two bands with dispersion relations given by $\xi_{\vb{k} \alpha}$ relative to the Fermi level. $\sigma \in \lbrace \uparrow, \downarrow \rbrace$ labels the spin. We assume attractive intraband interactions $g_{\alpha \alpha} < 0$. Since the single-particle action $S_0$ conserves charge in each band separately, any phase factor on $g_{12} = g_{21}^*$ may be freely gauged away, so we choose $g_{12} = g_{21} \le 0$ without loss of generality. This model was studied by Leggett \cite{leggett1966number}; we present an analysis in modern language to set the stage for the more technically involved, but conceptually analogous case with momentum-dependent interactions. The Gaussian pair-fluctuation (GPF) approach we apply is standard \cite{sharapov2002effective}, so we skip most details and present only the important results.

\subsection{Mean-field theory}
The GPF approach is a leading-order expansion around mean-field theory, and begins with a standard Hubbard-Stratonovich decoupling in the Cooper channel. Denoting the corresponding complex auxiliary bosons by $\Delta_{\alpha}$ and introducing the Nambu spinors $\Psi_{k \alpha} = \begin{pmatrix} c_{k, \alpha \uparrow} & \overline{c}_{-k, \alpha \downarrow} \end{pmatrix}^T$, the resulting action is
\begin{multline}
    S [\Psi^{\dag}, \Psi, \overline{\Delta}, \Delta]
    =
    - \frac{L^2}{T} \sum_{q}
    \begin{pmatrix}
        \overline{\Delta}_{q 1} & \overline{\Delta}_{q 2}
    \end{pmatrix} \hat{g}^{-1} \begin{pmatrix}
        \Delta_{q 1} \\
        \Delta_{q 2}
    \end{pmatrix}
    \\
    - \sum_{\alpha k q}
    \Psi^{\dag}_{\alpha} \mathds{G}^{-1}_{\alpha} \Psi_{\alpha}.
\end{multline}
The inverse fermion propagator is
\begin{equation}
    \mathds{G}^{-1}_{\alpha; k, k-q}
    =
    \begin{pmatrix}
        (i \omega_n - \xi_{\vb{k} \alpha}) \delta_{q, 0}
        & \Delta_{q, \alpha} \\
        \overline{\Delta}_{-q, \alpha}
        & (i \omega_n + \xi_{\vb{k} \alpha}) \delta_{q, 0}
    \end{pmatrix}.
\end{equation}
The couplings are grouped into the matrix
\begin{equation}
    \hat{g} = \begin{pmatrix}
        g_{11} & g_{12} \\
        g_{12} & g_{22}
    \end{pmatrix}.
\end{equation}
We assume that $g_{\alpha \alpha} < 0$ and $\det \hat{g} > 0$ such that both eigenvalues are positive, i.e.,~the interactions are purely attractive. A repulsive eigenvalue would slightly modify the Hubbard-Stratonovich transformation; see Appendix \ref{appendix:general_case}. Note that for convenience, we use different Fourier-transform conventions for the fermions versus the pairing fields. We use the unitary normalization for the fermions,
\begin{subequations}
\begin{equation}
    \psi_{\alpha} (\tau, \vb{r})
    =
    \sqrt{\frac{T}{L^2}} 
    \sum_{k} e^{i (\vb{k} \vdot \vb{r} - \omega_n \tau)} c_{k \alpha}
\end{equation}
and
\begin{equation}
    c_{k \alpha}
    =
    \sqrt{\frac{T}{L^2}}
    \int \dd{\tau} \dd^2{\vb{r}} \,
    e^{-i (\vb{k} \vdot \vb{r} - \omega_n \tau)} \psi_{\alpha} (\tau, \vb{r}),
\end{equation}
\end{subequations}
but the nonunitary normalization for the pairing fields,
\begin{subequations}
\begin{equation}
    \Delta (\tau, \vb{r})
    =
    \sum_{q} e^{i (\vb{q} \vdot \vb{r} - \Omega_m \tau)} \Delta_{q}
\end{equation}
and
\begin{equation}
    \Delta_{q}
    =
    \frac{T}{L^2} \int \dd{\tau} \dd^2{\vb{r}} \,
    e^{-i (\vb{q} \vdot \vb{r} - \Omega_m \tau)} \Delta (\tau, \vb{r}).
\end{equation}
\end{subequations}
This choice is made so that in the momentum representation, the pairing terms look like $\Delta c^{\dag} c^{\dag} + \text{H.c.}$ without any normalization factor.

Integrating out the fermion yields the famous ``$\Tr \log$'' action for the HS fields,
\begin{multline}    \label{eq:toymodel_trlog}
    S [\overline{\Delta}, \Delta]
    =
    - \frac{L^2}{T} \sum_{q} \begin{pmatrix}
        \overline{\Delta}_{q 1} & \overline{\Delta}_{q 2}
    \end{pmatrix} \hat{g}^{-1} \begin{pmatrix}
        \Delta_{q 1} \\
        \Delta_{q 2}
    \end{pmatrix}
    \\
    -
    \Tr \log \mathds{G}^{-1}.
\end{multline}
The capital-T trace ($\Tr$) includes three-momentum indices, while the lowercase-t trace ($\tr$) used below is an ordinary matrix trace over Nambu indices only. The mean-field ground state is the saddle point $\delta S / \delta \overline{\Delta}_{\alpha} = 0$, corresponding to the coupled gap equations
\begin{equation}
    \Delta_{\alpha}
    =
    - \sum_{\alpha'} \int \frac{\dd^2 \vb{k'}}{(2 \pi)^2} \,
    g_{\alpha \alpha'} \tanh \left( \frac{E_{\vb{k'} \alpha'}}{2 T} \right)
    \frac{\Delta_{\alpha'}}{2 E_{\vb{k'} \alpha'}},
\end{equation}
or
\begin{subequations}
    \begin{multline}
        \frac{g_{22} \Delta_1 - g_{12} \Delta_2}{\det \hat{g}}
        \\
        =
        -\int \frac{\dd^2{\vb{k}}}{(2\pi)^2}
        \tanh \left( \frac{E_{\vb{k} 1}}{2 T} \right)
        \frac{\Delta_1}{2 E_{\vb{k} 1}}
    \end{multline}
    and
    \begin{multline}
        \frac{-g_{12} \Delta_1 + g_{11} \Delta_2}{\det \hat{g}}
        \\
        =
        -\int \frac{\dd^2{\vb{k}}}{(2\pi)^2}
        \tanh \left( \frac{E_{\vb{k} 2}}{2 T} \right)
        \frac{\Delta_2}{2 E_{\vb{k} 2}}.
    \end{multline}
\end{subequations}
The quasiparticle energies are $E_{\vb{k} \alpha} = \sqrt{\xi_{\vb{k} \alpha}^2 + |\Delta_{\alpha}|^2}$. Note that the solutions $\Delta_{\alpha}$ must be either in phase, or out of phase by $\pi$. So, we choose both to be real. Now we trade the momentum integrals for energy integrals via the densities of states $\nu_{\alpha}$, and introduce an energy cutoff $\Lambda$:
\begin{equation}
    \int \frac{\dd^2 \vb{k}}{(2 \pi)^2}
    \rightarrow \int_{-\Lambda}^{\Lambda} \dd{\xi} \nu_{\alpha}
    \int_{0}^{2 \pi} \frac{\dd{\varphi}}{2 \pi}.
\end{equation}
Then we can integrate the gap equations at $T \rightarrow 0$ as standard, yielding (for $\Lambda \gg |\Delta_{\alpha}|$)
\begin{subequations}
\begin{equation} \label{eq:toy_model_integrated_gap_equations}
    \frac{g_{22} \Delta_1 - g_{12} \Delta_2}{\det \hat{g}}
    =
    - \nu_1 \Delta_1 \log \left( 
        \frac{2 \Lambda}{|\Delta_1|} 
    \right)
\end{equation}
and
\begin{equation}
    \frac{-g_{12} \Delta_1 + g_{11} \Delta_2}{\det \hat{g}}
    =
    - \nu_2 \Delta_2 \log \left( 
        \frac{2 \Lambda}{|\Delta_2|} 
    \right).
\end{equation}

\end{subequations}

\subsection{Collective modes}
We express both pairing fields in terms of fluctuations $\phi$ around their mean-field values,
\begin{equation}
    \Delta_{q \alpha} = \Delta_{\alpha} \delta_{q, 0} + \phi_{q \alpha}.
\end{equation}
Note that from now on, we are using $\Delta_{\alpha}$ to refer to the mean-field gap for each band: all $q$-dependence is in the fluctuations $\phi$. We write $\mathds{G}^{-1}$ in terms of the mean-field inverse propagator $\mathcal{G}^{-1}$ and fluctuation vertex $\mathds{V}^{(\phi)}$ as $\mathds{G}^{-1}_{\alpha; k, k-q} = \mathcal{G}^{-1}_{k \alpha} \delta_{q, 0} - \mathds{V}^{(\phi)}_{\alpha; k, k-q}$:
\begin{multline}
    \mathcal{G}^{-1}_{k \alpha}
    =
    \begin{pmatrix}
        i \omega_n - \xi_{\vb{k} \alpha}
        & \Delta_{\alpha} \\
        \overline{\Delta}_{\alpha}
        & i \omega_n + \xi_{\vb{k} \alpha}
    \end{pmatrix}
    \\
    =
    \left[
        \frac{1}{(i \omega_n)^2 - E_{\vb{k} \alpha}^2}
        \begin{pmatrix}
            i \omega_n + \xi_{\vb{k} \alpha} & - \Delta_{\alpha} \\
            - \overline{\Delta}_{\alpha} & i \omega_n - \xi_{\vb{k} \alpha}
        \end{pmatrix}
    \right]^{-1}
    \\
    =
    \begin{pmatrix}
        G_{k \alpha} & F_{k \alpha} \\
        \overline{F}_{k \alpha} & -G_{-k, \alpha}
    \end{pmatrix}^{-1}
\end{multline}
and
\begin{equation}
    \mathds{V}^{(\phi)}_{\alpha; k, k-q}
    =
    \begin{pmatrix}
        0 & -\phi_{q \alpha} \\
        -\overline{\phi}_{-q, \alpha} & 0
    \end{pmatrix}.
\end{equation}

\begin{widetext}
The GPF action is the Gaussian approximation to the $\Tr \log$ action in Eq.~\eqref{eq:toymodel_trlog}:
\begin{multline}
    S [\overline{\phi}, \phi]
    =
    - \frac{L^2}{T} \sum_{q} \begin{pmatrix}
        \overline{\Delta}_{1} + \overline{\phi}_{q 1} & \overline{\Delta}_{2} + \overline{\phi}_{q 2} 
    \end{pmatrix} \hat{g}^{-1}
    \begin{pmatrix}
        \Delta_{1} + \phi_{q 1} \\
        \Delta_{2} + \phi_{q 2}
    \end{pmatrix}
    -
    \Tr \log \left[ 1 - \mathcal{G} \mathds{V}^{(\phi)} \right]
    \\
    \approx
    - \frac{L^2}{T} \sum_{q} \begin{pmatrix}
        \overline{\phi}_{q 1} & \overline{\phi}_{q 2} 
    \end{pmatrix} \hat{g}^{-1}
    \begin{pmatrix}
        \phi_{q 1} \\
        \phi_{q 2}
    \end{pmatrix}
    +
    \frac{1}{2} \Tr \left[
        \mathcal{G} \mathds{V}^{(\phi)} \mathcal{G} \mathds{V}^{(\phi)}
    \right]
    \\
    =
    - \frac{L^2}{T} \sum_{q} \begin{pmatrix}
        \overline{\phi}_{q 1} & \overline{\phi}_{q 2} 
    \end{pmatrix} \hat{g}^{-1}
    \begin{pmatrix}
        \phi_{q 1} \\
        \phi_{q 2}
    \end{pmatrix}
    +
    \frac{1}{2} \sum_{k q \alpha} \tr \left[
        \mathcal{G}_{k \alpha} \mathds{V}^{(\phi)}_{\alpha; k, k-q} 
        \mathcal{G}_{k-q, \alpha} \mathds{V}^{(\phi)}_{\alpha; k-q, k}
    \right]
    \\
    \equiv
    \frac{L^2}{T} \sum_{q} \begin{pmatrix}
        \mathcal{R}_{-q, 1}
        & \mathcal{R}_{-q, 2}
        & \mathcal{I}_{-q, 1}
        & \mathcal{I}_{-q, 2}
    \end{pmatrix}
    \mathcal{D}^{-1}_{q}
    \begin{pmatrix}
        \mathcal{R}_{q 1} \\
        \mathcal{R}_{q 2} \\
        \mathcal{I}_{q 1} \\
        \mathcal{I}_{q 2}
    \end{pmatrix}.
\end{multline}
Note that the linear terms necessarily cancel at the saddle point. Also note that we dropped a constant term $\sim \Delta^{\dag} g^{-1} \Delta$. In the last line we expressed the complex fluctuations in terms of their real and imaginary parts (describing amplitude and phase fluctuations respectively, since $\Delta_{\alpha} \in \mathds{R}$),
\begin{equation}
    \phi_{q \alpha} 
    = 
    \mathcal{R}_{q \alpha} + i \mathcal{I}_{q \alpha},
\end{equation}
and introduced the bosonic inverse propagator $\mathcal{D}^{-1}_{q}$ for Gaussian pair fluctuations:
\begin{equation}
    \mathcal{D}^{-1}_{\vb{q} \rightarrow 0} (i \Omega_m)
    =
    \begin{pmatrix}
        \tilde{g}^{-1} \frac{\Delta_2}{\Delta_1}
        + \nu_1 \gamma_{1} \cot \gamma_{1}
        &
        - \tilde{g}^{-1}
        & 0 & 0 \\
        - \tilde{g}^{-1}
        & \tilde{g}^{-1} \frac{\Delta_1}{\Delta_2}
        + \nu_2 \gamma_{2} \cot \gamma_{2}
        & 0 & 0 \\
        0 & 0
        & \tilde{g}^{-1} \frac{\Delta_2}{\Delta_1}
        - \nu_1 \gamma_{1} \tan \gamma_{1}
        & -\tilde{g}^{-1}    \\
        0 & 0
        & -\tilde{g}^{-1}
        & \tilde{g}^{-1} \frac{\Delta_1}{\Delta_2}
        - \nu_2 \gamma_{2} \tan \gamma_{2}
    \end{pmatrix}.
\end{equation}
\end{widetext}
Here
\begin{equation}
    \tilde{g}^{-1} = \frac{-g_{12}}{\det \hat{g}}
\end{equation}
(recall that $g_{\alpha \alpha'} \le 0$), and the variables $\gamma_{\alpha}$ are defined by
\begin{equation}
    \sin \gamma_{\alpha}
    =
    \frac{i \Omega_m}{2 |\Delta_{\alpha}|}.
\end{equation}
Along the way we took the $L \rightarrow \infty$ and $T \rightarrow 0$ limits, and used the gap equations Eq.~\eqref{eq:toy_model_integrated_gap_equations} to simplify some intermediate expressions. We also focus on the $\vb{q} \rightarrow 0$ case for simplicity. The necessary integrals are then
\begin{multline} \label{eq:amplitude_polarization}
    \Pi^{\mathcal{R R}}_{q \alpha} 
    =
    T \sum_{i \omega_n}
    \int \frac{\dd^{2} \vb{k}}{(2 \pi)^2}
    \bigg(
        - \frac{1}{2} G_{k \alpha} 
        (G_{-k+q, \alpha} + G_{-k-q, \alpha})
        \\
        + F_{k \alpha} F_{k-q, \alpha}
    \bigg)
    \\
    =
    \int \frac{\dd^2 \vb{k}}{(2 \pi)^2}
    \tanh \left( \frac{E_{\vb{k} \alpha}}{2 T} \right)
    \frac{1}{E_{\vb{k} \alpha}}
    \frac{2 \xi_{\vb{k} \alpha}^2 }{
        (i \Omega_m)^2 - 4 E_{\vb{k} \alpha}^2
    }
    \\
    \approx
    \int_{-\Lambda}^{\Lambda} \dd{\xi} \nu_{\alpha}
    \frac{1}{\sqrt{\xi^2 + \Delta_{\alpha}^2}}
    \frac{2 \xi^2}{
        (i \Omega_m)^2 - 4 \xi^2 - 4 \Delta_{\alpha}^2
    }
    \\
    \approx 
    \nu_{\alpha} \gamma_{\alpha} \cot \gamma_{\alpha}
    - \nu_{\alpha} \log \left( 
        \frac{2 \Lambda}{|\Delta_{\alpha}|} 
    \right),
\end{multline}
\begin{multline} \label{eq:phase_polarization}
    \Pi^{\mathcal{I I}}_{q \alpha}
    =
    T \sum_{i \omega_n}
    \int \frac{\dd^{2} \vb{k}}{(2 \pi)^2}
    \bigg(
        -\frac{1}{2} G_{k \alpha} 
        (G_{-k+q, \alpha} + G_{-k-q, \alpha})
        \\
        - F_{k \alpha} F_{k-q, \alpha}
    \bigg)
    \\
    =
    \int \frac{\dd^2 \vb{k}}{(2 \pi)^2}
    \tanh \left( \frac{E_{\vb{k} \alpha}}{2 T} \right)
    \frac{2 E_{\vb{k} \alpha}}{
        (i \Omega_m)^2 - 4 E_{\vb{k} \alpha}^2
    }
    \\
    \approx
    \int_{-\Lambda}^{\Lambda} \dd{\xi} \nu_{\alpha}
    \frac{2 \sqrt{\xi^2 + \Delta_{\alpha}^2}}{
        (i \Omega_m)^2 - 4 \xi^2 - 4 \Delta_{\alpha}^2
    }
    \\
    \approx 
    - \nu_{\alpha} \gamma_{\alpha} \tan \gamma_{\alpha}
    - \nu_{\alpha} \log \left( 
        \frac{2 \Lambda}{|\Delta_{\alpha}|} 
    \right),
\end{multline}
and
\begin{multline}
    \Pi^{\mathcal{R I}}_{q \alpha}
    =
    \frac{1}{2i} T \sum_{i \omega_n}
    \int \frac{\dd^2 \vb{k}}{(2 \pi)^2}
    \left(
        G_{k \alpha} G_{-k+q, \alpha}
        -
        G_{k \alpha} G_{-k-q, \alpha}
    \right)
    \\
    =
    \int \frac{\dd^2{\vb{k}}}{(2 \pi)^2}
    \tanh \left( \frac{E_{\vb{k} \alpha}}{2 T} \right)
    \frac{1}{E_{\vb{k} \alpha}}
    \frac{\xi_{\vb{k} \alpha} \Omega_m}{
        4 E_{\vb{k} \alpha}^2 - (i \Omega_m)^2
    }
    \\
    \approx
    \int_{-\Lambda}^{\Lambda} \dd{\xi} \nu_{\alpha}
    \frac{1}{\sqrt{\xi^2 + \Delta_{\alpha}^2}}
    \frac{\xi \Omega_m}{
        4 \xi^2 + 4 \Delta_{\alpha}^2 - (i \Omega_m)^2
    }
    =
    0.
\end{multline}
For Eqns.~\eqref{eq:amplitude_polarization} and \eqref{eq:phase_polarization}, the final approximation uses $\Lambda \gg |\Delta_{\alpha}|$. The cutoff can be eliminated from the expressions by using the integrated gap equations \eqref{eq:toy_model_integrated_gap_equations}.

Collective modes correspond to poles of the bosonic propagator $\mathcal{D}_{q}$, i.e.,~zeros of $\det \mathcal{D}^{-1}_{q}$, after analytically continuing $i \Omega_m \rightarrow \Omega + i 0^+$. Inside the gap ($|\Omega| < 2 |\Delta|$) the infinitesimal $0^+ > 0$ is immaterial, but outside the gap the branch cuts must be chosen with care to ensure analyticity in the upper half-plane. The amplitude modes solve
\begin{subequations} \label{eq:twoband_model_modes}
    \begin{multline} \label{eq:amplitude_modes}
        \frac{\tilde{g}^{-1}}{\Delta_1 \Delta_2}
        \left(
            \Delta_1^2 \nu_1 \gamma_{1} \cot \gamma_{1}
            +\Delta_2^2 \nu_2 \gamma_{2} \cot \gamma_{2}
        \right)
        \\ +
        \nu_1 \gamma_{1} \cot \gamma_{1}
        \nu_2 \gamma_{2} \cot \gamma_{2}
        =
        0,
    \end{multline}
and the phase modes (ABG and Leggett) solve
    \begin{multline} \label{eq:toy_phase_modes}
        \frac{\tilde{g}^{-1}}{\Delta_1 \Delta_2}
        \left(
            \Delta_1^2 \nu_1 \gamma_{1} \tan \gamma_{1}
            +\Delta_2^2 \nu_2 \gamma_{2} \tan \gamma_{2}
        \right)
        \\ -
        \nu_1 \gamma_{1} \tan \gamma_{1}
        \nu_2 \gamma_{2} \tan \gamma_{2}
        =
        0.
    \end{multline}
\end{subequations}
Equations \eqref{eq:twoband_model_modes} are transcendental, so in general we cannot solve them analytically. Instead we consider some instructive limits, focusing on the phase modes. Recall that we assumed $\det \hat{g} > 0$, so $|g_{12}| < \sqrt{g_{11} g_{22}}$. When $g_{12} \rightarrow 0$, $\tilde{g}^{-1} \rightarrow 0$, and the only solution of Eq.~\eqref{eq:toy_phase_modes} is $\gamma_{q, \alpha} = 0$, corresponding to $\Omega = 0$. This is to be expected: when the two condensates are completely uncoupled, they each have an independent ABG mode at $\Omega = 0$. Restoring finite $g_{12}$, now consider the case of completely symmetric bands, where $\nu_1 = \nu_2$, $g_{11} = g_{22} = g_0$, and hence $\Delta_1 = \Delta_2$ and $\gamma_{1} = \gamma_{2} = \gamma$. In this case, there is an in-gap solution (meaning $\gamma \in \mathds{R}$) whenever $|g_{12}|^2  < |g_0|$ (recall that $g_0 < 0$ by assumption), i.e.,~when there are two negative eigenvalues of $\hat{g}$. This solution (the Leggett mode) gets pushed up to $2 |\Delta|$ when $g_{12} \rightarrow g_{0}$, which is when the second eigenvalue of $\hat{g}$ approaches zero. 

\begin{figure}[!h]
\centering
\includegraphics[width=0.5\textwidth]{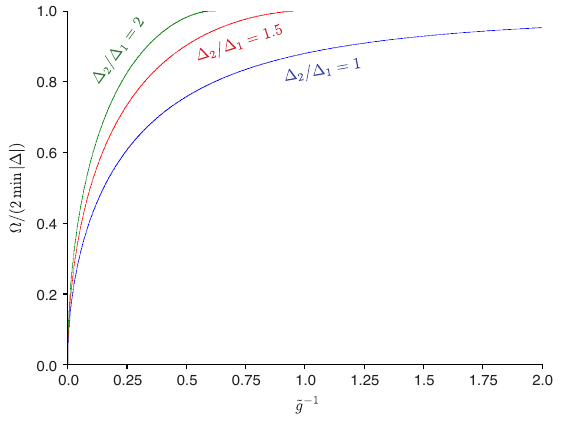}
\caption{\label{fig:bounds} Dependence of the Leggett mode frequency on $\tilde{g}^{-1}$ for three values of the gap ratio $\Delta_2/\Delta_1$ given $\nu_1 = \nu_2$, obtained by numerically solving Eq.~\eqref{eq:toy_phase_modes}. Note that in each case there is a critical value of $\tilde{g}^{-1}$ beyond which there are no in-gap modes, other than the ABG mode at Omega = 0.}
\end{figure}

The situation changes when the gaps are not equal. Here we assume $\Delta_1 < \Delta_2$ without loss of generality. Again, Eq.~\eqref{eq:toy_phase_modes} cannot be solved analytically. However, since the left-hand side (LHS) consists only of continuous functions for $|\Omega| < 2 
|\Delta_1|$, we can determine the \textit{parity} of the number of in-gap solutions by comparing the sign of the LHS just above $\Omega = 0$ to the sign just below $2 
|\Delta_1|$. If these signs are opposite, then there must be an odd number of zero crossings (i.e., phase modes) between $\Omega = 0$ and $\Omega = 2 |\Delta_1|$, not including the global phase mode which always lies at $\Omega = 0$. If the sign just above $0$ equals the sign just below $2 |\Delta_1|$, then there must be an even number of such modes. Expanding the LHS of Eq.~\eqref{eq:toy_phase_modes} just above $\Omega = 0$ and just below $\Omega = 2 |\Delta_1|$, one finds that the signs are opposite when and only when $\tilde{g}^{-1}$ is smaller than a critical value of $\tilde{g}^{-1}_{\text{crit}} = \nu_2 \arcsin(\Delta_1/\Delta_2) / \sqrt{1 - (\Delta_1 / \Delta_2)^2}$. Therefore, for $\tilde{g}^{-1} > \tilde{g}^{-1}_{\text{crit}}$, Eq.~\eqref{eq:toy_phase_modes} has either no nonzero solutions, or an even number of them. Physically, there is no clear reason for a third phase mode (besides the Leggett and ABG modes) to arise when $\tilde{g}^{-1}$ crosses the critical value, so we expect that in this regime there are no in-gap modes other than the ABG mode. In Fig.~\ref{fig:bounds} we show that there are indeed no nonzero solutions to Eq.~\eqref{eq:toy_phase_modes} in this regime. We show the dependence of zeros (corresponding to the Leggett mode) as a function of $\tilde{g}^{-1}$ for different values of $\Delta_2/\Delta_1$. We see that for unequal gap widths, there exists a critical $\tilde{g}^{-1}_{\text{crit}}$ such that for $\tilde{g}^{-1} > \tilde{g}^{-1}_{\text{crit}}$ there are no modes in between $\Omega = 0$ and the gap edge.

Assuming that $|g_{12}|^2 \ll g_{11} g_{22}$, we can expand Eq.~\eqref{eq:toy_phase_modes} in $\Omega$, yielding
\begin{multline}
    \Omega^2
    \frac{\tilde{g}^{-1} (\nu_1 + \nu_2)}{4}
    \\ -
    \frac{\Omega^4}{48 \Delta_1^2 \Delta_2^2}
    \left[
        3 \nu_1 \Delta_1 \nu_2 \Delta_2
        -
        2 \tilde{g}^{-1} 
        (
            \nu_1 \Delta_2^2
            + \nu_2 \Delta_1^2
        )
    \right]
    \\
    = 0.
\end{multline}
The nontrivial solution is
\begin{equation}    \label{eq:approximate_Leggett_freq}
    \Omega_{\text{L}}^2
    =
    \frac{
        12 \tilde{g}^{-1} (\nu_1 + \nu_2) \Delta_1^2 \Delta_2^2
    }{
        3 \nu_1 \Delta_1 \nu_2 \Delta_2
        -
        2 \tilde{g}^{-1} (\nu_1 \Delta_2^2 + \nu_2 \Delta_1^2)
    }.
\end{equation}

\subsection{Driving the Leggett mode}
\label{appendixsec:legget_driving}
With the collective modes identified, all that remains is to determine their response to driving fields. The basic idea is that a drive term $\sim \delta V (t) c^{\dag} c$, upon integrating out the fermions, will yield a source term $\sim \delta V \phi$ for the Hubbard-Stratonovich fluctuations. Integrating out those fluctuations then yields a quadratic free-energy functional $\mathcal{F} [\delta V] = - T \log \mathcal{Z} [\delta V]$, in which the coefficient of $\delta V^2$ is the capacitance. $\mathcal{Z}$ is the partition function, $\mathcal{Z} [\delta V] = \int \mathcal{D}[\Psi^{\dag}, \Psi, \overline{\Delta}, \Delta] e^{-S}$.

As discussed in the main text, the driving Hamiltonian is
\begin{equation}
    H_{\delta V}
    =
    \sum_{\vb{k}} u \, \delta V(t)
    \begin{pmatrix}
        c^{\dag}_{1, \vb{k}} & c^{\dag}_{2, \vb{k}}
    \end{pmatrix}
    \begin{pmatrix}
        1 & 0 \\
        0 & -1
    \end{pmatrix}
    \begin{pmatrix}
        c_{1, \vb{k}}   \\
        c_{2, \vb{k}}
    \end{pmatrix},
\end{equation}
with $u = V_{\rm DC} / \sqrt{V_{\rm DC}^2 + J^2}$ in the bilayer model. Wick rotating to imaginary time and moving to the Nambu representation, the corresponding action term is
\begin{multline}
    S_{\delta V} [\Psi^{\dag}, \Psi]
    =
    u \sum_{k q} \delta V_q \bigg\lbrace
        \Psi^{\dag}_{k 1} \begin{pmatrix}
            1 & 0 \\
            0 & -1
        \end{pmatrix} \Psi_{k-q, 1}
        \\
        -
        \Psi^{\dag}_{k 2} \begin{pmatrix}
            1 & 0 \\
            0 & -1
        \end{pmatrix} \Psi_{k-q, 2}
    \bigg\rbrace.
\end{multline}
Here we assumed that $\delta V(t) = \delta V (-t)$ for simplicity, such that $\delta V_q = \delta V_{-q}$ (since $\delta V (t)$ is real). This contributes an additional vertex to the full fermion inverse propagator $\mathds{G}^{-1} = \mathcal{G}^{-1} - \mathds{V}^{(\phi)} - \mathds{V}^{(\delta V)}$,
\begin{equation}
    \mathds{V}^{(\delta V)}_{\alpha; k, k-q}
    =
    u \delta V_{q} (-1)^{\alpha} \begin{pmatrix}
        1 & 0 \\
        0 & -1
    \end{pmatrix}.
\end{equation}

Integrating out the fermion and expanding the $\Tr \log$ out to second order in all vertices, two new terms result. The $\delta V^2$ term,
\begin{equation}
    \frac{1}{T} \mathcal{F}_{\rm QP} [ \delta V ]
    =
    \frac{1}{2} \Tr \left[ 
        \mathcal{G} \mathds{V}^{(\delta V)}
        \mathcal{G} \mathds{V}^{(\delta V)}
    \right],
\end{equation}
corresponds to the contribution of quasiparticles to the free-energy functional for $\delta V$, while the $\phi \delta V$ term,
\begin{equation}
    S_{\rm drive} [\overline{\phi}, \phi, \delta V]
    =
    \Tr \left[
        \mathcal{G} \mathds{V}^{(\phi)}
        \mathcal{G} \mathds{V}^{(\delta V)}
    \right],
\end{equation}
corresponds to driving of the collective modes. It is useful to immediately go to amplitude and phase variables, which amounts to expanding
\begin{multline}
    V^{(\phi)}_{\alpha; k, k-q}
    =
    \begin{pmatrix}
        0 & -\phi_{q \alpha}    \\
        - \overline{\phi}_{-q, \alpha} & 0
    \end{pmatrix}
    \\
    =
    - \begin{pmatrix}
        0 & \mathcal{R}_{q \alpha} + i \mathcal{I}_{q \alpha} \\
        \mathcal{R}_{q \alpha} - i \mathcal{I}_{q \alpha} & 0
    \end{pmatrix}
    \\
    \equiv
    \mathds{V}^{(\mathcal{R})}_{\alpha; k, k-q} 
    + 
    \mathds{V}^{(\mathcal{I})}_{\alpha; k, k-q}.
\end{multline}

One finds
\begin{multline}
    \Tr \left[
        \mathcal{G} \mathds{V}^{(\mathcal{R})}
        \mathcal{G} \mathds{V}^{(\delta V)}
    \right]
    \\
    =
    \sum_{k q \alpha} 2 (-1)^{\alpha + 1} u
    G_{k \alpha} \left\lbrace
        F_{k+q, \alpha} + F_{k-q, \alpha}
    \right\rbrace
    \delta V_{-q} \mathcal{R}_{q \alpha}
\end{multline}
and
\begin{multline}
    \Tr \left[
        \mathcal{G} \mathds{V}^{(\mathcal{I})}
        \mathcal{G} \mathds{V}^{(\delta V)}
    \right]
    \\
    =
    \sum_{k q \alpha} 2 i (-1)^{\alpha} u
    G_{k \alpha} \left\lbrace
        F_{k+q, \alpha} - F_{k-q, \alpha}
    \right\rbrace
    \delta V_{-q} \mathcal{I}_{q \alpha}.
\end{multline}
We again focus on $\vb{q} \rightarrow 0$ for simplicity (in keeping with the form chosen for $H_{\delta V}$); $q = (i \Omega_m, 0)$ is then just a notationally convenient way of writing the Matsubara frequency. The amplitude sector decouples from the drive:
\begin{multline}
    \frac{T}{L^2} \sum_{k} 
    G_{k \alpha} (F_{k+q, \alpha} + F_{k-q, \alpha})
    \\
    \xrightarrow{L \rightarrow \infty} 
    \int \frac{\dd^2 \vb{k}}{(2 \pi)^2}
    \tanh \left( \frac{ E_{\vb{k} \alpha}}{2 T} \right)
    \frac{1}{E_{\vb{k} \alpha}}
    \frac{2 \Delta_{\alpha} \xi_{\vb{k} \alpha}}{
        (i \Omega_m)^2 - 4 E_{\vb{k} \alpha}^2
    }
    \\
    \approx
    \int_{-\Lambda}^{\Lambda} \dd{\xi} \nu_{\alpha}
    \frac{1}{\sqrt{\xi^2 + \Delta_{\alpha}^2}}
    \frac{2 \Delta_{\alpha} \xi}{
         (i \Omega_m)^2 - 4 \xi^2 - 4 \Delta_{\alpha}^2
    }
    \\
    = 0.
\end{multline}
The phase sector feels the drive:
\begin{multline}
    i \frac{T}{L^2} \sum_{k} 
    G_{k \alpha} (F_{k+q, \alpha} - F_{k-q, \alpha})
    \\
    \xrightarrow{L \rightarrow \infty} 
    \int \frac{\dd^2 \vb{k}}{(2 \pi)^2}
    \tanh \left( \frac{ E_{\vb{k} \alpha}}{2 T} \right)
    \frac{1}{E_{\vb{k} \alpha}}
    \frac{\Delta_{\alpha} \Omega_m}{
        (i \Omega_m)^2 - 4 E_{\vb{k} \alpha}^2
    }
    \\
    \approx
    \int_{-\Lambda}^{\Lambda} \dd{\xi} \nu_{\alpha}
    \frac{1}{\sqrt{\xi^2 + \Delta_{\alpha}^2}}
    \frac{\Delta_{\alpha} \Omega_m}{
        (i \Omega_m)^2 - 4 \xi^2 - 4 \Delta_{\alpha}^2
    }
    \\
    =
    \nu_{\alpha} \frac{i \Omega_m}{|\Omega_m|}
    \sgn (\Delta_{\alpha}) \gamma_{\alpha} \sec \gamma_{\alpha},
\end{multline}
where we took $\Lambda \rightarrow \infty$ in the last equality.

Neglecting the amplitude modes (since they aren't affected by $\delta V$), including driving, the action for the phase modes at $\vb{q} \rightarrow 0$ is
\begin{multline}
    S_{\rm eff} [\mathcal{I}, \delta V]
    \\
    = 
    \frac{L^2}{T} \sum_{q}
    \bigg\lbrace
        \begin{pmatrix}
            \mathcal{I}_{-q, 1} & \mathcal{I}_{-q, 2}
        \end{pmatrix} \mathcal{D}^{-1}_{\mathcal{I}, \vb{q} \rightarrow 0} (i \Omega_m)
        \begin{pmatrix}
            \mathcal{I}_{q 1} \\
            \mathcal{I}_{q 2}
        \end{pmatrix}
        \\
        + \begin{pmatrix}
            j_{-q, 1} & j_{-q, 2}
        \end{pmatrix} \begin{pmatrix}
            \mathcal{I}_{q 1} \\
            \mathcal{I}_{q 2}
        \end{pmatrix}
        \\
        + \begin{pmatrix}
            \mathcal{I}_{-q, 1} 
            & \mathcal{I}_{-q, 2}
        \end{pmatrix}
        \begin{pmatrix}
            j_{q 1} \\
            j_{q 2}
        \end{pmatrix} 
    \bigg\rbrace,
\end{multline}
where the sources are
\begin{subequations}
\begin{equation}
    j_{-q, 1} 
    = 
    u \delta V_{-q} 
    \sgn(\Omega_m) \sgn(\Delta_1) \nu_1 
    \gamma_{1} \sec \gamma_{1},
\end{equation}
\begin{equation}
    j_{-q, 2}
    = 
    - u \delta V_{-q} 
    \sgn(\Omega_m) \sgn(\Delta_2) \nu_2
    \gamma_{2} \sec \gamma_{2},
\end{equation}
\begin{equation}
    j_{q 1}
    =
    u \delta V_{q} 
    \sgn(\Omega_m) \sgn (\Delta_1)
    \nu_1 \gamma_{1} \sec \gamma_{1},
\end{equation}
and
\begin{equation}
    j_{q 2}
    =
    -u \delta V_{q} 
    \sgn(\Omega_m) \sgn (\Delta_2)
    \nu_2 \gamma_{q, 2} \sec \gamma_{q, 2}.
\end{equation}
\end{subequations}

Performing the Gaussian integration over $\mathcal{I}_{\alpha}$, we obtain the collective-mode contribution to the free energy functional:
\begin{widetext}
\begin{multline}
    \frac{1}{T} \mathcal{F}_{\phi} [\delta V]
    =
    - \frac{L^2}{T} \sum_{q} 
    \begin{pmatrix}
        j_{-q, 1} & j_{-q, 2}
    \end{pmatrix} 
    \mathcal{D}_{\mathcal{I}, \vb{q} \rightarrow 0} (i \Omega_m)
    \begin{pmatrix}
        j_{q 1} \\
        j_{q 2}
    \end{pmatrix}
    \\
    =
    - \frac{L^2}{T} u^2 \sum_{q} \delta V_{-q} \delta V_{q}
    \,
    \begin{pmatrix}
        \sgn(\Delta_1) \nu_1 \gamma_{1} \sec \gamma_{1}
        & -\sgn(\Delta_2) \nu_2 \gamma_{2} \sec \gamma_{2}
    \end{pmatrix}
    \mathcal{D}_{\mathcal{I}, \vb{q} \rightarrow 0} (i \Omega_m)
    \begin{pmatrix}
        \sgn(\Delta_1) \nu_1 \gamma_{1} \sec \gamma_{1} \\
        -\sgn(\Delta_2) \nu_2 \gamma_{2} \sec \gamma_{2}
    \end{pmatrix}.
\end{multline}
The phase-mode propagator is 
\begin{multline}
    \mathcal{D}_{\mathcal{I}, \vb{q}\rightarrow 0} (i \Omega_m)
    =
    \begin{pmatrix}
        \tilde{g}^{-1} \frac{\Delta_2}{\Delta_1}
        - \nu_1 \gamma_{1} \tan \gamma_{1}
        & -\tilde{g}^{-1}    \\
        - \tilde{g}^{-1}
        & \tilde{g}^{-1} \frac{\Delta_1}{\Delta_2}
        - \nu_2 \gamma_{2} \tan \gamma_{2}
    \end{pmatrix}^{-1}
    \\
    =
    \frac{1}{
        \left( 
            \tilde{g}^{-1} \frac{\Delta_2}{\Delta_1} 
            - \nu_1 \gamma_{1} \tan \gamma_{1}
        \right) \left( 
            \tilde{g}^{-1} \frac{\Delta_1}{\Delta_2} 
            - \nu_2 \gamma_{2} \tan \gamma_{2}
        \right)
        -
        (\tilde{g}^{-1})^2
    } \begin{pmatrix}
        \tilde{g}^{-1} \frac{\Delta_1}{\Delta_2}
        - \nu_2 \gamma_{2} \tan \gamma_{2}
        & \tilde{g}^{-1}    \\
        \tilde{g}^{-1}
        & \tilde{g}^{-1} \frac{\Delta_2}{\Delta_1}
        - \nu_1 \gamma_{1} \tan \gamma_{1}
    \end{pmatrix}.
\end{multline}
Analytically continuing $i \Omega_m \rightarrow \Omega + i 0^{+}$, the full expression for the collective-mode response is
\begin{multline}
\label{eq:full_leggett_response}
    \mathcal{C}_{\phi} (\Omega)
    \\
    =
    \frac{2 V_{\text{DC}}^2}{V_{\text{DC}}^2 + J^2}
    \frac{
        \tilde{g}^{-1}
        \left(
            \nu_1 \Delta_1 \gamma_1 \sec \gamma_1
            -
            \nu_2 \Delta_2 \gamma_2 \sec \gamma_2
        \right)^2
        -
        \nu_1 \Delta_1 \gamma_1
        \nu_2 \Delta_2 \gamma_2
        \left(
            \nu_1 \gamma_1 \sec^2 \gamma_1 \tan \gamma_2
            +
            \nu_2 \gamma_2 \sec^2 \gamma_2 \tan \gamma_1
        \right)
    }{
        \tilde{g}^{-1}
        \left(
            \nu_1 \Delta_1^2 \gamma_1 \tan \gamma_1
            +
            \nu_2 \Delta_2^2 \gamma_2 \tan \gamma_2
        \right)
        -
        \nu_1 \Delta_1 \gamma_1 \tan \gamma_1
        \nu_2 \Delta_2 \gamma_2 \tan \gamma_2
    }.
\end{multline}
The result is rather compact in the case of completely symmetric bands, where we can drop all subscripts $\alpha \in \lbrace 1, 2 \rbrace$ and find
\begin{equation}
    \mathcal{C}_{\phi} [\delta V]
    =
    \frac{4 V_{\text{DC}}^2}{V_{\text{DC}}^2 + J^2}
    \frac{(\nu \gamma \sec \gamma)^2}{
        \nu \gamma \tan \gamma
        - 2 \tilde{g}^{-1}
    }.
\end{equation}
The Leggett mode satisfies $\nu \gamma_{\text{L}} \tan \gamma_{\text{L}} = 2 \tilde{g}^{-1}$. Around that pole, we can approximate
\begin{equation}
    \mathcal{C}_{\phi} (\Omega)
    \approx
    \frac{4 V_{\text{DC}}^2}{V_{\text{DC}}^2 + J^2}
    \frac{\nu (\gamma_{\text{L}} \sec \gamma_{\text{L}})^2}{
        \tan \gamma_{\text{L}}
        + \gamma_{\text{L}} \sec^2 \gamma_{\text{L}}
    }
    \frac{1}{\gamma - \gamma_{\text{L}}}
    \approx
    \frac{4 V_{\text{DC}}^2}{V_{\text{DC}}^2 + J^2}
    \frac{\nu \gamma_{\text{L}}^2}{
        \sin \gamma_{\text{L}}
        + \gamma_{\text{L}} \sec \gamma_{\text{L}}
    }
    \frac{2 \Delta}{\Omega - \Omega_{\text{L}} + i 0^{+}}.
\end{equation}
In the asymmetric case, assuming $|g_{12}^{(s)}|^2 \ll g_{11}^{(s)} g_{22}^{(s)}$, we can expand the numerator and denominator for small $\Omega$:
\begin{equation}
    \mathcal{C} (\Omega)
    \approx
    \frac{V_{\text{DC}}^2}{V_{\text{DC}}^2 + J^2}
    \frac{
        \frac{1}{2} \tilde{g}^{-1} (\nu_1 - \nu_2)^2
        -
        \frac{\Omega^2}{24 \Delta_1^2 \Delta_2^2}
        \left[
            3 \nu_1 \Delta_1 \nu_2 \Delta_2 (
                \nu_1 + \nu_2
            )
            -
            4 \tilde{g}^{-1} (\nu_1 - \nu_2)
            \left(
               \nu_1 \Delta_2^2
               - \Delta_1^2 \nu_2
            \right)
        \right]
    }{
        \frac{1}{4} \tilde{g}^{-1} (\nu_1 + \nu_2)
        -
        \frac{(\Omega + i 0^+)^2}{48 \Delta_1^2 \Delta_2^2}
        \left[
            3 \nu_1 \Delta_1 \nu_2 \Delta_2
            -
            2 \tilde{g}^{-1} (\nu_1 \Delta_2^2 + \Delta_1^2 \nu_2)
        \right]
    }.
\end{equation}

Of course, the observable capacitance also has the quasiparticle contribution:
\begin{multline}
\label{eq:toy_2band_qp_response}
    \mathcal{F}_{\rm QP} [\delta V]
    =
    T \frac{1}{2} \Tr \left[
        \mathcal{G} \mathds{V}^{(\delta V)}
        \mathcal{G} \mathds{V}^{(\delta V)}
    \right]
    =
    T \sum_{k q \alpha} u^2 \delta V_{-q} \delta V_{q}
    \bigg\lbrace
        \frac{1}{2} G_{k \alpha} 
        (G_{k-q, \alpha} + G_{k+q, \alpha})
        - F_{k \alpha} F_{k-q, \alpha}
    \bigg\rbrace
    \\
    \xrightarrow[L \rightarrow \infty]{T \rightarrow 0}
    L^2 \sum_{q \alpha} u^2 \delta V_{-q} \delta V_{q}
    \int \frac{\dd^2 \vb{k}}{( 2 \pi )^2}
    \frac{1}{E_{\vb{k} \alpha}}
    \frac{2 \Delta_{\alpha}^2}{(i \Omega_m)^2 - 4 E_{\vb{k} \alpha}^2}
    \\
    \approx
    L^2 \sum_{q \alpha} u^2 \delta V_{-q} \delta V_{q}
    \int_{-\infty}^{\infty} \nu_{\alpha} \dd{\xi}
    \frac{1}{\sqrt{\xi^2 + \Delta_{\alpha}^2}}
    \frac{2 \Delta_{\alpha}^2}{(i \Omega_m)^2 - 4 \xi^2 - 4 \Delta_{\alpha}^2} 
    \\
    =
    - L^2 \sum_{q \alpha} u^2 \delta V_{-q} \delta V_{q}
    \sgn(\Omega_m) \nu_{\alpha} \gamma_{q, \alpha} 
    \csc \gamma_{q, \alpha}
    \sec \gamma_{q, \alpha}.
\end{multline}
Anticipating the analytic continuation to the upper half-plane, we can set $\sgn(\Omega_m) = 1$. For the case of symmetric bands, we obtain for the total free-energy functional
\begin{equation}
    \mathcal{F} [\delta V]
    =
    \mathcal{F}_{\phi} [\delta V] + \mathcal{F}_{\rm QP} [\delta V]
    =
    L^2 \sum_{q} \delta V_{-q} \delta V_{q} \,
    \frac{V_{\text{DC}}^2}{V_{\text{DC}}^2 + J^2}
    2 \nu \gamma \sec^2 \gamma 
    \left(
        \frac{\nu \gamma}{\nu \gamma \tan \gamma - 2 \tilde{g}^{-1}}
        -
        \cot \gamma
    \right).
\end{equation}
We finally analytically continue, and read off the result for the total capacitance per unit area:
\begin{equation}
    \mathcal{C} (\Omega)
    =
    \frac{V_{\text{DC}}^2}{V_{\text{DC}}^2 + J^2}
    4 \nu \gamma \sec^2 \gamma
    \left(
        \frac{\nu \gamma}{
            \nu \gamma \tan \gamma - 2 \tilde{g}^{-1}
        }
        -
        \cot \gamma
    \right).
\end{equation}
The branch cut of the $\arcsin$ must be chosen to ensure that the result is analytic on the upper half-plane. 
Absorption is described by $- \Im \mathcal{C}$, which is plotted in Fig.~1(d) of the main text.
\end{widetext}

\section{Effect of Fermi-liquid terms on the Leggett mode}
\label{appendix:leggett_fl}
Since number and phase are canonically conjugate variables, one expects that Fermi-liquid terms of the form $H_{\text{FL}} = g_{\text{FL}} \left( n_1 - n_2 - \langle n_1 - n_2 \rangle_0 \right)^2$ might shift the Leggett mode frequency. Here $\langle \dots \rangle_0$ denotes an equilibrium average. In this Appendix we show explicitly that attractive Fermi liquid terms of this form indeed can stabilize the Leggett mode inside the superconducting gap even when the pairing interactions alone would push that mode to the band edge (or, by continuity, beyond), as Leggett anticipated but did not calculate~\cite{leggett1966number}. We adopt the toy model considered in Appendix \ref{appendix:toy_leggett}. The additional term in the action due to the Fermi-liquid (FL) interaction in question takes the simple form
\begin{multline}
    \delta S_{\text{FL}}
    \\
    =
    \frac{T}{L^2}
    \sum_{\substack{
        k k' q \\ \sigma \sigma'
    }}
    g_{\text{FL}}
    \left(
        \overline{c}_{k+q, 1 \sigma}  c_{k, 1 \sigma}
        -
        \overline{c}_{k+q, 2 \sigma}  c_{k, 2 \sigma}
        -
        \langle
            \delta n_{\vb{k}}
        \rangle_0 \delta_{q, 0}
    \right) 
    \\ 
    \left(
        \overline{c}_{k'-q, 1 \sigma'}  c_{k', 1 \sigma'}
        -
        \overline{c}_{k'-q, 2 \sigma'}  c_{k', 2 \sigma'}
        -
        \langle
            \delta n_{\vb{k'}}
        \rangle_0 \delta_{q, 0}
    \right),
\end{multline}
where $g_{\text{FL}} < 0$, $\delta n_{\vb{k}} = n_{\vb{k} 1} - n_{\vb{k} 2}$, and $\langle \dots \rangle_0$ denotes an equilibrium average; subtracting the averages means that these interactions have no impact on the saddle point. For simplicity we will assume equal densities of states in two layers $\nu_1 = \nu_2$, and equal intralayer pairing interaction strengths $g_{11} = g_{22}$. Also for simplicity, we focus of $\vb{q} = 0$, so $q = (i \Omega_m, \vb{q}=0)$ is again a shorthand for the bosonic Matsubara frequency $i \Omega_m$.

We then perform the Hubbard-Stratonovich transformation simultaneously in the Cooper and Fermi-liquid channels. While the former is introduced in Appendix \ref{appendix:toy_leggett}, the latter is performed via the boson $\rho_{\alpha,q} \sim \sum_{k \sigma}  c^{\dag}_{k-q, \alpha ,\sigma}
    {c}_{k, \alpha,\sigma}$.
The resulting action is
\begin{multline}
    S [\Psi^{\dag}, \Psi, \overline{\Delta}, \Delta,\rho]
    \\
    =
    - \frac{L^2}{T} \sum_{q,\alpha,\alpha'}
    \left(
        \overline{\Delta}_{q,\alpha} 
        \hat{g}^{-1}_{\alpha\alpha'} 
        \Delta_{q \alpha'}  
        +  
        {\rho}_{-q,\alpha} 
        \hat{g}^{-1}_{\text{FL}; \alpha\alpha'} 
        \rho_{q \alpha'} 
    \right)
    \\
    - 
    \sum_{\alpha k q}
    \Psi^{\dag}_{\alpha} \mathds{G}^{-1}_{\alpha} \Psi_{\alpha},
\end{multline}
where 
\begin{align}
\hat{g}_{\text{FL}} = g_{\text{FL}}\begin{pmatrix} 1 & -1 \\ -1 & 1\end{pmatrix}.
\end{align}
We note that since $\hat{g}_{\text{FL}}$ is singular, the HS transformation is not well defined. However, we however can overcome this difficulty by adding a small negative diagonal part to $\hat{g}_{\text{FL}}: \hat{g}_{\text{FL}}\rightarrow \hat{g}_{\text{FL}}+\varepsilon \times \mathbb{I}_{2\times 2}$ with $\varepsilon < 0$, and then take the limit $\varepsilon\rightarrow 0$ in the final answer.
The inverse fermion propagator is
\begin{multline}
    \mathds{G}^{-1}_{\alpha; k, k-q}
    \\
    =
    \begin{pmatrix}
        (i \omega_n - \xi_{\vb{k} \alpha}) \delta_{q, 0} +\rho_{q,\alpha}
        & \Delta_{q, \alpha} \\
        \overline{\Delta}_{-q, \alpha}
        & (i \omega_n + \xi_{\vb{k} \alpha}) \delta_{q, 0}-\rho_{q,\alpha}
    \end{pmatrix}.
\end{multline}
We then integrate out the fermions to get the effective action $S_{\text{eff}}[\overline{\Delta},\Delta,\rho]$. As mentioned, the saddle point is the same as for the case without Fermi-liquid interactions. The vertex for density fluctuations is
\begin{align}
\mathbb{V}^{(\rho)}_{\alpha;k,k-q}= 2 \begin{pmatrix} \rho_{q,\alpha} & 0 \\ 0 & - \rho_{q,\alpha}\end{pmatrix}.
\end{align}
We then obtain the effective action in the same manner as in Appendix \ref{appendix:toy_leggett}, yielding
\begin{widetext}
\begin{equation}
    S[\mathcal{R},\mathcal{I},\rho] 
    =
    -\sum_{q} \begin{pmatrix}
        \mathcal{R}_{-q, 1}
        & \mathcal{R}_{-q, 2}
        & \mathcal{I}_{-q, 1}
        & \mathcal{I}_{-q, 2} & \rho_{-q,1} & \rho_{-q,2}
    \end{pmatrix}
    \mathcal{D}^{-1}_{q}
    \begin{pmatrix}
        \mathcal{R}_{q, 1} \\
        \mathcal{R}_{q, 2} \\
        \mathcal{I}_{q, 1} \\
        \mathcal{I}_{q, 2}\\
        \rho_{q, 1}\\
        \rho_{q,2}
    \end{pmatrix},
\end{equation}
where the bosonic inverse propagator $\mathcal{D}_{q}^{-1}$ can be written in block form as
\begin{equation}
    \mathcal{D}_{q}^{-1}
    =
    \begin{pmatrix} 
        \hat{g}^{-1} + \Pi^{\mathcal{RR}}\times\mathbb{I}_{2\times 2} 
        & \mathbf{0}  
        & \mathbf{0}
        \\
        \mathbf{0} 
        & \hat{g}^{-1} + \Pi^{\mathcal{II}}\times\mathbb{I}_{2\times 2} 
        & \Pi^{\mathcal{I}\rho}\times\mathbb{I}_{2\times 2} 
        \\ 
        \mathbf{0} 
        & -\Pi^{\mathcal{I}\rho}\times\mathbb{I}_{2\times 2} 
        & \hat{g}_{\text{FL}}^{-1} + \Pi^{\rho\rho}\times \mathbb{I}_{2\times 2} 
    \end{pmatrix},
\end{equation}
\end{widetext}
with $\Pi^{\mathcal{II}}$ and $\Pi^{\mathcal{RR}}$ as given in Appendix \ref{appendix:toy_leggett}, $\Pi^{\rho\rho}=-8\nu\gamma_q/\sin 2\gamma_q$, and $\Pi^{\mathcal{I}\rho}=-2 i\nu\gamma_q/\cos\gamma_q$. The $2 \times 2$ blocks act in band space.

We can see that the amplitude fluctuations $\mathcal{R}$ do not mix with the density fluctuations $\rho$. The collective mode frequencies are again given by the zeros of the determinant of the inverse propagator, which for the coupled $\{\mathcal{I},\rho\}$ sector correspond to
\begin{multline}
    \gamma_q \tan\gamma_q 
    \left(
        2 \tilde{g}^{-1}
        - \nu \gamma_q \tan \gamma_q
        + \frac{
            16 \gamma_q \tilde{g}^{-1} g_{\text{FL}}\nu
        }{
            \sin \gamma_q \cos \gamma_q
        }
    \right)
    \\
    = 0,
\end{multline}
where $\gamma_q=\arcsin(\Omega/(2\Delta))$. Recall that $\tilde{g}^{-1}=-g_{12}/\mathrm{det}\hat{g}$ (where we chose all $g_{\alpha \alpha'} < 0$), and that for $g_{\text{FL}} = 0$, when the determinant of the pairing interaction matrix $\det \hat{g}$ approaches zero, the Leggett mode is pushed to the band edge, $\Omega_{\text{L}} \rightarrow 2\Delta$. However, for attractive $g_{\text{FL}} < 0$ the Leggett mode remains inside the gap even at $\det \hat{g}\rightarrow 0$. Specifically, in this limit, the Legett mode frequency solves the equation
\begin{align}
g_{\text{FL}}\nu = -\frac{\sin 2\gamma_q}{16\gamma_q}.
\end{align}
As can be seen, provided that $0 < -g_{\text{FL}}\nu < 1 / 8$, the Fermi-liquid interactions pull the Leggett mode back into the gap even though $g_{12} \rightarrow g_{0}$ such that the second eigenvalue of $\hat{g}$ approaches zero.

We have thus established that Fermi-liquid interactions can keep the Leggett mode a finite distance inside the gap when the pairing interactions alone would place it $\epsilon$-close to the gap edge. By continuity, we conclude that even when pairing interactions alone would not yield an in-gap Leggett mode at all, Fermi-liquid terms could stabilize such an in-gap mode; nothing drastic should happen when $g_{12}$ crosses its would-be critical value if the Leggett mode retains a finite distance from the gap edge at that value of $g_{12}$. To establish this rigorously, one would need to modify the Cooper-channel Hubbard-Stratonovich transformation to handle a repulsive eigenvalue (as we do for a more general model in Appendix \ref{appendix:general_case}); since the Fermi-liquid effects discussed in this Appendix are not the main purpose of the present paper, we instead invoke the above continuity argument for brevity.

\section{Toy model for clapping modes}
\label{appendix:toy_clapping}
The minimal model containing clapping modes is that of a single-band chiral $p_x + i p_y$ superconductor, corresponding to the toy model in the main text with $g_{\alpha \alpha'}^{(s)} = 0$. The interactions are
\begin{equation}
    \mathcal{V}_{\vb{k k'}}
    =
    2 g \cos (\varphi_{\vb{k}} - \varphi_{\vb{k'}})
    =
    \sum_{\ell = 1, 2}
    g \chi^{(\ell)}_{\vb{k}} \chi^{(\ell)}_{\vb{k'}},
\end{equation}
where $g < 0$, $\varphi_{\vb{k}}$ is the angle of $\vb{k}$ relative to the $k_x$ axis, i.e.,~$\vb{k} = |\vb{k}| (\cos \varphi_{\vb{k}}, \sin \varphi_{\vb{k}})$, and the form factors are
\begin{equation}
    \chi^{(1)}_{\vb{k}}
    =
    \sqrt{2} \cos (\varphi_{\vb{k}})
    \quad \text{and} \quad
    \chi^{(2)}_{\vb{k}}
    =
    \sqrt{2} \sin (\varphi_{\vb{k}}).
\end{equation}
The Cooper-channel Hubbard-Stratonovich decoupling then involves two auxiliary bosons $\Delta^{(\ell)}_{q}$, yielding the action
\begin{multline}
    S
    =
    S_0
    +
    \frac{L^2}{T} \sum_{q}
    \sum_{\ell = 1, 2}
    \frac{|\Delta^{(\ell)}_{q}|^2}{|g|}
    \\
    -
    \sum_{k q \ell}
    \left(
        \Delta^{(\ell)}_{q} \chi^{(\ell)}_{\vb{k}}
        \overline{c}_{k, \uparrow} \overline{c}_{-k+q, \downarrow}
        +
        \overline{\Delta}^{(\ell)}_{q} \chi^{(\ell)}_{\vb{k}}
        c_{-k+q, \downarrow} c_{k, \uparrow}
    \right)
    \\
    =
    S_0
    +
    \frac{L^2}{T} \sum_{q}
    \sum_{\ell = 1, 2}
    \frac{|\Delta^{(\ell)}_{q}|^2}{|g|}
    \\ 
    -
    \sum_{k q \ell}
    \Psi^{\dag}_{k}
    \begin{pmatrix}
        0 & \Delta^{(\ell)}_{q} \chi^{(\ell)}_{\vb{k}}  \\
        \overline{\Delta}^{(\ell)}_{-q} \chi^{(\ell)}_{\vb{k-q}} & 0
    \end{pmatrix}
    \Psi_{k-q}.
\end{multline}
$S_0$ is the free action for a single parabolic band.

\subsection{Mean-field theory}
It is useful to introduce the chiral combinations
\begin{equation}
    \chi^{(\pm)}_{\vb{k}}
    =
    \frac{
        \chi^{(1)}_{\vb{k}}
        \pm i \chi^{(2)}_{\vb{k}}
    }{\sqrt{2}}
    =
    e^{\pm i \varphi_{\vb{k}}}.
\end{equation}
The gap equation is
\begin{equation}
    \Delta_{\vb{k}}
    =
    - g \int \frac{\dd^2 \vb{k}}{(2 \pi)^2}
    \cos (\varphi_{\vb{k}} - \varphi_{\vb{k'}})
    \frac{\Delta_{\vb{k'}}}{E_{\vb{k'}}}.
\end{equation}
With the chiral $p$-wave ansatz $\Delta_{\vb{k}} = \Delta_0 \chi^{(+)}_{\vb{k}} = \Delta_0 e^{i \varphi_{\vb{k}}}$ and introducing an energy cutoff $\Lambda$, the mean-field result is standard:
\begin{equation}
    \Delta_0 = 2 \Lambda e^{1 / (g \nu)}.
\end{equation}
The $\chi^{(+)}$ and $\chi^{(-)}$ channels are degenerate; we assume that the $\chi^{(+)}$ channel condenses. Fluctuations in the $\chi^{(-)}$ channel then give rise to the clapping modes.

\subsection{Collective modes and compressibility}
Schematically, the total pairing field is $\Delta = \Delta^{(1)} \chi^{(1)} + \Delta^{(2)} \chi^{(2)} = \Delta^{(+)} \chi^{(+)} + \Delta^{(-)} \chi^{(-)}$. In terms of spacetime coordinates, we parametrize the fluctuations following Ref.~\cite{poniatowski2022spectroscopic}:
\begin{subequations}
\begin{equation}
    \Delta^{(+)} (x)
    =
    e^{i \theta (x)}
    \left( \Delta_0 + h(x) \right)
\end{equation}
and
\begin{equation}
    \Delta^{(-)} (x)
    =
    e^{i \theta (x)}
    \left(
        a(x) + i b(x)
    \right).
\end{equation}
\end{subequations}
$x = (\tau, \vb{r})$ contains the spacetime coordinates. $\theta$ is the ABG mode, $\Delta_0 \ge 0$ is the mean-field gap, $h$ is the Higgs mode, and $a$ and $b$ are the clapping modes. After minimal coupling to the electromagnetic [$U(1)$] gauge field $A = (A^0, \vb{A})$, we can gauge out slow variations of the pairing phase $\theta$, combining it with $\mathcal{A}$ into the gauge-invariant field $\Ay = (\Ay^0, \bAy) = (A^0 + \partial_{\tau} \theta, \, \vb{A} - \grad \theta)$. Before the gauge transformation, expressed in coordinate space, the couplings between the Hubbard-Stratonovich bosons and the fermion are
\begin{widetext}
\begin{equation}
    S_{\text{pair}}
    =
    - \sum_{\ell} \int_{x x'}
    \Psi^{\dag} (x)
    \begin{pmatrix}
        0 & \chi^{(\ell)} (x - x') \Delta^{(\ell)} (x') \\
        \overline{\Delta}^{(\ell)} (x) \chi^{(\ell)} (x - x') & 0
    \end{pmatrix}
    \Psi (x').
\end{equation}
Assuming that the phase $\theta$ varies slowly compared to the scale over which $\chi$ vanishes, such that $e^{i \theta (x)} \chi (x - x') \approx e^{i \theta (x')} \chi (x - x')$---in other words, pairing is essentially local compared to the scale of phase variations---we can remove $\theta$ using the unitary $U_{x x'} = \delta(x - x') \exp (-i \theta (x) \tau^z / 2)$, which implements the needed gauge transformation as $\mathds{G}^{-1} \rightarrow U \mathds{G}^{-1} U^{\dag}$. 

After the gauge transformation, one has
\begin{subequations}
\begin{equation}
    \Delta^{(1)}
    =
    \frac{1}{\sqrt{2}} \left(
        \Delta^{(+)} + \Delta^{(-)}
    \right)
    =
    \frac{1}{\sqrt{2}} \left(
        \Delta_0 + h + a + i b
    \right)
\end{equation}
and
\begin{equation}
    \Delta^{(2)}
    =
    \frac{i}{\sqrt{2}} \left(
        \Delta^{(+)} - \Delta^{(-)}
    \right)
    =
    \frac{i}{\sqrt{2}} \left(
        \Delta_0 + h - a - i b
    \right)
\end{equation}
\end{subequations}
The action is then
\begin{equation}
    S
    =
    \frac{L^2}{T} \sum_{\ell}
    \frac{|\Delta^{(\ell)}_{q}|^2}{|g|}
    -
    \sum_{k} \Psi^{\dag}_{k} \mathcal{G}^{-1}_{k} \Psi_{k}
    \\
    +
    \sum_{k q} \Psi^{\dag}_{k}
    \left(
        \mathds{V}^{(\Ay^0)}_{k, k-q}
        +
        \mathds{V}^{(\bAy)}_{k, k-q}
        +
        \mathds{V}^{(h)}_{k, k-q}
        +
        \mathds{V}^{(a)}_{k, k-q}
        +
        \mathds{V}^{(b)}_{k, k-q}
    \right)
    \Psi_{k-q}.
\end{equation}
The mean-field inverse propagator is
\begin{equation}
    \mathcal{G}_{k}^{-1}
    =
    \begin{pmatrix}
        i \omega_n - \xi_{\vb{k}}
        & \Delta_0 e^{i \varphi_{\vb{k}}} \\
        \Delta_0 e^{-i \varphi_{\vb{k}}}
        & i \omega_n + \xi_{\vb{k}}
    \end{pmatrix}.
\end{equation}
The vertices for the gauge variables are
\begin{equation}
    \mathds{V}^{(\Ay^0)}_{k, k-q}
    =
    i \Ay^{0}_{q} \begin{pmatrix}
        1 & 0 \\
        0 & -1
    \end{pmatrix}
    \qquad \text{and} \qquad
    \mathds{V}^{(\bAy)}_{k, k-q}
    =
    - \frac{1}{m} 
    \bAy_{q} \vdot \left( 
        \vb{k} - \frac{1}{2} \vb{q}
    \right) \begin{pmatrix}
        1 & 0 \\
        0 & 1
    \end{pmatrix}
    +
    \frac{1}{2m} \sum_{p}
    \bAy_{q-p} \vdot \bAy_{p}
    \begin{pmatrix}
        1 & 0 \\
        0 & -1
    \end{pmatrix},
\end{equation}
and the vertices for the Higgs and clapping modes are
\begin{equation}
    \mathds{V}^{(h)}_{k, k-q}
    =
    h_q \begin{pmatrix}
        0 & \chi^{(+)}_{\vb{k}} \\
        \chi^{(-)}_{\vb{k-q}} & 0
    \end{pmatrix},
    \qquad
    \mathds{V}^{(a)}_{k, k-q}
    =
    a_q \begin{pmatrix}
        0 & \chi^{(-)}_{\vb{k}} \\
        \chi^{(+)}_{\vb{k-q}} & 0
    \end{pmatrix},
    \qquad \text{and} \qquad
    \mathds{V}^{(b)}_{k, k-q}
    =
    i b_q \begin{pmatrix}
        0 & \chi^{(-)}_{\vb{k}} \\
        -\chi^{(+)}_{\vb{k-q}} & 0
    \end{pmatrix}.
\end{equation}
\end{widetext}

For brevity, we focus only on the part of the effective action which directly relates to the compressibility, and neglect the spatial part of the overall-phase mode, and the Anderson-Higgs mechanism (hence ignoring $\bAy$ altogether), leaving
\begin{multline}
    \frac{T}{L^2} S_{\text{eff}}
    \\
    =
    \sum_{q}
    \Big(
        \Pi^{00}_{q} \Ay^0_{-q} \Ay^0_{q}
        - \mathcal{D}^{-1}_{a, q} a_{-q} a_{q}
        - \mathcal{D}^{-1}_{b, q} b_{-q} b_{q}
        \\
        + \Pi^{0 a}_{q} \Ay^0_{-q} a_{q}
        + \Pi^{0 b}_{q} \Ay^0_{-q} b_{q}
        + \Pi^{a b}_{q} a_{-q} b_{q}
    \Big),
\end{multline}
where
\begin{subequations}
\begin{multline}
    \Pi^{0 0}_{q} \Ay^0_{-q} \Ay^0_{q}
    \\
    =
    \frac{1}{2} \frac{T}{L^2} \sum_{k} \tr{
        \mathcal{G}_{k} \mathds{V}^{(\Ay^0)}_{k, k-q}
        \mathcal{G}_{k-q} \mathds{V}^{(\Ay^0)}_{k-q, k}
    },
\end{multline}
\begin{multline}
    - \mathcal{D}^{-1}_{a, q} a_{-q} a_{q}
    =
    \frac{1}{|g|} a_{-q} a_{q}
    \\
    +
    \frac{1}{2} \frac{T}{L^2} \sum_{k} \tr{
        \mathcal{G}_{k} \mathds{V}^{(a)}_{k, k-q}
        \mathcal{G}_{k-q} \mathds{V}^{(a)}_{k-q, k}
    },
\end{multline}
\begin{multline}
    - \mathcal{D}^{-1}_{b, q} b_{-q} b_{q}
    =
    \frac{1}{|g|} b_{-q} b_{q}
    \\
    +
    \frac{1}{2} \frac{T}{L^2} \sum_{k} \tr{
        \mathcal{G}_{k} \mathds{V}^{(b)}_{k, k-q}
        \mathcal{G}_{k-q} \mathds{V}^{(b)}_{k-q, k}
    },
\end{multline}
\begin{equation}
    \Pi^{0 a}_{q} \Ay^0_{-q} a_{q}
    =
    \sum_{k} \frac{T}{L^2} \tr{
        \mathcal{G}_{k} \mathds{V}^{(a)}_{k, k-q}
        \mathcal{G}_{k-q} \mathds{V}^{(\Ay^0)}_{k-q, k}
    },
\end{equation}
\begin{equation}
    \Pi^{0 b}_{q} \Ay^0_{-q} b_{q}
    =
    \sum_{k} \frac{T}{L^2} \tr{
        \mathcal{G}_{k} \mathds{V}^{(b)}_{k, k-q}
        \mathcal{G}_{k-q} \mathds{V}^{(\Ay^0)}_{k-q, k}
    },
\end{equation}
and
\begin{equation}
    \Pi^{a b}_{q} a_{-q} b_{q}
    =
    \sum_{k} \frac{T}{L^2} \tr{
        \mathcal{G}_{k} \mathds{V}^{(b)}_{k, k-q}
        \mathcal{G}_{k-q} \mathds{V}^{(a)}_{k-q, k}
    }.
\end{equation}
\end{subequations}
Strictly speaking, the collective-mode propagator is a matrix in the full mode space (spanned by $a$, $b$, and $h$). However, since we will evaluate that propagator at $\vb{q} = 0$ (as explained below), we can neglect the couplings linking $a$ and $b$ to the Higgs mode; these couplings can easily be shown to vanish at $\vb{q} = 0$ in this simple model.

We are interested in the collective-mode contribution to the compressibility which comes from integrating out $a$ and $b$,
\begin{equation}
    \delta \Pi^{00}_{q}
    =
    \frac{1}{4}
    \Pi^{0a}_{-q} \mathcal{D}_{a, q} \Pi^{0 a}_{q}
    + \frac{1}{4}
    \Pi^{0b}_{-q} \mathcal{D}_{b, q} \Pi^{0 b}_{q},
\end{equation}
to lowest nonvanishing order in $\vb{q}$. In the rotationally symmetric models we consider (either continuous rotations or $C_{3v}$), the couplings vanish at $O(|\vb{q}|^0)$. This means that, neglecting dispersion of the collective modes, we can simply evaluate the propagators at $\vb{q} = 0$, and the couplings $\Pi^{0 X}_{q}$ at leading nonvanishing order. 

The couplings $\Pi^{0X}$ are
\begin{widetext}
\begin{multline}
    \Pi^{0 a}_{q}
    =
    -i \frac{T}{L^2} \sum_{k}
    \left\lbrace
        \chi^{(+)}_{\vb{k-q}/2} 
        \left( 
            G_{k-q/2} F_{k+q/2}
            + F_{k-q/2} G_{-k-q/2}
        \right)
        +
        \chi^{(-)}_{\vb{k+q}/2}
        \left(
            \overline{F}_{k-q/2}
            G_{k+q/2}
            +
            G_{-k+q/2}
            \overline{F}_{k+q/2}
        \right)
    \right\rbrace
    \\
    =
    i \frac{T}{L^2} \sum_{k} \Bigg\lbrace
        \chi^{(+)}_{\vb{k-q}/2}
        \frac{i \omega_n - i \Omega_m / 2 + \xi_{\vb{k-q}/2}}{
            (i \omega_n - i \Omega_m/2)^2 - E_{\vb{k-q}/2}^2
        }
        \frac{\Delta_{\vb{k+q}/2}}{
            (i \omega_n + i \Omega_m/2)^2 - E_{\vb{k+q}/2}^2
        }
        \\
        + \chi^{(+)}_{\vb{k-q}/2}
        \frac{\Delta_{\vb{k-q}/2}}{
            (i \omega_n - i \Omega_m/2)^2 - E_{\vb{k-q}/2}^2
        }
        \frac{-i \omega_n - i \Omega_m / 2 + \xi_{\vb{-k-q}/2}}{
            (-i \omega_n - i \Omega_m/2)^2 - E_{\vb{-k-q}/2}^2
        }
        \\
        + \chi^{(-)}_{\vb{k+q}/2}
        \frac{\overline{\Delta}_{\vb{k-q}/2}}{
            (i \omega_n - i \Omega_m/2)^2 - E_{\vb{k-q}/2}^2
        }
        \frac{i \omega_n + i \Omega_m / 2 + \xi_{\vb{k+q}/2}}{
            (i \omega_n + i \Omega_m/2)^2 - E_{\vb{k+q}/2}^2
        }
        \\
        + \chi^{(-)}_{\vb{k+q}/2}
        \frac{
            - i \omega_n + i \Omega_m/2 + \xi_{\vb{-k+q}/2}
        }{
            (-i \omega_n + i \Omega_m/2)^2
            - E_{\vb{-k+q}/2}^2
        }
        \frac{\overline{\Delta}_{\vb{k+q}/2}}{
            (i \omega_n + i \Omega_m/2)^2 - E_{\vb{k+q}/2}^2
        }
    \Bigg\rbrace
    \\
    =
    \Delta_0 \frac{T}{L^2} \sum_{k}
    \frac{i}{
        \left[ (i \omega_n - i \Omega_m / 2)^2 - E_{\vb{k-q}/2}^2 \right]
        \left[ (i \omega_n + i \Omega_m / 2)^2 - E_{\vb{k+q}/2}^2 \right]
    }
    \\
    \times
    \bigg\lbrace
        \chi^{(+)}_{\vb{k-q}/2}
        \left[
            \chi^{(+)}_{\vb{k+q}/2}
            \left(
                i \omega_n - i \Omega_m / 2 + \xi_{\vb{k-q}/2}
            \right)
            +
            \chi^{(+)}_{\vb{k-q}/2} 
            \left(
                -i \omega_n - i \Omega_m/2 + \xi_{\vb{k+q}/2}
            \right)
        \right]
        \\
        +
        \chi^{(-)}_{\vb{k+q}/2}
        \left[
            \chi^{(-)}_{\vb{k-q}/2}
            \left(
                i \omega_n + i \Omega_m / 2 + \xi_{\vb{k+q}/2}
            \right)
            +
            \chi^{(-)}_{\vb{k+q}/2} 
            \left(
                -i \omega_n + i \Omega_m/2 + \xi_{\vb{k-q}/2}
            \right)
        \right]
    \bigg\rbrace
    \\
    \xrightarrow[L \rightarrow \infty]{T \rightarrow 0}
    -i \Delta_0 \int \frac{\dd^2{\vb{k}}}{(2 \pi)^2}
    \frac{1}{
        2 E_{\vb{k-q}/2} E_{\vb{k+q}/2}
        \left[ (i \Omega_m)^2 - (E_{\vb{k-q}/2} + E_{\vb{k+q}/2})^2 \right]
    } 
    \\
    \times \bigg\lbrace
        (E_{\vb{k-q}/2} + E_{\vb{k+q}/2})
        \left[
            \xi_{\vb{k-q}/2} \left(
                (\chi^{(-)}_{\vb{k+q}/2})^2
                + \chi^{(+)}_{\vb{k-q}/2} \chi^{(+)}_{\vb{k+q}/2}
            \right)
            +
            \xi_{\vb{k+q}/2} \left(
                (\chi^{(+)}_{\vb{k-q}/2})^2
                + \chi^{(-)}_{\vb{k-q}/2} \chi^{(-)}_{\vb{k+q}/2}
            \right)
        \right]
        \\
        + i \Omega_m \left[
            E_{\vb{k-q}/2} \left(
                (\chi^{(-)}_{\vb{k+q}/2})^2
                -
                \chi^{(+)}_{\vb{k-q}/2}
                \chi^{(+)}_{\vb{k+q}/2}
            \right)
            + E_{\vb{k+q}/2} \left(
                \chi^{(-)}_{\vb{k-q}/2}
                \chi^{(-)}_{\vb{k+q}/2}
                -
                (\chi^{(+)}_{\vb{k-q}/2})^2
            \right)
        \right]
    \bigg\rbrace
\end{multline}
and
\begin{multline}
    \Pi^{0 b}_{q}
    =
    \frac{T}{L^2} \sum_{k} \left\lbrace
        - \chi^{(+)}_{\vb{k-q}/2}
        \left(
            G_{k-q/2} F_{k+q/2} + F_{k-q/2} G_{-k-q/2}
        \right)
        + \chi^{(-)}_{\vb{k+q}/2}
        \left(
            \overline{F}_{k-q/2} G_{k+q/2}
            + G_{-k+q/2} \overline{F}_{k+q/2}
        \right)
    \right\rbrace
    \\
    =
    \frac{T}{L^2} \sum_{k}
    \Bigg\lbrace
        \chi^{(+)}_{\vb{k-q}/2}
        \frac{i \omega_n - i \Omega_m / 2 + \xi_{\vb{k-q}/2}}{
            (i \omega_n - i \Omega_m/2)^2 - E_{\vb{k-q}/2}^2
        }
        \frac{\Delta_{\vb{k+q}/2}}{
            (i \omega_n + i \Omega_m/2)^2 - E_{\vb{k+q}/2}^2
        }
        \\
        + \chi^{(+)}_{\vb{k-q}/2}
        \frac{\Delta_{\vb{k-q}/2}}{
            (i \omega_n - i \Omega_m/2)^2 - E_{\vb{k-q}/2}^2
        }
        \frac{-i \omega_n - i \Omega_m / 2 + \xi_{\vb{-k-q}/2}}{
            (-i \omega_n - i \Omega_m/2)^2 - E_{\vb{-k-q}/2}^2
        }
        \\
        -\chi^{(-)}_{\vb{k+q}/2}
        \frac{\overline{\Delta}_{\vb{k-q}/2}}{
            (i \omega_n - i \Omega_m/2)^2 - E_{\vb{k-q}/2}^2
        }
        \frac{i \omega_n + i \Omega_m / 2 + \xi_{\vb{k+q}/2}}{
            (i \omega_n + i \Omega_m/2)^2 - E_{\vb{k+q}/2}^2
        }
        \\
        - \chi^{(-)}_{\vb{k+q}/2}
        \frac{
            - i \omega_n + i \Omega_m/2 + \xi_{\vb{-k+q}/2}
        }{
            (-i \omega_n + i \Omega_m/2)^2
            - E_{\vb{-k+q}/2}^2
        }
        \frac{\overline{\Delta}_{\vb{k+q}/2}}{
            (i \omega_n + i \Omega_m/2)^2 - E_{\vb{k+q}/2}^2
        }
    \Bigg\rbrace
    \\
    =
    \Delta_0 \frac{T}{L^2} \sum_{k}
    \frac{1}{
        \left[ (i \omega_n - i \Omega_m / 2)^2 - E_{\vb{k-q}/2}^2 \right]
        \left[ (i \omega_n + i \Omega_m / 2)^2 - E_{\vb{k+q}/2}^2 \right]
    }
    \\
    \times
    \bigg\lbrace
        \chi^{(+)}_{\vb{k-q}/2}
        \left[
            \chi^{(+)}_{\vb{k+q}/2}
            \left(
                i \omega_n - i \Omega_m / 2 + \xi_{\vb{k-q}/2}
            \right)
            +
            \chi^{(+)}_{\vb{k-q}/2} 
            \left(
                -i \omega_n - i \Omega_m/2 + \xi_{\vb{k+q}/2}
            \right)
        \right]
        \\
        -
        \chi^{(-)}_{\vb{k+q}/2}
        \left[
            \chi^{(-)}_{\vb{k-q}/2}
            \left(
                i \omega_n + i \Omega_m / 2 + \xi_{\vb{k+q}/2}
            \right)
            +
            \chi^{(-)}_{\vb{k+q}/2} 
            \left(
                -i \omega_n + i \Omega_m/2 + \xi_{\vb{k-q}/2}
            \right)
        \right]
    \bigg\rbrace
    \\
    \xrightarrow[L \rightarrow \infty]{T \rightarrow 0}
    \Delta_0 \int \frac{\dd^2{\vb{k}}}{(2 \pi)^2}
    \frac{1}{
        2 E_{\vb{k-q}/2} E_{\vb{k+q}/2}
        \left[ (i \Omega_m)^2 - (E_{\vb{k-q}/2} + E_{\vb{k+q}/2})^2 \right]
    } 
    \\
    \times \bigg\lbrace
        (E_{\vb{k-q}/2} + E_{\vb{k+q}/2})
        \left[
            \xi_{\vb{k-q}/2} \left(
                (\chi^{(-)}_{\vb{k+q}/2})^2
                -
                \chi^{(+)}_{\vb{k-q}/2}
                \chi^{(+)}_{\vb{k+q}/2}
            \right)
            +
            \xi_{\vb{k+q}/2} \left(
                \chi^{(-)}_{\vb{k-q}/2}
                \chi^{(-)}_{\vb{k+q}/2}
                -
                (\chi^{(+)}_{\vb{k-q}/2})^2
            \right)
        \right]
        \\
        + i \Omega_m \left[
            E_{\vb{k-q}/2} \left(
                (\chi^{(-)}_{\vb{k+q}/2})^2
                +
                \chi^{(+)}_{\vb{k-q}/2}
                \chi^{(+)}_{\vb{k+q}/2}
            \right)
            +
            E_{\vb{k+q}/2} \left(
                (\chi^{(+)}_{\vb{k-q}/2})^2
                +
                \chi^{(-)}_{\vb{k-q}/2}
                \chi^{(-)}_{\vb{k+q}/2}
            \right)
        \right]
    \bigg\rbrace.
\end{multline}

We expand to leading nonvanishing order in $\vb{q}$. Working at $T \rightarrow 0$ so that
\begin{equation}
    T \sum_{\omega_n}
    \rightarrow 
    \int_{-\infty}^{\infty}
    \frac{\dd{\omega}}{2 \pi}
\end{equation}
and performing the Fermi surface integrals as standard using
\begin{equation}
    \int \frac{\dd^2 \vb{k}}{(2 \pi)^2}
    \rightarrow
    \nu \int_{- \Lambda}^{\Lambda} \dd{\xi}
    \int_{0}^{2 \pi} \frac{\dd{\varphi}}{2 \pi},
\end{equation}
one finds
\begin{multline}
    \Pi^{0 a}_{q}
    \approx
    - \frac{i \nu \Delta_0 \vb{q}^2}{16 m \mu}
    \Bigg\lbrace
        - \frac{
            8 \sin(2 \varphi_{\vb{q}}) 
            (2 \mu^2 - \Omega_m^2 - 4 \Delta_0^2)
        }{\Omega_m (\Omega_m^2 + 4 \Delta_0^2)}
        +
        \frac{48 \mu \cos(2 \varphi_{\vb{q}})}{\Omega_m^2}
        \\
        + 
        2 i \arcsin \left( \frac{i \Omega_m}{2 \Delta_0} \right)
        \frac{1}{
            |\Omega_m|^3 \left( \Omega_m^2 + 4 \Delta_0^2 \right)^{3/2}
        } \Big[
            \Omega_m \sin (2 \varphi_{\vb{q}})
            \left(
                64 \Delta_0^4 - 16 \mu^2 \Omega_m^2 + \Omega_m^4
                + 4 \Delta_0^2 [ 5 \Omega_m^2 - 8 \mu^2 ]
            \right)
            \\
            + 12 \mu \cos (2 \varphi_{\vb{q}})
            \left(
                32 \Delta_0^4 
                + 12 \Delta_0^2 \Omega_m^2
                + \Omega_m^4
            \right)
        \Big]
    \Bigg\rbrace
\end{multline}
and
\begin{multline}
    \Pi^{0b}_{q}
    \approx
    -\frac{i \nu \Delta_0 \vb{q}^2}{16 m \mu}
    \Bigg\lbrace
        \frac{
            8 \cos (2 \varphi_{\vb{q}})
            (2 \mu^2 - \Omega_m^2 - 4 \Delta_0^2)
        }{\Omega_m (\Omega_m^2 + 4 \Delta_0^2)}
        +
        \frac{48 \mu \sin(2 \varphi_{\vb{q}})}{
            \Omega_m^2
        }
        \\
        +
        2 i \arcsin \left(
            \frac{i \Omega_m}{2 \Delta_0}
        \right)
        \frac{1}{
            |\Omega_m|^3
            (\Omega_m^2 + 4 \Delta_0^2)^{3/2}
        } \Big[
            - \Omega_m \cos(2 \varphi_{\vb{q}})
            \left(
                64 \Delta_0^4 - 16 \mu^2 \Omega_m^2
                + \Omega_m^4
                + 4 \Delta_0^2 [5 \Omega_m^2 - 8 \mu^2]
            \right)
            \\
            + 12 \mu \sin (2 \varphi_{\vb{q}})
            \left(
                32 \Delta_0^4 
                + 12 \Delta_0^2 \Omega_m^2
                + \Omega_m^4
            \right)
        \Big]
    \Bigg\rbrace
\end{multline}
where we took the large-cutoff limit $\Lambda \rightarrow \infty$.
Taking only leading terms in $\mu$, we obtain
\begin{subequations}
\label{eq:toy_clapping_couplings}
\begin{equation}
    \Pi^{0 a}_{q}
    \approx
    - \sin (2 \varphi_{\vb{q}}) f_{q}
    \qquad \text{and} \qquad
    \Pi^{0 b}_{q}
    \approx
    \cos(2 \varphi_{\vb{q}}) f_{q}
\end{equation}
with
\begin{equation}
    f_{q}
    =
    \frac{\nu \Delta_0 \vb{q}^2 \mu}{i \Omega_m m}
    \frac{1}{
        4 \Delta_0^2 - (i \Omega_m)^2
    }
    \left(
        1
        + 
        2 i \arcsin \left[
            \frac{i \Omega_m}{2 \Delta_0}
        \right]
        \frac{
            2 \Delta_0^2 - (i \Omega_m)^2
        }{
            |\Omega_m| 
            \sqrt{4 \Delta_0^2 - (i \Omega_m)^2}
        }
    \right).
\end{equation}
\end{subequations}
For the clapping mode propagators ($X \in \lbrace a, b \rbrace$), we have
\begin{equation}
    \mathcal{D}_{X q}
    =
    - \frac{1}{|g|} - \Pi^{XX}_{q}
\end{equation}
with
\begin{equation}
    \Pi^{aa}_{q}
    =
    \frac{1}{2} \frac{T}{L^2} \sum_{k}
    \tr{
        \mathcal{G}_{k}
        \begin{pmatrix}
            0 & \chi^{(-)}_{\vb{k}} \\
            \chi^{(+)}_{\vb{k-q}} & 0
        \end{pmatrix}
        \mathcal{G}_{k-q}
        \begin{pmatrix}
            0 & \chi^{(-)}_{\vb{k-q}} \\
            \chi^{(+)}_{\vb{k}} & 0
        \end{pmatrix}
    }
\end{equation}
and
\begin{equation}
    \Pi^{bb}_{q}
    =
    - \frac{1}{2} \frac{T}{L^2} \sum_{k}
    \tr{
        \mathcal{G}_{k}
        \begin{pmatrix}
            0 & \chi^{(-)}_{\vb{k}} \\
            -\chi^{(+)}_{\vb{k-q}} & 0
        \end{pmatrix}
        \mathcal{G}_{k-q}
        \begin{pmatrix}
            0 & \chi^{(-)}_{\vb{k-q}} \\
            -\chi^{(+)}_{\vb{k}} & 0
        \end{pmatrix}
    }.
\end{equation}
Explicitly,
\begin{multline}
    \Pi^{a a}_{\vb{q} = 0}
    =
    \frac{1}{2} \frac{T}{L^2} \sum_{k}
    \tr{
        \mathcal{G}_{k} \left(
            \chi^{(+)}_{\vb{k}} \tau^{-}
            + \chi^{(-)}_{\vb{k}} \tau^{+}
        \right)
        \mathcal{G}_{k-q} \left(
            \chi^{(+)}_{\vb{k}} \tau^{-}
            + \chi^{(-)}_{\vb{k}} \tau^{+}
        \right)
    }
    \\
    =
    \frac{1}{2} \frac{T}{L^2} \sum_{k}
    \left\lbrace
        \left( \chi^{(+)}_{\vb{k}} \right)^2 
        F_{k} F_{k-q}
        + 
        \left( \chi^{(-)}_{\vb{k}} \right)^2 
        \overline{F}_{k} \overline{F}_{k-q}
        -
        \chi^{(+)}_{\vb{k}} \chi^{(-)}_{\vb{k}}
        \left(
            G_{k} G_{-k+q} + G_{-k} G_{k-q}
        \right)
    \right\rbrace
    \\
    =
    \frac{T}{L^2} \sum_{k}
    \frac{
        \Delta_0^2 \cos(4 \varphi_{\vb{k}})
        -
        \frac{1}{2} \left( i \omega_n + \xi_{\vb{k}} \right) 
        \left( -i \omega_n + i \Omega_m + \xi_{\vb{k}} \right)
        - \frac{1}{2} \left( -i \omega_n + \xi_{\vb{k}} \right)
        \left( i \omega_n - i \Omega_m + \xi_{\vb{k}} \right)
    }{
        \left[
            (i \omega_n)^2 - E_{\vb{k}}^2
        \right] \left[
            (i \omega_n - i \Omega_m)^2 - E_{\vb{k}}^2
        \right]
    }
    \\
    =
    \frac{T}{L^2} \sum_{k}
    \frac{i \omega_n (i \omega_n - i \Omega_m) - \xi_{\vb{k}}^2}{
        \left[
            (i \omega_n)^2 - E_{\vb{k}}^2
        \right] \left[
            (i \omega_n - i \Omega_m)^2 - E_{\vb{k}}^2
        \right]
    }
    \\
    \xrightarrow[L \rightarrow \infty]{T \rightarrow 0}
    \int \frac{\dd^2{\vb{k}}}{(2 \pi)^2}
    \frac{1}{E_{\vb{k}}}
    \frac{2 \xi_{\vb{k}}^2 + \Delta_0^2}{
        (i \Omega_m)^2 - 4 E_{\vb{k}}^2
    }
    \approx
    \nu \int_{-\Lambda}^{\Lambda} \dd{\xi}
    \frac{1}{\sqrt{\xi^2 + \Delta_0^2}} \frac{2 \xi^2 + \Delta_0^2}{(i \Omega_m)^2 - 4 (\xi^2 + \Delta_0^2)}
    \\
    \approx
    \nu \gamma \cot \gamma 
    -
    \gamma \frac{\nu}{2} 
    \frac{2 \Delta_0}{i |\Omega_m|}
    \frac{2 \Delta_0}{\sqrt{4 \Delta_0^2 - (i \Omega_m)^2}}
    - \nu \log \left( \frac{2 \Lambda}{\Delta_0} \right)
    \\
    =
    \nu \gamma \cot \gamma 
    -
    \frac{\nu}{2} \gamma \sec \gamma
    \frac{2 \Delta_0}{i |\Omega_m|}
    - \nu \log \left( \frac{2 \Lambda}{\Delta_0} \right),
\end{multline}
with $\gamma = \arcsin (i \Omega_m / 2 |\Delta|)$ before analytic continuation. For the $b$ mode,
\begin{multline}
    \Pi^{b b}_{\vb{q} = 0}
    =
    -\frac{1}{2} \frac{T}{L^2} \sum_{k}
    \tr{
        \mathcal{G}_{k} \left(
            \chi^{(+)}_{\vb{k}} \tau^{-}
            - \chi^{(-)}_{\vb{k}} \tau^{+}
        \right)
        \mathcal{G}_{k-q} \left(
            \chi^{(+)}_{\vb{k}} \tau^{-}
            - \chi^{(-)}_{\vb{k}} \tau^{+}
        \right)
    }
    \\
    =
    -\frac{1}{2} \frac{T}{L^2} \sum_{k}
    \left\lbrace
        \left( \chi^{(+)}_{\vb{k}} \right)^2 
        F_{k} F_{k-q}
        + 
        \left( \chi^{(-)}_{\vb{k}} \right)^2 
        \overline{F}_{k} \overline{F}_{k-q}
        +
        \chi^{(+)}_{\vb{k}} \chi^{(-)}_{\vb{k}}
        \left(
            G_{k} G_{-k+q} + G_{-k} G_{k-q}
        \right)
    \right\rbrace
    \\
    =
    - \frac{T}{L^2} \sum_{k}
    \frac{
        \Delta_0^2 \cos(4 \varphi_{\vb{k}})
        +
        \frac{1}{2} \left( i \omega_n + \xi_{\vb{k}} \right) 
        \left( -i \omega_n + i \Omega_m + \xi_{\vb{k}} \right)
        + \frac{1}{2} \left( -i \omega_n + \xi_{\vb{k}} \right)
        \left( i \omega_n - i \Omega_m + \xi_{\vb{k}} \right)
    }{
        \left[
            (i \omega_n)^2 - E_{\vb{k}}^2
        \right] \left[
            (i \omega_n - i \Omega_m)^2 - E_{\vb{k}}^2
        \right]
    }
    \\
    =
    \frac{T}{L^2} \sum_{k}
    \frac{i \omega_n (i \omega_n - i \Omega_m) - \xi_{\vb{k}}^2}{
        \left[
            (i \omega_n)^2 - E_{\vb{k}}^2
        \right] \left[
            (i \omega_n - i \Omega_m)^2 - E_{\vb{k}}^2
        \right]
    }
    = \Pi^{aa}_{\vb{q} = 0}.
\end{multline}
\end{widetext}
Using the mean-field equation, the $\log$ terms cancel against the $1/|g|$ terms from the HS transformation, yielding the clapping mode propagator:
\begin{multline}    \label{eq:toy_clapping_propagator}
    \mathcal{D}^{-1}_{a, \vb{q} = 0} (i \Omega_m)
    =
    \mathcal{D}^{-1}_{b, \vb{q} = 0} (i \Omega_m)
    =
    \mathcal{D}^{-1}_{\vb{q} = 0} (i \Omega_m)
    \\
    =
    - \nu \gamma \cot \gamma 
    +
    \frac{\nu}{2} \gamma \sec \gamma
    \frac{2 \Delta_0}{i |\Omega_m|}.
\end{multline}
Finally, we obtain the compressibility by integrating out $a$ and $b$, and then analytically continuing $i \Omega_m \rightarrow \Omega + i 0^{+}$ in the upper half-plane:
\begin{multline}
    \delta \Pi^{00}_{q}
    =
    \frac{1}{4} \Pi^{0 a}_{-q} \mathcal{D}_{q} \Pi^{0 a}_{q}
    + \frac{1}{4} \Pi^{0 b}_{-q} \mathcal{D}_{q} \Pi^{0 b}_{q}
    \\
    =
    \frac{\nu}{2} \left(
        \frac{\pi}{4} \xi_{\text{BCS}} \vb{q}
    \right)^4
    \csc^2 \gamma \sec^4 \gamma
    \\ \times
    \frac{1 - \gamma^2 (\cot \gamma - \tan \gamma)^2}{
        \gamma (\cot \gamma - \tan \gamma)
    },
\end{multline}
where
\begin{equation}
    \xi_{\text{BCS}}
    =
    \frac{v_{\text{F}}}{\pi \Delta_0}
    =
    \frac{k_{\text{F}}}{\pi m \Delta_0}
    =
    \sqrt{\frac{2 \mu}{m}} \frac{1}{\pi \Delta_0},
\end{equation}
and after analytic continuation
\begin{equation}
    \gamma = \arcsin \left( \frac{\Omega + i 0^{+}}{2 \Delta_0} \right).
\end{equation}

\subsection{Trigonal warping}
\label{appendixsec:trigonal_warping}
To consider the effect of trigonal warping, we introduce a second valley. The valleys are labeled by $\lambda \in \lbrace +, - \rbrace$, and $\vb{k}$ labels momentum relative to each valley; we assume symmetry under mapping $\lambda, \vb{k} \rightarrow \overline{\lambda}, -\vb{k}$. Pairing occurs between valleys. The Cooper-channel interactions are
\begin{equation}
    S_{\text{int}}
    =
    \frac{T}{L^2} \sum_{k k' q \lambda}
    \mathcal{V}_{\vb{k k'}}
    \overline{c}_{k \lambda \uparrow}
    \overline{c}_{-k+q, \overline{\lambda} \downarrow}
    c_{-k'+q, \overline{\lambda} \downarrow}
    c_{k' \lambda \uparrow}.
\end{equation}

As discussed in the main text, the idea is that reducing rotational symmetry from complete $U(1)$ circular symmetry down to discrete $C_{3v}$ symmetry relaxes the conservation of angular momentum down to conservation mod 3, enabling access to the clapping modes at $O(|\vb{q}|^2)$. In particular, the couplings $\Pi^{0X}$ between the clapping modes $X \in \lbrace a, b \rbrace$ and the scalar potential can now show up at $O(|\vb{q}|)$ instead of $O(|\vb{q}|^2)$.

The first thing to try is the first nontrivial harmonic of a trigonally warped dispersion relation, $\xi_{\vb{k}} \rightarrow \xi_{\vb{k}} + \eta \cos (3 \varphi_{\vb{k}})$. Repeating the same analysis, we find that at $O(\eta |\vb{q}|)$, the couplings $\Pi^{0X}$ vanish in the large-cutoff limit $\Lambda \rightarrow \infty$. However, trigonally warped band dispersion will lead to a trigonally warped mean-field gap function $\Delta_{\vb{k}}$. Therefore, to capture the leading dependence of the compressibility on the warping, we consider the trigonally warped gap function $\Delta_{\vb{k}} = \left( \Delta_0 + \tilde{\Delta} \cos (3 \varphi_{\vb{k}}) \right) e^{i \varphi_{\vb{k}}}$. Of course, even considering only the first nontrivial harmonic, the phase variation would also be modified, but since our goal is simply to point out the qualitative change in the compressibility's momentum dependence [$O(|\vb{q}|^2)$ versus $O(|\vb{q}|^4)$], we consider only the first nontrivial harmonic of the gap amplitude, which we will show is sufficient.

The analysis is completely analogous to the circular case, only particles and holes are taken from opposite valleys [i.e., $\Psi^{\dag}_{\vb{k} \lambda} = \begin{pmatrix} c^{\dag}_{\vb{k} \lambda \uparrow} & c_{-\vb{k} \overline{\lambda} \downarrow} \end{pmatrix}$], and we Taylor expand the integrands for $\Pi^{0X}$ to first order in the warping parameter $\tilde{\Delta}$. Since we only want the leading-order behavior, we evaluate the collective-mode propagators $\mathcal{D}_{a} = \mathcal{D}_b$ at zero warping ($\tilde{\Delta} = 0$), since the couplings $\Pi^{0X}$ vanish at $O(\tilde{\Delta}^0 \vb{q}^2)$. To leading order in $|\vb{q}|$ and $\tilde{\Delta}$, we obtain
\begin{multline} \label{eq:trigonal_clapping_compressibility}
    \delta \Pi^{00}_{q}
    =
    \frac{\nu}{2}
    \left(
        \frac{1}{4}
        \frac{\tilde{\Delta}}{\Delta_0}
        \frac{\vb{q}}{k_{\text{F}}}
    \right)^{2}
    \frac{\sec^2 \gamma \tan^2 \gamma}{\gamma}
    \frac{1 - \gamma^2 \tan^2 \gamma}{\cot \gamma - \tan \gamma}
    \\
    \approx
    \nu \left(
        \frac{1}{4}
        \frac{\tilde{\Delta}}{\Delta_0}
        \frac{\vb{q}}{k_{\text{F}}}
    \right)^{2}
    \frac{1 - (\pi/4)^2}{\pi}
    \frac{- \sqrt{2} \Delta_0}{\Omega - \sqrt{2} \Delta_0 + i 0^+}
\end{multline}
where the approximation holds in the vicinity of the positive-frequency pole. The momentum-dependent factor can be re-expressed as
\begin{equation}
    \left( 
        \frac{1}{4} 
        \frac{\tilde{\Delta}}{\Delta_0} 
        \frac{\vb{q}}{k_{\text{F}}}
    \right)^2
    =
    \frac{1}{2}
    \left(
        \frac{\pi}{4} \frac{\tilde{\Delta}}{\mu} \xi_{\text{BCS}} \vb{q}
    \right)^2.
\end{equation}
Note that while trigonal warping indeed saves a factor of $(\xi_{\text{BCS}} \vb{q})^2$ as expected, we pay the price of the factor $(\tilde{\Delta} / \mu)^2$; even in the best-case scenario where $\tilde{\Delta} \approx \Delta_0$, this term is suppressed by a factor $1 / (\xi_{\text{BCS}} k_{\text{F}})^2$, which is likely to be quite small.

\section{Collective modes for a general static interaction}
\label{appendix:general_case}
\subsection{Setup}
In preparation for a our trilayer graphene calculation, we now extend the preceding analyses to the case of arbitrary static interactions. The same procedure could be applied to retarded interactions as well, by promoting the indices of the interaction potential to include timelike components [$\vb{k} \rightarrow k = (i \omega_n, \vb{k})$]. First, we perform a spectral decomposition of the interaction potential in terms of form factors (i.e., eigenfunctions),
\begin{equation}
    \mathcal{V}_{\vb{k k'}}
    =
    \sum_{\ell}
    g_{\ell} \chi^{(\ell)}_{\vb{k}} \chi^{(\ell)}_{\vb{k'}}.
\end{equation}
The form factors $\chi^{(\ell)}_{\vb{k}}$ may always be taken to be real, and we choose them to be normalized according to
\begin{equation}
    \int_{\vb{k} \in \text{FS}}
    \chi^{(\ell)}_{\vb{k}}
    \chi^{(\ell')}_{\vb{k}}
    =
    \delta_{\ell \ell'}.
\end{equation}
Next, we introduce one HS boson $\Delta^{(\ell)}$ for each nonnull eigenfunction:
\begin{widetext}
\begin{equation}
    S
    =
    \frac{L^2}{T} \sum_{q \ell}
    \frac{|\Delta^{(\ell)}_{q}|^2}{|g_{\ell}|}
    -
    \sum_{k q}
    \Psi^{\dag}_{k}
    \begin{pmatrix}
        (i \omega_n - \xi_{\vb{k}}) \delta_{q, 0}
        & \Phi^{(-)}_{\vb{k}, q} + i \Phi^{(+)}_{\vb{k}, q}
        \\
        \overline{\Phi^{(-)}}_{\vb{k-q}, -q}
        + i \overline{\Phi^{(+)}}_{\vb{k-q}, -q}
        & (i \omega_n + \xi_{-\vb{k}}) \delta_{q, 0}
    \end{pmatrix}
    \Psi_{k-q}
\end{equation}
\end{widetext}
where the attractive ($g_{\ell} < 0$) and repulsive ($g_{\ell} > 0$) channels respectively add up to
\begin{equation}
    \Phi^{(\mp)}_{\vb{k}, q}
    =
    \sum_{\ell: g_{\ell} \lessgtr 0}
    \Delta^{(\ell)}_{q} \chi^{(\ell)}_{\vb{k}}.
\end{equation}
Note the factor of $i$ in front of the repulsive ($\Phi^{(+)}$) terms. As usual, mean-field theory is the saddle point of $S$. Note that at the saddle point, $\overline{\Phi^{(+)}}$ is \textit{not} the complex conjugate of $\Phi^{(+)}$, but is rather $-1$ times the conjugate; this may be understood in terms of a deformation of the functional integration contour for the repulsive channels \cite{dalal2023repulsion}. The saddle-point equation is standard:
\begin{multline}
    \Delta_{\vb{k}}
    =
    \frac{T}{L^2} \sum_{k'}
    \mathcal{V}_{\vb{k k'}} 
    \frac{\Delta_{\vb{k'}}}{(i \omega_{n'})^2 - E_{\vb{k'}}^2}
    \\
    =
    -\int \frac{\dd^2 \vb{k}}{(2 \pi)^2}
    \mathcal{V}_{\vb{k k'}}
    \tanh \left(
        \frac{E_{\vb{k'}}}{2 T}
    \right)
    \frac{\Delta_{\vb{k'}}}{2 E_{\vb{k'}}},
\end{multline}
where $\Delta_{\vb{k}} = \Phi^{(-)}_{\vb{k}, q=0} + i \Phi^{(+)}_{\vb{k}, q=0}$ at the saddle point.

We then expand each HS boson in terms of fluctuations $\mathcal{R}^{(\ell)}_{q} + i \mathcal{I}^{(\ell)}_{q}$ around its mean-field value. We have one vertex for each fluctuation channel,
\begin{equation}
    \mathds{V}^{(\mathcal{R}, \ell)}_{k, k-q}
    =
    \begin{cases}
        -\mathcal{R}^{(\ell)}_{q}
        \begin{pmatrix}
            0 & \chi^{(\ell)}_{\vb{k}} \\
            \chi^{(\ell)}_{\vb{k-q}} & 0
        \end{pmatrix}
        & \text{if } g_{\ell} < 0,
        \\
        -i \mathcal{R}^{(\ell)}_{q}
        \begin{pmatrix}
            0 & \chi^{(\ell)}_{\vb{k}} \\
            \chi^{(\ell)}_{\vb{k-q}} & 0
        \end{pmatrix}
        & \text{if } g_{\ell} > 0,
    \end{cases}
\end{equation}
and
\begin{equation}
    \mathds{V}^{(\mathcal{I}, \ell)}_{k, k-q}
    =
    \begin{cases}
        \mathcal{I}^{(\ell)}_{q}
        \begin{pmatrix}
            0 & -i \chi^{(\ell)}_{\vb{k}} \\
            i \chi^{(\ell)}_{\vb{k-q}} & 0
        \end{pmatrix}
        \text{if } g_{\ell} < 0,
        \\
        \mathcal{I}^{(\ell)}_{q}
        \begin{pmatrix}
            0 & \chi^{(\ell)}_{\vb{k}} \\
            -\chi^{(\ell)}_{\vb{k-q}} & 0
        \end{pmatrix}
        \text{if } g_{\ell} > 0.
    \end{cases}
\end{equation}
The collective-mode propagator $\mathcal{D}$ is then a matrix in the space of pairing channels $\ell$ and fluctuation types $\mathcal{R} / \mathcal{I}$; writing $X, Y \in \lbrace \mathcal{R}, \mathcal{I} \rbrace$,
\begin{equation}
    \left( \mathcal{D}^{-1}_{q} \right)^{X Y}_{\ell \ell'}
    =
    -\frac{1}{|g_{\ell}|} \delta_{X Y} \delta_{\ell \ell'}
    -
    \Pi^{X Y}_{q; \ell \ell'}
\end{equation}
where the polarization components correspond to
\begin{multline}
    X^{(\ell)}_{-q} \Pi^{X Y}_{q; \ell \ell'} Y^{(\ell')}_{q}
    \\
    =
    \frac{1}{2} \frac{T}{L^2} \sum_{k} \tr{
        \mathcal{G}_{k}
        \mathds{V}^{(Y, \ell')}_{k, k-q}
        \mathcal{G}_{k-q}
        \mathds{V}^{(X, \ell)}_{k-q, k}
    }.
\end{multline}
Note that \textit{all} terms (not just the diagonal ones) carry a factor of $1/2$, since we write the effective action in the symmetric form
\begin{equation}
    S_{\text{eff}}
    =
    -\sum_{\ell \ell' X Y}
    \sum_{q}
    X^{(\ell)}_{-q} 
    \left( \mathcal{D}^{-1}_{q} \right)^{X Y}_{\ell \ell'}
    Y^{(\ell')}_{q}
\end{equation}
where the sums on $X, Y, \ell,$ and $\ell'$ are not restricted (e.g., we include both $\mathcal{R I}$ and $\mathcal{I R}$ terms).

Introducing
\begin{equation}
    \rho_{\ell \ell'}
    =
    \begin{cases}
        0 & g_{\ell} < 0 \text{ and } g_{\ell'} < 0, \\
        1 & g_{\ell} < 0 \text{ and } g_{\ell'} > 0 \text{ or vice-versa}, \\
        2 & g_{\ell} > 0 \text{ and } g_{\ell'} > 0,
    \end{cases}
\end{equation}
one finds
\begin{widetext}
\begin{subequations}
\begin{equation}
    \Pi^{\mathcal{RR}}_{q; \ell \ell'}
    =
    i^{\rho_{\ell \ell'}} \frac{1}{2} \frac{T}{L^2} \sum_{k}
    \left\lbrace
        - \chi^{(\ell)}_{\vb{k}}
        \chi^{(\ell')}_{\vb{k}}
        G_{k} (
            G_{-k+q} + G_{-k-q}
        )
        +
        \chi^{(\ell)}_{\vb{k}}
        \chi^{(\ell')}_{\vb{k-q}}
        F_{k} F_{k-q}
        +
        \chi^{(\ell)}_{\vb{k-q}}
        \chi^{(\ell')}_{\vb{k}}
        \overline{F}_{k} \overline{F}_{k-q}
    \right\rbrace,
\end{equation}
\begin{equation}
    \Pi^{\mathcal{I I}}_{q; \ell \ell'}
    =
    i^{\rho_{\ell \ell'}} \frac{1}{2} \frac{T}{L^2} \sum_{k} 
    \left\lbrace
        -\chi^{(\ell_{-})}_{\vb{k}} \chi^{(\ell'_{-})}_{\vb{k}}
        G_{k} (G_{-k+q} + G_{-k-q})
        -
        \chi^{(\ell_{-})}_{\vb{k}} \chi^{(\ell'_{-})}_{\vb{k-q}}
        F_{k} F_{k-q}
        -
        \chi^{(\ell_{-})}_{\vb{k-q}} \chi^{(\ell'_{-})}_{\vb{k}}
        \overline{F}_{k} \overline{F}_{k-q}  
    \right\rbrace,
\end{equation}
\begin{equation}
    \Pi^{\mathcal{R I}}_{q; \ell \ell'}
    =
    i^{\rho_{\ell \ell'} - 1} \frac{1}{2} \frac{T}{L^2} \sum_{k} 
    \left\lbrace
        \chi^{(\ell)}_{\vb{k}} \chi^{(\ell')}_{\vb{k}}
        G_{k} (G_{-k+q} - G_{-k-q})
        +
        \chi^{(\ell)}_{\vb{k}} \chi^{(\ell')}_{\vb{k-q}}
        F_{k} F_{k-q}
        -
        \chi^{(\ell)}_{\vb{k-q}} \chi^{(\ell')}_{\vb{k}}
        \overline{F}_{k} \overline{F}_{k-q}   
    \right\rbrace,
\end{equation}
and
\begin{equation}
    \Pi^{\mathcal{I R}}_{q; \ell \ell'}
    =
    i^{\rho_{\ell \ell'} - 1} \frac{1}{2} \frac{T}{L^2} \sum_{k} 
    \left\lbrace
        -\chi^{(\ell)}_{\vb{k}} \chi^{(\ell')}_{\vb{k}}
        G_{k} (G_{-k+q} - G_{-k-q})
        +
        \chi^{(\ell)}_{\vb{k}} \chi^{(\ell')}_{\vb{k-q}}
        F_{k} F_{k-q}
        -
        \chi^{(\ell)}_{\vb{k-q}} \chi^{(\ell')}_{\vb{k}}
        \overline{F}_{k} \overline{F}_{k-q}   
    \right\rbrace.
\end{equation}
\end{subequations}
The Matsubara sums are easy to evaluate. At $T \rightarrow 0$, they yield
\begin{subequations}    \label{eq:polarization_sums}
\begin{multline}
    \lim_{T \rightarrow 0} T \sum_{i \omega_n} G_{k} G_{-k+q}
    =
    \frac{
        (E_{\vb{k}} - \xi_{\vb{k}})
        (E_{\vb{k}} + \xi_{\vb{-k+q}} + i \Omega_m)
    }{
        2 E_{\vb{k}} \left[
            (i \Omega_m + E_{\vb{k}})^2 - E_{\vb{-k+q}}^2
        \right]
    }
    +
    \frac{
        (E_{\vb{-k+q}} + \xi_{\vb{-k+q}})
        (E_{\vb{-k+q}} - \xi_{\vb{k}} - i \Omega_m)
    }{
        2 E_{\vb{-k+q}} \left[
            (i \Omega_m - E_{\vb{-k+q}})^2 - E_{\vb{k}}^2
        \right]
    }
    \\
    =
    \frac{(E_{\vb{k}} + E_{\vb{-k+q}}) (
        E_{\vb{k}} E_{\vb{-k+q}} + \xi_{\vb{k}} \xi_{\vb{-k+q}}
    ) + i \Omega_m (\xi_{\vb{k}} E_{\vb{-k+q}} + E_{\vb{k}} \xi_{\vb{-k+q}})
    }{
        2 E_{\vb{k}} E_{\vb{-k+q}} \left[
            (E_{\vb{k}} + E_{\vb{-k+q}})^2 - (i \Omega_m)^2
        \right]
    }
    \\
    =
    \frac{(E_{\vb{k}} + E_{\vb{k-q}}) (
        E_{\vb{k}} E_{\vb{k-q}} + \xi_{\vb{k}} \xi_{\vb{k-q}}
    ) + i \Omega_m (\xi_{\vb{k}} E_{\vb{k-q}} + E_{\vb{k}} \xi_{\vb{k-q}})
    }{
        2 E_{\vb{k}} E_{\vb{k-q}} \left[
            (E_{\vb{k}} + E_{\vb{k-q}})^2 - (i \Omega_m)^2
        \right]
    }
\end{multline}
and
\begin{multline}
    \lim_{T \rightarrow 0} T \sum_{i \omega_n} F_{k} F_{k-q}
    =
    -\frac{\Delta_{\vb{k}} \Delta_{\vb{k-q}}}{
        2 E_{\vb{k}} \left[
            (i \Omega_m + E_{\vb{k}})^2 - E_{\vb{k-q}}^2
        \right]
    }
    -\frac{\Delta_{\vb{k}} \Delta_{\vb{k-q}}}{
        2 E_{\vb{k-q}} \left[
            (i \Omega_m - E_{\vb{k-q}})^2 - E_{\vb{k}}^2
        \right]
    }
    \\
    =
    \frac{\Delta_{\vb{k}} \Delta_{\vb{k-q}} (E_{\vb{k}} + E_{\vb{k-q}})}{
        2 E_{\vb{k}} E_{\vb{k-q}} \left[
            (E_{\vb{k}} + E_{\vb{k-q}})^2 - (i \Omega_m)^2
        \right]
    }.
\end{multline}
\end{subequations}
\end{widetext}

To compute a response function, we also need the coupling between the pair fluctuations and the perturbing field. Assuming the perturbing drive field $d$ couples to the fermion via a vertex $\mathds{V}^{(d)}$, integrating out the fermion yields source terms
\begin{multline}
    \Pi^{d X}_{q; \ell} d_{-q} X^{(\ell)}_{q}
    \\
    =
    \frac{T}{L^2} \sum_{k}
    \tr{
        \mathcal{G}_{k}
        \mathds{V}^{(X, \ell)}_{k, k-q}
        \mathcal{G}_{k-q}
        \mathds{V}^{(d)}_{k-q, k}
    }.
\end{multline}
The desired response function $\chi_{q}$ is then obtained via integration of the pair fluctuations, 
\begin{equation}
    \chi_{q}
    =
    \Pi^{dd}_{q}
    +
    \delta \Pi^{dd}_{q}
\end{equation}
where
\begin{equation}
    \delta \Pi^{dd}_{q}
    =
    \frac{1}{4} \sum_{X Y \ell \ell'}
    \Pi^{d X}_{-q; \ell} (\mathcal{D}_{q})^{X Y}_{\ell \ell'}
    \Pi^{d Y}_{q; \ell'}
\end{equation}
and
\begin{equation}
    \Pi^{d d}_{q} d_{-q} d_{q}
    =
    \frac{1}{2} \frac{T}{L^2} \sum_{k}
    \tr{
        \mathcal{G}_{k}
        \mathds{V}^{(d)}_{k, k-q}
        \mathcal{G}_{k-q}
        \mathds{V}^{(d)}_{k-q, k}
    }
\end{equation}
corresponds to the quasiparticle contribution.

\subsection{Projection to the Fermi surface}
We now reduce the problem to a set of Fermi surface integrals. Since the Fermi velocity may vary over the Fermi surface, denoting the location on the Fermi surface as $k_{\|}$, we replace
\begin{equation}
    \int \frac{\dd^2 \vb{k}}{(2 \pi)^2}
    \rightarrow
    \int_{\text{FS}} \frac{\dd{k_{\|}}}{(2 \pi)^2 v_{k_{\|}}}
    \int_{-\Lambda}^{\Lambda} \dd{\xi}
\end{equation}
where
\begin{equation}
    v_{k_{\|}}
    =
    \left|
        \frac{\partial \xi_{\vb{k}}}{\partial k_{\perp}}
    \right|,
\end{equation}
with $k_{\perp}$ the momentum component normal to the Fermi surface. Throughout, we assume that $\mathcal{V}_{\vb{k k'}}$, $\chi^{(\ell)}_{\vb{k}}$, and $\Delta_{\vb{k}}$ vary at most weakly with $k_{\perp}$, allowing us to write them as functions of $k_{\|}$ only.

The gap equation becomes [at $T \rightarrow 0$, so that $\tanh (E / 2T) \rightarrow 1$]
\begin{multline}
    \Delta_{k_{\|}}
    =
    - \int_{\text{FS}} 
    \frac{\dd{k'_{\|}}}{(2 \pi)^2 v_{k'_{\|}}}
    \mathcal{V}_{k_{\|} k'_{\|}}
    \Delta_{k'_{\|}}
    \\
    \int_{-\Lambda}^{\Lambda} \dd{\xi}
    \frac{1}{2 \sqrt{\xi^2 + |\Delta_{k'_{\|}}|^2}}
    \\
    =
    - \int_{\text{FS}} 
    \frac{\dd{k'_{\|}}}{(2 \pi)^2 v_{k'_{\|}}}
    \mathcal{V}_{k_{\|} k'_{\|}}
    \Delta_{k'_{\|}}
    \\
    \frac{1}{2}
    \log \left(
        1 + \frac{
            2 \Lambda \left[ \Lambda + \sqrt{\Lambda^2 + |\Delta_{k'_{\|}}|^2} \right]
        }{|\Delta_{k'_{\|}}|^2}
    \right)
    \\
    \approx
    - \int_{\text{FS}} 
    \frac{\dd{k'_{\|}}}{(2 \pi)^2 v_{k'_{\|}}}
    \mathcal{V}_{k_{\|} k'_{\|}}
    \Delta_{k'_{\|}}
    \log \left(
        \frac{2 \Lambda}{
            |\Delta_{k'_{\|}}|
        }
    \right),
\end{multline}
where the approximation assumes the cutoff is much larger than the gap, $\Lambda \gg |\Delta_{k_{\|}}|$.

Next we tackle the polarization components. To express these as Fermi surface quantities, we must expand in $\vb{q}$. To that end, we express
\begin{subequations}
\begin{equation}
    \xi_{\vb{k-q}}
    \approx
    \xi_{\vb{k}} - \vb{v}_{\vb{k}} \vdot \vb{q},
\end{equation}
\begin{equation}
    \chi^{(\ell)}_{\vb{k-q}}
    \approx
    \chi^{(\ell)}_{\vb{k}}
    -
    \vb{u}^{(\ell)}_{\vb{k}} \vdot \vb{q},
\end{equation}
and
\begin{equation}
    \Delta_{\vb{k-q}}
    \approx
    \Delta_{\vb{k}} 
    - \vb{u}^{(\Delta)}_{\vb{k}} \vdot \vb{q},
\end{equation}
\end{subequations}
where $\vb{v} = \grad_{\vb{k}} \xi$, $\vb{u}^{(\ell)} = \grad_{\vb{k}} \chi^{(\ell)}$, and $\vb{u}^{(\Delta)} = \grad_{\vb{k}} \Delta$.
In principle, all three terms contribute to the $\vb{q}$ dependence. However, in practice, the Fermi velocity $\vb{v}$ is likely to be large compared to the ``gap velocity'' $\vb{u}^{(\Delta)}$. If we assume that the gap varies at most slowly in the $k_{\perp}$ direction (normal to the Fermi surface), and varies considerably only parallel to the Fermi surface, then $|\grad_{\vb{k}} \Delta| \approx |\partial_{k_{\|}} \Delta| \sim (1/k_{\text{F}}) |\partial_{\varphi} \Delta|$ assuming that $k_{\text{F}}$ does not vary too drastically as a function of the angle $\varphi$. Therefore, assuming $|\partial_{\varphi} \Delta| \lesssim |\Delta|$ and denoting $v = |\vb{v}|$, we compare
\begin{equation}    \label{eq:gap_velocity_approx}
    |\vb{u}^{(\Delta)}|
    \lesssim \frac{|\Delta|}{k_{\text{F}}} 
    \sim \frac{|\Delta|^2 \xi_{\text{BCS}}}{\mu}
    \quad \text{ vs. } \quad
    v \sim |\Delta| \xi_{\text{BCS}},
\end{equation}
so we expect $\vb{q}$ dependence arising from the gap velocity to be smaller by a factor of $|\Delta| / \mu$ relative to the dependence arising from the Fermi velocity. Therefore, since this factor is likely to be very small, in practice we neglect the gap velocity entirely. We make a similar approximation for $\chi^{(\ell)}$. We assumed that $\chi^{(\ell)}_{\vb{k}}$ is essentially independent of the normal direction to the Fermi surface. Again using the assumption that $k_{\text{F}}$ does not vary too much with $\varphi$, we then have $\vb{q} \vdot \grad_{\vb{k}} \chi^{(\ell)} \sim (1/k_{\text{F}}) \vb{q} \vdot \hat{\boldsymbol{\varphi}} \partial_{\varphi} \chi^{(\ell)}$ at the Fermi surface. This contains the small parameter $|\vb{q}| / k_{\text{F}} \sim |\vb{q}| v / \mu \sim \xi_{\text{BCS}} |\vb{q}| |\Delta| / \mu$. We capture $\vb{q}$ dependence to zeroth order in $|\Delta|/\mu$, so we then approximate $\chi^{(\ell)}_{\vb{k-q}} \approx \chi^{(\ell)}_{\vb{k}}$, $\Delta_{\vb{k-q}} \approx \Delta_{\vb{k}}$, and $\xi_{\vb{k-q}} \approx \xi_{\vb{k}} - \vb{v}_{\vb{k}} \vdot \vb{q}$ in all integrals.

Neglecting terms of $O(\vb{v} \vdot \vb{q} / \Lambda)$, we ultimately find below that the polarization components are 
\begin{subequations}    \label{eq:polarizations}
    \begin{multline}
        \Pi^{\mathcal{RR}}_{q; \ell \ell'}
        =
        -i^{\rho_{\ell \ell'}} \int_{\text{FS}}
        \frac{\dd{k_{\|}}}{4 \pi^2 v}
        \chi^{(\ell)} \chi^{(\ell')}
        \bigg[
            \tilde{\Lambda}
            \\
            -
            \left(
                \cos^2 \theta - \sin^2 z
            \right)
            \frac{z}{\sin z \cos z}
        \bigg],
    \end{multline}
    \begin{multline}
        \Pi^{\mathcal{II}}_{q; \ell \ell'}
        =
        -i^{\rho_{\ell \ell'}} \int_{\text{FS}}
        \frac{\dd{k_{\|}}}{4 \pi^2 v}
        \chi^{(\ell)} \chi^{(\ell')}
        \bigg[
            \tilde{\Lambda}
            \\
            -
            \left(
                \sin^2 \theta - \sin^2 z
            \right)
            \frac{z}{\sin z \cos z}
        \bigg],
    \end{multline}
    and
    \begin{multline}
        \Pi^{\mathcal{RI}}_{q; \ell \ell'}
        =
        \Pi^{\mathcal{IR}}_{q; \ell \ell'}
        \\
        =
        i^{\rho_{\ell \ell'}}
        \int_{\text{FS}} \frac{\dd{k_{\|}}}{4 \pi^2 v}
        \chi^{(\ell)} \chi^{(\ell')}
        \sin (2 \theta)
        \frac{z}{2 \sin z \cos z},
    \end{multline}
\end{subequations}
where
\begin{equation}
    \theta = \arg \Delta,
\end{equation}
\begin{multline}
    z
    = 
    \arcsin \left(
        \frac{\sqrt{(i \Omega_m)^2 - (\vb{v} \vdot \vb{q})^2}}{2 |\Delta|}
    \right)
    \\
    \xrightarrow{i \Omega_m \rightarrow \Omega + i 0^{+}}
    \arcsin \left(
        \frac{\sqrt{(\Omega + i 0^{+})^2 - (\vb{v} \vdot \vb{q})^2}}{2 |\Delta|}
    \right),
\end{multline}
and
\begin{equation}
    \tilde{\Lambda}
    =
    \arcsinh \left(
        \frac{\Lambda}{|\Delta|}
    \right).
\end{equation}
The drive couplings are
\begin{widetext}
\begin{subequations}
    \begin{multline}
        \Pi^{d \mathcal{R}}_{q; \ell}
        =
        i^{\rho_{\ell}}
        \int_{\text{FS}} 
        \frac{\dd{k_{\|}}}{4 \pi^2 v}
        \bra{u} \hat{o} \ket{u}
        \chi^{(\ell)} \sin \theta 
        \frac{i \Omega_m}{|\Delta|}
        \frac{i z}{
            \sin z \cos z
        }
        \\
        \xrightarrow{i \Omega_m \rightarrow \Omega + i 0^{+}}
        i^{\rho_{\ell}}
        \int_{\text{FS}} 
        \frac{\dd{k_{\|}}}{4 \pi^2 v}
        \bra{u} \hat{o} \ket{u}
        \chi^{(\ell)} \sin \theta 
        \frac{\Omega + i 0^{+}}{|\Delta|}
        \frac{i z}{
            \sin z \cos z
        }
    \end{multline}
    and
    \begin{multline}
        \Pi^{d \mathcal{I}}_{q; \ell} 
        =
        -i^{\rho_{\ell}}
        \int_{\text{FS}} 
        \frac{\dd{k_{\|}}}{4 \pi^2 v}
        \bra{u} \hat{o} \ket{u}
        \chi^{(\ell)} \cos \theta 
        \frac{i \Omega_m}{|\Delta|}
        \frac{i z}{
            \sin z \cos z
        }
        \\
        \xrightarrow{i \Omega_m \rightarrow \Omega + i 0^{+}}
        -i^{\rho_{\ell}}
        \int_{\text{FS}} 
        \frac{\dd{k_{\|}}}{4 \pi^2 v}
        \bra{u} \hat{o} \ket{u}
        \chi^{(\ell)} \cos \theta 
        \frac{\Omega + i 0^{+}}{|\Delta|}
        \frac{i z}{
            \sin z \cos z
        },
    \end{multline}
\end{subequations}
with
\begin{equation}
    \rho_{\ell}
    =
    \begin{cases}
        0 & g_{\ell} < 0,   \\
        1 & g_{\ell} > 0.
    \end{cases}
\end{equation}
The quasiparticle response is
\begin{equation}    \label{eq:qp_resp}
    \Pi^{dd}_{q}
    =
    - \int_{\text{FS}} 
    \frac{\dd{k_{\|}}}{4 \pi^2 v}
    \bra{u} \hat{o} \ket{u}^2
    \frac{(i \Omega_m)^2}{4 |\Delta|^2}
    \frac{2 z}{
        \sin^3 z \cos z
    }
\end{equation}
prior to analytic continuation ($i \Omega_m \rightarrow \Omega + i 0^{+})$. The remaining Fermi surface integration can be performed numerically.

\subsubsection{Evaluating the polarization components}
First we evaluate
\begin{multline}
    \lim_{T \rightarrow 0}
    \frac{T}{L^2} \sum_{k} 
    \chi^{(\ell)}_{\vb{k}} \chi^{(\ell')}_{\vb{k}} G_{k} G_{-k+q}
    \\
    \approx
    \int_{\text{FS}} \frac{\dd{k_{\|}}}{4 \pi^2 v}
    \chi^{(\ell)} \chi^{(\ell')}
    \int_{-\Lambda}^{\Lambda} \dd{\xi}
    \left\lbrace
        \frac{
            (E - \xi)
            (E + \xi - \vb{v} \vdot \vb{q} + i \Omega_m)
        }{
            2 E \left[
                (i \Omega_m + E)^2 - (E')^2
            \right]
        }
        +
        \frac{
            (E' + \xi - \vb{v} \vdot \vb{q})
            (E' - \xi - i \Omega_m)
        }{
            2 E' \left[
                (i \Omega_m - E')^2 - E^2
            \right]
        }
    \right\rbrace
\end{multline}
where $E = \sqrt{\xi^2 + |\Delta|^2}$, and $E' = \sqrt{(\xi - \vb{v} \vdot \vb{q})^2 + |\Delta|^2}$, and we dropped subscripts on all quantities evaluated at $k_{\|}$ for brevity. We can then shift the integration variable $\xi \rightarrow \xi + \vb{v} \vdot \vb{q}$ in the second term, and after dropping terms of $O(\vb{v} \vdot \vb{q} / \Lambda)$, we obtain
\begin{multline}
    \lim_{T \rightarrow 0}
    \frac{T}{L^2} \sum_{k} 
    \chi^{(\ell)}_{\vb{k}} \chi^{(\ell')}_{\vb{k}} G_{k} G_{-k+q}
    \approx
    \int_{\text{FS}} \frac{\dd{k_{\|}}}{4 \pi^2 v}
    \chi^{(\ell)} \chi^{(\ell')}
    \int_{-\Lambda}^{\Lambda} \dd{\xi}
    \Bigg\lbrace
        \frac{
            (E - \xi)
            (E + \xi - \vb{v} \vdot \vb{q} + i \Omega_m)
        }{
            2 E \left[
                (i \Omega_m)^2 + 2 i \Omega_m E
                + 2 \xi (\vb{v} \vdot \vb{q}) 
                - (\vb{v} \vdot \vb{q})^2
            \right]
        }
    \\
        +
        \frac{
            (E + \xi)
            (E - \xi - \vb{v} \vdot \vb{q} - i \Omega_m)
        }{
            2 E \left[
                (i \Omega_m)^2 - 2 i \Omega_m E 
                - 2 \xi (\vb{v} \vdot \vb{q})
                - (\vb{v} \vdot \vb{q})^2
            \right]
        }
    \Bigg\rbrace
    \\
    =
    \int_{\text{FS}} \frac{\dd{k_{\|}}}{4 \pi^2 v}
    \chi^{(\ell)} \chi^{(\ell')}
    \int_{-\Lambda}^{\Lambda} \dd{\xi}
    \Re{
        \frac{1}{E}
        \frac{
            |\Delta|^2 + (E - \xi) (i \Omega_m - \vb{v} \vdot \vb{q})
        }{
            (i \Omega_m)^2 + 2 i \Omega_m E
            + 2 \xi (\vb{v} \vdot \vb{q}) 
            - (\vb{v} \vdot \vb{q})^2
        }
    }.
\end{multline}
Performing the same shift,
\begin{multline}
    \lim_{T \rightarrow 0} \frac{T}{L^2} \sum_{k}
    \chi^{(\ell)}_{\vb{k}} \chi^{(\ell')}_{\vb{k-q}}
    F_{k} F_{k-q}
    \approx
    - \int_{\text{FS}} \frac{\dd{k_{\|}}}{4 \pi^2 v}
    \chi^{(\ell)} \chi^{(\ell')}
    \int_{-\Lambda}^{\Lambda} \dd{\xi}
    \Bigg\lbrace
        \frac{\Delta^2}{
            2 E \left[
                (i \Omega_m)^2 + 2 i \Omega_m E
                + 2 \xi (\vb{v} \vdot \vb{q}) - (\vb{v} \vdot \vb{q})^2
            \right]
        }
    \\
        +
        \frac{\Delta^2}{
            2 E \left[
                (i \Omega_m)^2 - 2 i \Omega_m E
                - 2 \xi (\vb{v} \vdot \vb{q}) - (\vb{v} \vdot \vb{q})^2
            \right]
        }
    \Bigg\rbrace
    \\
    =
    - \int_{\text{FS}} \frac{\dd{k_{\|}}}{4 \pi^2 v}
    \chi^{(\ell)} \chi^{(\ell')}
    \int_{-\Lambda}^{\Lambda} \dd{\xi}
    \Delta^2 \Re{
        \frac{1}{E}
        \frac{1}{
            (i \Omega_m)^2 + 2 i \Omega_m E
            + 2 \xi (\vb{v} \vdot \vb{q}) - (\vb{v} \vdot \vb{q})^2
        }
    }.
\end{multline}

We then combine these expressions to find the polarization components, writing $\Delta = |\Delta| e^{i \theta}$:
\begin{multline}
    \Pi^{\mathcal{RR}}_{q; \ell \ell'}
    \approx
    - i^{\rho_{\ell \ell'}} \frac{1}{2}
    \int_{\text{FS}} \frac{\dd{k_{\|}}}{4 \pi^2 v}
    \chi^{(\ell)} \chi^{(\ell')}
    \int_{-\Lambda}^{\Lambda} \dd{\xi}
    \frac{1}{E}
    \Bigg[
        \Re{
            \frac{
                |\Delta|^2 + (E - \xi)(i \Omega_m - \vb{v} \vdot \vb{q})
            }{
                (i \Omega_m)^2 + 2 i \Omega_m E
                + 2 \xi (\vb{v} \vdot \vb{q}) - (\vb{v} \vdot \vb{q})^2
            }
        }
        \\
        +
        \Re{
            \frac{
                |\Delta|^2 - (E - \xi)(i \Omega_m - \vb{v} \vdot \vb{q})
            }{
                (i \Omega_m)^2 - 2 i \Omega_m E
                - 2 \xi (\vb{v} \vdot \vb{q}) - (\vb{v} \vdot \vb{q})^2
            }
        }
        + 2 \Re{\Delta^2} \Re{
            \frac{1}{
                (i \Omega_m)^2 + 2 i \Omega_m E
                + 2 \xi (\vb{v} \vdot \vb{q}) - (\vb{v} \vdot \vb{q})^2
            }
        }
    \Bigg]
    \\
    =
    - i^{\rho_{\ell \ell'}}
    \int_{\text{FS}} \frac{\dd{k_{\|}}}{4 \pi^2 v}
    \chi^{(\ell)} \chi^{(\ell')}
    \int_{-\Lambda}^{\Lambda} \dd{\xi}
    \Re{
        \frac{1}{E} 
        \frac{
            |\Delta|^2 (1 + \cos 2 \theta) 
            + \xi (\vb{v} \vdot \vb{q}) + i \Omega_m E
        }{
            (i \Omega_m)^2 + 2 i \Omega_m E
            + 2 \xi (\vb{v} \vdot \vb{q}) - (\vb{v} \vdot \vb{q})^2
        }
    },
\end{multline}
\begin{equation}
    \Pi^{\mathcal{II}}_{q; \ell \ell'}
    \approx
    - i^{\rho_{\ell \ell'}}
    \int_{\text{FS}} \frac{\dd{k_{\|}}}{4 \pi^2 v}
    \chi^{(\ell)} \chi^{(\ell')}
    \int_{-\Lambda}^{\Lambda} \dd{\xi}
    \Re{
        \frac{1}{E} 
        \frac{
            |\Delta|^2 (1 - \cos 2 \theta) 
            + \xi (\vb{v} \vdot \vb{q}) + i \Omega_m E
        }{
            (i \Omega_m)^2 + 2 i \Omega_m E
            + 2 \xi (\vb{v} \vdot \vb{q}) - (\vb{v} \vdot \vb{q})^2
        }
    }
\end{equation}

\begin{multline}
    \Pi^{\mathcal{RI}}_{q; \ell \ell'}
    \approx
    i^{\rho_{\ell \ell'} - 1} \frac{1}{2}
    \int_{\text{FS}} \frac{\dd{k_{\|}}}{4 \pi^2 v}
    \chi^{(\ell)} \chi^{(\ell')}
    \int_{-\Lambda}^{\Lambda} \dd{\xi}
    \frac{1}{E}
    \Bigg[
        \Re{
            \frac{
                |\Delta|^2 + (E - \xi) (i \Omega_m - \vb{v} \vdot \vb{q})
            }{
                (i \Omega_m)^2 + 2 i \Omega_m E
                + 2 \xi (\vb{v} \vdot \vb{q}) 
                - (\vb{v} \vdot \vb{q})^2
            }
        }
    \\
        -
        \Re{
            \frac{
                |\Delta|^2 - (E - \xi) (i \Omega_m - \vb{v} \vdot \vb{q})
            }{
                (i \Omega_m)^2 - 2 i \Omega_m E
                - 2 \xi (\vb{v} \vdot \vb{q}) 
                - (\vb{v} \vdot \vb{q})^2
            }
        }
    \\
        - 2 i \Im {\Delta^2}
        \Re{
            \frac{1}{
                (i \Omega_m)^2 + 2 i \Omega_m E
                + 2 \xi (\vb{v} \vdot \vb{q}) - (\vb{v} \vdot \vb{q})^2
            }
        }
    \Bigg]
    \\
    =
    - i^{\rho_{\ell \ell'} - 1}
    \int_{\text{FS}} \frac{\dd{k_{\|}}}{4 \pi^2 v}
    \chi^{(\ell)} \chi^{(\ell')}
    \int_{-\Lambda}^{\Lambda} \dd{\xi}
    \frac{1}{E}
    \Bigg[
        \Re{
            \frac{
                i \Omega_m \xi + E (\vb{v} \vdot \vb{q})
            }{
                (i \Omega_m)^2 + 2 i \Omega_m E
                + 2 \xi (\vb{v} \vdot \vb{q}) 
                - (\vb{v} \vdot \vb{q})^2
            }
        }
    \\
        + i |\Delta|^2 \sin 2 \theta
        \Re{
            \frac{1}{
                (i \Omega_m)^2 + 2 i \Omega_m E
                + 2 \xi (\vb{v} \vdot \vb{q}) - (\vb{v} \vdot \vb{q})^2
            }
        }
    \Bigg]
\end{multline}
and
\begin{multline}
    \Pi^{\mathcal{IR}}_{q; \ell \ell'}
    \approx
    - i^{\rho_{\ell \ell'} - 1}
    \int_{\text{FS}} \frac{\dd{k_{\|}}}{4 \pi^2 v}
    \chi^{(\ell)} \chi^{(\ell')}
    \int_{-\Lambda}^{\Lambda} \dd{\xi}
    \frac{1}{E}
    \Bigg[
        - \Re{
            \frac{
                i \Omega_m \xi + E (\vb{v} \vdot \vb{q})
            }{
                (i \Omega_m)^2 + 2 i \Omega_m E
                + 2 \xi (\vb{v} \vdot \vb{q}) 
                - (\vb{v} \vdot \vb{q})^2
            }
        }
    \\
        + i |\Delta|^2 \sin 2 \theta
        \Re{
            \frac{1}{
                (i \Omega_m)^2 + 2 i \Omega_m E
                + 2 \xi (\vb{v} \vdot \vb{q}) - (\vb{v} \vdot \vb{q})^2
            }
        }
    \Bigg].
\end{multline}
\end{widetext}

To perform the $\xi$ integration, we use the change of variables
\begin{subequations}
\begin{equation}
    \frac{\xi}{|\Delta_{k_{\|}}|}
    =
    \sinh (x_{k_{\|}}),
\end{equation}
\begin{equation}
    \frac{E}{|\Delta|}
    =
    \sqrt{\sinh(x)^2 + 1}
    = \cosh(x),
\end{equation}
\begin{equation}
    \frac{\vb{v} \vdot \vb{q}}{|\Delta|}
    =
    i a \sinh (i b),
\end{equation}
\begin{equation}
    \frac{i \Omega_m}{|\Delta|} = i a \cosh (i b),
\end{equation}
\end{subequations}
where
\begin{equation}
    a 
    =
    \frac{1}{|\Delta|}
    \sqrt{\Omega_m^2 + (\vb{v} \vdot \vb{q})^2}
    \in \mathds{R}.
\end{equation}
Note that since $\sinh(i b) \in i \mathds{R}$ and $\cosh(i b) \in \mathds{R}$, it follows that $b \in \mathds{R}$. (Note that in this Appendix, we never use $a$ and $b$ to refer to clapping modes, and use those letters only to label these variables.)
Then,
\begin{widetext}
\begin{multline}
    \Pi^{\mathcal{RR}}_{q; \ell \ell'}
    \approx
    - i^{\rho_{\ell \ell'}}
    \int_{\text{FS}} \frac{\dd{k_{\|}}}{4 \pi^2 v}
    \chi^{(\ell)} \chi^{(\ell')}
    \int_{-\Lambda}^{\Lambda} \dd{\xi}
    \Re{
        \frac{1}{|\Delta| \cosh(x)}
        \frac{
            1 + \cos 2 \theta + i a \cosh (x + i b)
        }{
            2 i a \cosh(x + i b) - a^2
        }
    }
    \\
    =
    - i^{\rho_{\ell \ell'}}
    \int_{\text{FS}} \frac{\dd{k_{\|}}}{4 \pi^2 v}
    \chi^{(\ell)} \chi^{(\ell')}
    \int_{-\tilde{\Lambda}}^{\tilde{\Lambda}} \dd{x}
    \Re{
        \frac{
            1 + \cos 2 \theta + i a \cosh (x + i b)
        }{
            2 i a \cosh(x+y) - a^2
        }
    }
    \\
    =
    - i^{\rho_{\ell \ell'}}
    \int_{\text{FS}} \frac{\dd{k_{\|}}}{4 \pi^2 v}
    \chi^{(\ell)} \chi^{(\ell')}
    \int_{-\tilde{\Lambda}}^{\tilde{\Lambda}} \dd{x}
    \Re{
        \frac{
            1 + \cos 2 \theta + \frac{1}{2} a^2
        }{
            2 i a \cosh(x + i b) - a^2
        }
        + \frac{1}{2}
    }
    \\
    \approx
    - i^{\rho_{\ell \ell'}}
    \int_{\text{FS}} \frac{\dd{k_{\|}}}{4 \pi^2 v}
    \chi^{(\ell)} \chi^{(\ell')}
    \left[
        \tilde{\Lambda}
        +
        \left(
            1 + \cos 2 \theta + \frac{1}{2} a^2
        \right)
        \int_{-\infty}^{\infty} \dd{x}
        \Re{\frac{1}{2 i a \cosh(x + i b) - a^2}}
    \right],
\end{multline}
\end{widetext}
where $\tilde{\Lambda} = \arcsinh(\Lambda / |\Delta|)$, and the final approximation evaluates the convergent part of the integral in the $\Lambda \rightarrow \infty$ limit.
For the remaining $x$ integral, note that even though $i b \in i \mathds{R}$ (so a naive variable shift $x \rightarrow x - i b$ is not valid without ensuring complex analyticity), one has 
\begin{multline}
    \int_{-\infty}^{\infty} \dd{x}
    \frac{1}{2 i a \cosh(x + i b) - a^2}
    \\
    =
    - \frac{4 \arctanh \left( \frac{2i + a}{\sqrt{4 + a^2}}\right)}{a \sqrt{4 + a^2}}
\end{multline}
which is independent of $b$, so we can set $b = 0$ inside the integral and take the real part explicitly to get a simpler expression:
\begin{multline}
    \int_{-\infty}^{\infty} \dd{x}
    \Re{
        \frac{1}{2 i a \cosh(x + i b) - a^2}
    }
    \\
    =
    \frac{1}{2 a} \int_{-\infty}^{\infty} \dd{x}
    \left(
        \frac{1}{-a + 2 i \cosh(x)}
        - \frac{1}{a + 2 i \cosh(x)}
    \right)
    \\
    =
    -\int_{-\infty}^{\infty} \dd{x}
    \frac{1}{a^2 + 4 \cosh^2 (x)}
    \\
    =
    -\frac{\arcsinh(a/2)}{a \sqrt{1 + (a/2)^2}}.
\end{multline}
Therefore,
\begin{multline}
    \Pi^{\mathcal{RR}}_{q; \ell \ell'}
    =
    -i^{\rho_{\ell \ell'}}
    \int_{\text{FS}} \frac{\dd{k_{\|}}}{4 \pi^2 v}
    \chi^{(\ell)} \chi^{(\ell')}
    \Bigg[
        \tilde{\Lambda}
        \\
        - \left(
            1 + \cos 2 \theta + \frac{1}{2} a^2
        \right)
        \frac{\arcsinh(a/2)}{a \sqrt{1 + (a/2)^2}}
    \Bigg].
\end{multline}
We now define
\begin{multline}
    z = i \arcsinh \left(
        \frac{a}{2}
    \right)
    \\
    =
    i \arcsinh \left(
        \frac{
            -i \sqrt{(i \Omega_m)^2 - (\vb{v} \vdot \vb{q})^2}
        }{2 |\Delta|}
    \right)
    \\
    =
    \arcsin \left(
        \frac{
            \sqrt{(i \Omega_m)^2 - (\vb{v} \vdot \vb{q})^2}
        }{2 |\Delta|}
    \right).
\end{multline}
We can then express
\begin{equation}
    a = 2 \sinh(-i z) = - 2 i \sin z
\end{equation}
and
\begin{equation}
    \sqrt{1 + (a/2)^2}
    =
    \sqrt{1 - \sin^2 z}
    =
    \sqrt{\cos^2 z}
    =
    \cos z.
\end{equation}
The final equality uses
\begin{multline}
    \cos z
    =
    \cos \left[
        \arcsin \left(
            \frac{
                \sqrt{(i \Omega_m)^2 - (\vb{v} \vdot \vb{q})^2}
            }{2 |\Delta|}
        \right)
    \right]
    \\
    =
    \sqrt{
        1
        +
        \frac{\Omega_m^2 + (\vb{v} \vdot \vb{q})^2}{4|\Delta|^2}
    }
    > 0.
\end{multline}

Then,
\begin{multline}
    \Pi^{\mathcal{R R}}_{q; \ell \ell'}
    =
    -i^{\rho_{\ell \ell'}} \int_{\text{FS}}
    \frac{\dd{k_{\|}}}{4 \pi^2 v}
    \chi^{(\ell)} \chi^{(\ell')}
    \Big[
        \tilde{\Lambda}
        \\
        -
        \left(
            \cos^2 \theta - \sin^2 z
        \right)
        \frac{z}{\sin z \cos z}
    \Big].
\end{multline}
Similarly,
\begin{multline}
    \Pi^{\mathcal{II}}_{q; \ell \ell'}
    \approx
    -i^{\rho_{\ell \ell'}} \int_{\text{FS}}
    \frac{\dd{k_{\|}}}{4 \pi^2 v}
    \chi^{(\ell)} \chi^{(\ell')}
    \Big[
        \tilde{\Lambda}
        \\
        -
        \left(
            \sin^2 \theta - \sin^2 z
        \right)
        \frac{z}{\sin z \cos z}
    \Big].
\end{multline}

For the cross terms,
\begin{widetext}
\begin{multline}
    \Pi^{\mathcal{RI}}_{q; \ell \ell'}
    \\
    \approx
    - i^{\rho_{\ell \ell'} - 1}
    \int_{\text{FS}} \frac{\dd{k_{\|}}}{4 \pi^2 v}
    \chi^{(\ell)} \chi^{(\ell')}
    \int_{-\tilde{\Lambda}}^{\tilde{\Lambda}} \dd{x}
    \Bigg[
        \Re{
            \frac{ \sinh(x + i b) }{i a + 2 \cosh(x + i b)}
        }
        + i \sin 2 \theta
        \Re{
            \frac{1}{2 i a \cosh(x + i b) - a^2}
        }
    \Bigg]
    \\
    \approx
    - i^{\rho_{\ell \ell'} - 1}
    \int_{\text{FS}} \frac{\dd{k_{\|}}}{4 \pi^2 v}
    \chi^{(\ell)} \chi^{(\ell')}
    \Bigg[
        \Re{
            \int_{-\tilde{\Lambda}}^{\tilde{\Lambda}} \dd{x}
            \frac{ \sinh(x + i b) }{i a + 2 \cosh(x + i b)}
        }
        + i \sin 2 \theta
        \Re{
            \int_{-\infty}^{\infty} \dd{x}
            \frac{1}{2 i a \cosh(x + i b) - a^2}
        }
    \Bigg]
    \\
    =
    i^{\rho_{\ell \ell'} - 1}
    \int_{\text{FS}} \frac{\dd{k_{\|}}}{4 \pi^2 v}
    \chi^{(\ell)} \chi^{(\ell')}
    \Bigg[
        \sin 2 \theta
        \frac{i z}{2 \sin z \cos z}
        -
        \Re{
            \int_{-\tilde{\Lambda}}^{\tilde{\Lambda}} \dd{x}
            \frac{ \sinh(x + i b) }{i a + 2 \cosh(x + i b)}
        } 
    \Bigg]
\end{multline}
and
\begin{equation}
    \Pi^{\mathcal{IR}}_{q; \ell \ell'}
    \approx
    i^{\rho_{\ell \ell'} - 1}
    \int_{\text{FS}} \frac{\dd{k_{\|}}}{4 \pi^2 v}
    \chi^{(\ell)} \chi^{(\ell')}
    \Bigg[
        \sin 2 \theta
        \frac{i z}{2 \sin z \cos z}
        +
        \Re{
            \int_{-\tilde{\Lambda}}^{\tilde{\Lambda}} \dd{x}
            \frac{ \sinh(x + i b) }{i a + 2 \cosh(x + i b)}
        } 
    \Bigg].
\end{equation}
\end{widetext}
The remaining $x$ integral evaluates to
\begin{multline}
    I
    =
    \int_{-\tilde{\Lambda}}^{\tilde{\Lambda}} \dd{x}
    \frac{\sinh(x + ib)}{i a + 2 \cosh(x + i b)}
    \\
    =
    \frac{1}{2} \bigg\lbrace
        \Log \left[
            a - 2 i \cos(b - i \tilde{\Lambda})
        \right]
        \\
        - \Log \left[
            a - 2 i \cos(b + i \tilde{\Lambda})
        \right]
    \bigg\rbrace
\end{multline}
where $\Log$ is the principal branch of the complex $\log$. We can then express
\begin{equation}
    \cos (b \mp i \tilde{\Lambda})
    =
    \cos (b) \cosh( \tilde{\Lambda} )
    \pm
    i \sin (b) \sinh( \tilde{\Lambda} ).
\end{equation}
In the large-cutoff limit $\Lambda \rightarrow \infty$, $\tilde{\Lambda} \rightarrow \infty$ as well, so $\cosh (\tilde{\Lambda}) \approx \sinh(\tilde{\Lambda}) \approx e^{\tilde{\Lambda}}/2 \equiv \lambda$. Then,
\begin{multline}
    I
    =
    \frac{1}{2}
    \Big\lbrace
        \Log \left[
            a^2 - 2 i a \lambda \left(
                \cos(b) + i \sin(b)
            \right)
        \right]
        \\
        -
        \Log \left[
            a^2 - 2 i a \lambda \left(
                \cos(b) - i \sin(b)
            \right)
        \right]
    \Big\rbrace.
\end{multline}
Noting that $a^2 = (\Omega_m^2 + (\vb{v} \vdot \vb{q})^2) / |\Delta|^2$, $i a \cos (b) = i \Omega_m / |\Delta|$, and $a \sin(b) = - (\vb{v} \vdot \vb{q}) / |\Delta|$, it follows that
\begin{multline}
    I
    =
    \frac{1}{2} \bigg\lbrace
        \Log \left[
            \frac{\Omega_m^2 + (\vb{v} \vdot \vb{q})^2}{|\Delta|^2}
            - 2 \lambda \frac{i \Omega_m}{|\Delta|}
            - 2 \lambda \frac{\vb{v} \vdot \vb{q}}{|\Delta|}
        \right]
        \\
        -
        \Log \left[
            \frac{\Omega_m^2 + (\vb{v} \vdot \vb{q})^2}{|\Delta|^2}
            - 2 \lambda \frac{i \Omega_m}{|\Delta|}
            + 2 \lambda \frac{\vb{v} \vdot \vb{q}}{|\Delta|}
        \right]
    \bigg\rbrace.
\end{multline}
Since the cutoff is large ($\lambda \gg 1$), we can approximate
\begin{multline}
    I
    \approx
    \frac{1}{2} \bigg\lbrace
        \Log \left[
            -2 \lambda
            \frac{i \Omega_m}{|\Delta|}
            - 2 \lambda 
            \frac{\vb{v} \vdot \vb{q}}{|\Delta|}
        \right]
        \\
        - \Log \left[
            -2 \lambda
            \frac{i \Omega_m}{|\Delta|}
            + 2 \lambda 
            \frac{\vb{v} \vdot \vb{q}}{|\Delta|}
        \right]
    \bigg\rbrace.
\end{multline}
Notice that the complex arguments of the $\Log$s have the same magnitude, meaning the difference of $\Log$s is purely imaginary. Therefore, taking the real part yields zero, and we conclude that
\begin{multline}
    \Pi^{\mathcal{RI}}_{q; \ell \ell'}
    =
    \Pi^{\mathcal{IR}}_{q; \ell \ell'}
    \\
    =
    i^{\rho_{\ell \ell'}}
    \int_{\text{FS}} \frac{\dd{k_{\|}}}{4 \pi^2 v}
    \chi^{(\ell)} \chi^{(\ell')}
    \sin (2 \theta)
    \frac{z}{2 \sin z \cos z}.
\end{multline}

\subsubsection{Evaluating the driving terms}
We now consider the response to driving. We assume that the drive field $d$ couples locally in real space as
\begin{multline}
    H_{\text{drive}} (t)
    =
    \int \dd^2{\vb{r}} \sum_{j}
    d(t, \vb{r}) o_{j}
    \psi^{\dag}_{j} \psi_{j}
    \\
    =
    \sum_{\vb{k q} j}
    d_{\vb{q}} (t) o_{j} f^{\dag}_{\vb{k} j} f_{\vb{k-q}, j}.
\end{multline}
$j$ indexes the orbitals in the unit cell, spin, and any other internal degree of freedom; $o_{j}$ encodes how the field $d$ affects the energy of local state $j$. For our RTG calculation, the relevant index will be the layer.

Suppose the noninteracting part of the problem is diagonalized in terms of the band operators $c_{\vb{k} \alpha \sigma}$, where $\alpha$ labels the bands. Assuming no spin-orbit coupling, the band operators correspond to wavefunctions $\ket{u_{\vb{k} \alpha}}$, i.e., $u_{\vb{k} \alpha j} = \braket{j}{u_{\vb{k} \alpha}}$, with
\begin{equation}
    f_{\vb{k} j \sigma}
    =
    \sum_{\alpha} u_{\vb{k} j \alpha} c_{\vb{k} \alpha \sigma}.
\end{equation}
In the band basis, the driving Hamiltonian is then
\begin{multline}
    H_{\text{drive}}
    =
    \sum_{\substack{
        \vb{k q} \\ j \alpha \alpha' \sigma
    }}
    d_{\vb{q}} (t)
    u^{*}_{\vb{k} \alpha j}
    o_{j} 
    u_{\vb{k-q}, \alpha' j}
    c^{\dag}_{\vb{k} \alpha \sigma} c_{\vb{k-q}, \alpha' \sigma}
    \\
    \equiv
    \sum_{\vb{k q} \alpha \alpha' \sigma}
    d_{\vb{q}} (t)
    \bra{u_{\vb{k} \alpha}} \hat{o} \ket{u_{\vb{k-q}, \alpha'}}
    c^{\dag}_{\vb{k} \alpha \sigma} c_{\vb{k-q}, \alpha' \sigma}.
\end{multline}
$d (t, \vb{r}) \hat{o}$ is the one-particle operator corresponding to driving; we assumed $\hat{o}$ is diagonal in the basis of orbitals within the unit cell. The last equation is more general, and would also hold for a drive which induces transitions between orbitals.

For the RTG calculation, we will project to a single active band, in which case the driving terms have one fewer summation:
\begin{multline}
    H_{\text{drive}}
    =
    \sum_{\vb{k q} \sigma}
    d_{\vb{q}} (t)
    \bra{u_{\vb{k}}} \hat{o} \ket{u_{\vb{k-q}}}
    c^{\dag}_{\vb{k} \sigma} c_{\vb{k-q}, \sigma}
    \\
    =
    \sum_{\vb{k q}}
    d_{\vb{q}} (t)
    \Psi^{\dag}_{\vb{k}}
    \begin{pmatrix}
        \bra{u_{\vb{k}}} \hat{o} \ket{u_{\vb{k-q}}} & 0 \\
        0 & -\bra{u_{\vb{-k+q}}} \hat{o} \ket{u_{-\vb{k}}}
    \end{pmatrix}
    \Psi_{\vb{k-q}}.
\end{multline}
For simplicity, we assume that the wavefunctions respect $\ket{u_{\vb{k}}} = \ket{u_{-\vb{k}}}$ such that
\begin{equation}
    H_{\text{drive}}
    =
    d_{\vb{q}} (t)
    \bra{u_{\vb{k}}} \hat{o} \ket{u_{\vb{k-q}}}
    \Psi^{\dag}_{\vb{k}} \begin{pmatrix}
        1 & 0 \\
        0 & -1
    \end{pmatrix} \Psi_{\vb{k-q}}.
\end{equation}
Wick rotating to imaginary time and Fourier transforming to Matsubara frequency, we obtain the vertex in the action,
\begin{equation}
    \mathds{V}^{(d)}_{k, k-q}
    =
    - d_{q} \bra{u_{\vb{k}}} \hat{o} \ket{u_{\vb{k-q}}} \begin{pmatrix}
        1 & 0 \\
        0 & -1
    \end{pmatrix}.
\end{equation}
Integrating out the fermion then yields two kinds of terms. The quadratic term corresponds to the quasiparticle response,
\begin{widetext}
\begin{multline}
    \frac{1}{L^2} \mathcal{F}_{\text{QP}} [d]
    =
    \frac{1}{2} \frac{T}{L^2} \Tr{
        \mathcal{G} \mathds{V}^{(d)} \mathcal{G} \mathds{V}^{(d)}
    }
    =
    \sum_{q} \Pi^{dd}_{q} d_{-q} d_{q}
    \\
    =
    \frac{1}{2} \frac{T}{L^2} \sum_{k q}
    d_{q} d_{-q}
    |\bra{u_{\vb{k}}} \hat{o} \ket{u_{\vb{k-q}}}|^2
    \tr{
        \mathcal{G}_{k}
        \begin{pmatrix}
            1 & 0 \\
            0 & -1
        \end{pmatrix}
        \mathcal{G}_{k-q}
        \begin{pmatrix}
            1 & 0 \\
            0 & -1
        \end{pmatrix}
    }
    \\
    =
    \frac{1}{2} \frac{T}{L^2} \sum_{k q}
    d_{q} d_{-q}
    |\bra{u_{\vb{k}}} \hat{o} \ket{u_{\vb{k-q}}}|^2
    \left(
        G_{k} G_{k-q}
        + G_{-k} G_{-k+q}
        - F_{k} \overline{F}_{k-q}
        - \overline{F}_{k} F_{k-q}
    \right),
\end{multline}
and the linear terms correspond to driving of the collective modes:
\begin{multline}
    \frac{1}{L^2} 
    \mathcal{F}_{\phi} [d, \mathcal{R}, \mathcal{I}]
    =
    \frac{T}{L^2} \Tr{
        \mathcal{G} \left(
            \mathds{V}^{(\mathcal{R})} + \mathds{V}^{(\mathcal{I})}
        \right)
        \mathcal{G} \mathds{V}^{(d)}
    }
    =
    \frac{T}{L^2} \sum_{k q \ell} \tr{
        \mathcal{G}_{k} \left(
            \mathds{V}^{(\mathcal{R}, \ell)}_{k, k-q} 
            + \mathds{V}^{(\mathcal{I}, \ell)}_{k, k-q}
        \right)
        \mathcal{G}_{k} \mathds{V}^{(d)}_{k-q, k}
    }
    \\
    =
    \sum_{q \ell}
    \left(
        d_{-q} \Pi^{d \mathcal{R}}_{q; \ell} 
        \mathcal{R}_{q}^{(\ell)}
        +
        d_{-q} \Pi^{d \mathcal{I}}_{q; \ell} 
        \mathcal{I}_{q}^{(\ell)}
    \right)
    \\
    =
    \frac{T}{L^2} \sum_{k q \ell}
    i^{\rho_{\ell}} d_{-q}
    \bra{u_{\vb{k-q}}} \hat{o} \ket{u_{\vb{k}}}
    \bigg\lbrace
        \mathcal{R}_{q}^{(\ell)} \left[
            \chi^{(\ell)}_{\vb{k-q}} \left(
                F_{k} G_{k-q} + G_{-k} F_{k-q}
            \right)
            + 
            \chi^{(\ell)}_{\vb{k}} \left(
                G_{k} \overline{F}_{k-q} + \overline{F}_{k} G_{-k+q}
            \right)
        \right]
        \\
        + i \mathcal{I}_{q}^{(\ell)} \left[
            - \chi^{(\ell)}_{\vb{k-q}} \left(
                F_{k} G_{k-q} + G_{-k} F_{k-q}
            \right)
            + \chi^{(\ell)}_{k} \left(
                G_{k} \overline{F}_{k-q} + \overline{F}_{k} G_{-k+q}
            \right)
        \right]
    \bigg\rbrace.
\end{multline}
Here we defined $\rho_{\ell} = 0$ if $g_{\ell} < 0$, and $\rho_{\ell} = 1$ if $g_{\ell} > 0$.

The Matsubara sums are again standard, yielding
\begin{equation}
    \lim_{T \rightarrow 0} T \sum_{\omega_n} G_{k} G_{k-q}
    =
    -\frac{1}{2 E_{\vb{k}}}
    \frac{
        (E_{\vb{k}} - \xi_{\vb{k}})
        (i \Omega_m + E_{\vb{k}} - \xi_{\vb{k-q}})
    }{(E_{\vb{k}} + i \Omega_m)^2 - E_{\vb{k-q}}^2}
    -
    \frac{1}{2 E_{\vb{k-q}}}
    \frac{
        (E_{\vb{k-q}} - \xi_{\vb{k-q}})
        (i \Omega_m + \xi_{\vb{k}} - E_{\vb{k-q}})
    }{E_{\vb{k}}^2 - (E_{\vb{k-q}} - i \Omega_m)^2},
\end{equation}
\begin{equation}
    \lim_{T \rightarrow 0} T \sum_{\omega_n} G_{-k} G_{-k+q}
    =
    -\frac{1}{2 E_{\vb{k}}}
    \frac{
        (E_{\vb{k}} + \xi_{\vb{k}})
        (i \Omega_m + E_{\vb{k}} + \xi_{\vb{k-q}})
    }{(E_{\vb{k}} + i \Omega_m)^2 - E_{\vb{k-q}}^2}
    -
    \frac{1}{2 E_{\vb{k-q}}}
    \frac{
        (E_{\vb{k-q}} + \xi_{\vb{k-q}})
        (i \Omega_m - \xi_{\vb{k}} - E_{\vb{k-q}})
    }{E_{\vb{k}}^2 - (E_{\vb{k-q}} - i \Omega_m)^2},
\end{equation}
\begin{equation}
    \lim_{T \rightarrow 0} T \sum_{\omega_n} F_{k} \overline{F}_{k-q}
    =
    -\frac{1}{2 E_{\vb{k}}}
    \frac{\Delta_{\vb{k}} \overline{\Delta}_{\vb{k-q}}}{
        (E_{\vb{k}} + i \Omega_m)^2 - E_{\vb{k-q}}^2
    }
    +
    \frac{1}{2 E_{\vb{k-q}}}
    \frac{\Delta_{\vb{k}} \overline{\Delta}_{\vb{k-q}}}{
        E_{\vb{k}}^2 - (E_{\vb{k-q}} - i \Omega_m)^2
    }
    =
    \left(
        \lim_{T \rightarrow 0} \sum_{\omega_n} \overline{F}_{k} F_{k-q}
    \right)^{*},
\end{equation}
\begin{equation}
    \lim_{T \rightarrow 0} T \sum_{\omega_n} F_{k} G_{k-q}
    =
    - \frac{\Delta_{\vb{k}}}{2 E_{\vb{k}}}
    \frac{E_{\vb{k}} + i \Omega_m - \xi_{\vb{k-q}}}{
        (E_{\vb{k}} + i \Omega_m)^2 - E_{\vb{k-q}}^2
    }
    +
    \frac{\Delta_{\vb{k}}}{2 E_{\vb{k-q}}}
    \frac{E_{\vb{k-q}} - \xi_{\vb{k-q}}}{
        E_{\vb{k}}^2 - (E_{\vb{k-q}} - i \Omega_m)^2
    },
\end{equation}
\begin{equation}
    \lim_{T \rightarrow 0} T \sum_{\omega_n} G_{-k} F_{k-q}
    =
    \frac{\Delta_{\vb{k-q}}}{2 E_{\vb{k}}}
    \frac{E_{\vb{k}} + \xi_{\vb{k}}}{
        (E_{\vb{k}} + i \Omega_m)^2 - E_{\vb{k-q}}^2
    }
    -
    \frac{\Delta_{\vb{k-q}}}{2 E_{\vb{k-q}}}
    \frac{\xi_{\vb{k}} - i \Omega_m + E_{\vb{k-q}}}{
        E_{\vb{k}}^2 - (E_{\vb{k-q}} - i \Omega_m)^2
    },
\end{equation}
\begin{equation}
    \lim_{T \rightarrow 0} T \sum_{\omega_n} G_{k} \overline{F}_{k-q}
    =
    -
    \frac{\overline{\Delta}_{\vb{k-q}}}{2 E_{\vb{k}}}
    \frac{E_{\vb{k}} - \xi_{\vb{k}}}{
        (E_{\vb{k}} + i \Omega_m)^2 - E_{\vb{k-q}}^2
    }
    - \frac{\overline{\Delta}_{\vb{k-q}}}{2 E_{\vb{k-q}}}
    \frac{\xi_{\vb{k}} + i \Omega_m - E_{\vb{k-q}}}{
        E_{\vb{k}}^2 - (E_{\vb{k-q}} - i \Omega_m)^2 
    },
\end{equation}
and
\begin{equation}
    \lim_{T \rightarrow 0} T \sum_{\omega_n} \overline{F}_{k} G_{-k+q}
    =
    \frac{\overline{\Delta}_{\vb{k}}}{2 E_{\vb{k}}}
    \frac{E_{\vb{k}} + i \Omega_m + \xi_{\vb{k-q}}}{
        (E_{\vb{k}} + i \Omega_m)^2 - E_{\vb{k-q}}^2
    }
    -
    \frac{\overline{\Delta}_{\vb{k}}}{2 E_{\vb{k-q}}}
    \frac{E_{\vb{k-q}} + \xi_{\vb{k-q}}}{
        E_{\vb{k}}^2 - (E_{\vb{k-q}} - i \Omega_m)^2
    }.
\end{equation}
\end{widetext}
We now project to the Fermi surface using the same approximations as for the polarization components ($\chi^{(\ell)}_{\vb{k-q}} \approx \chi^{(\ell)}_{\vb{k}}$, $\Delta_{\vb{k-q}} \approx \Delta_{\vb{k}}$, and $\xi_{\vb{k-q}} \approx \xi_{\vb{k}} - \vb{v}_{\vb{k}} \vdot \vb{q}$), with the additional approximation that
\begin{equation}
    \bra{u_{\vb{k-q}}} \hat{o} \ket{u_{\vb{k}}}
    \approx
    \bra{u_{\vb{k}}} \hat{o} \ket{u_{\vb{k}}},
\end{equation}
which we further assume is essentially constant normal to the Fermi surface (i.e., $\ket{u_{\vb{k}}} \approx \ket{u_{k_{\|}}}$).
We then need the integrals
\begin{multline}
    \frac{1}{2} \int \frac{\dd^2 \vb{k}}{4 \pi^2}
    \bra{u_{\vb{k}}} \hat{o} \ket{u_{\vb{k}}}^2
    \big(
        G_{k} G_{k-q}
        + G_{-k} G_{-k+q}
        \\
        - F_{k} \overline{F}_{k-q}
        - \overline{F}_{k} F_{k-q}
    \big)
\end{multline}
for the quasiparticle response, and
\begin{multline}
    \int \frac{\dd^2 \vb{k}}{4 \pi^2}
    \bra{u_{\vb{k}}} \hat{o} \ket{u_{\vb{k}}}
    \chi^{(\ell)}_{\vb{k}}
    \big(
        \pm F_{k} G_{k-q} \pm G_{-k} F_{k-q}
        \\
        + G_{k} \overline{F}_{k-q} + \overline{F}_{k} G_{-k+q}
    \big)
\end{multline}
for the collective mode source terms.

For the quasiparticles,
\begin{widetext}
\begin{multline}
    \lim_{T \rightarrow 0} \sum_{\omega_n}
    \frac{1}{2} \int \frac{\dd^2 \vb{k}}{4 \pi^2}
    \bra{u_{\vb{k}}} \hat{o} \ket{u_{\vb{k}}}^2
    \left(
        G_{k} G_{k-q}
        + G_{-k} G_{-k+q}
        - F_{k} \overline{F}_{k-q}
        - \overline{F}_{k} F_{k-q}
    \right)
    \\
    \approx
    \int_{\text{FS}} \frac{\dd{k_{\|}}}{4 \pi^2 v}
    \bra{u} \hat{o} \ket{u}^2
    \int_{-\Lambda}^{\Lambda} \dd{\xi}
    \Bigg\lbrace
        -\frac{1}{2 E}
        \frac{
           E^2 + \xi (\xi - \vb{v} \vdot \vb{q})
           + i \Omega_m E
        }{(E + i \Omega_m)^2 - (E')^2}
        +
        \frac{1}{4 E}
        \frac{|\Delta|^2}{(E + i \Omega_m)^2 - (E')^2}
        \\
        +
        \frac{1}{4 E}
        \frac{|\Delta|^2}{(E - i \Omega_m)^2 - (E')^2}
        +
        \frac{1}{2 E'}
        \frac{
            (E')^2 + \xi (\xi - \vb{v} \vdot \vb{q})
            - i \Omega_m E'
        }{E^2 - (E' - i \Omega_m)^2}
        \\
        -
        \frac{1}{4 E'}
        \frac{|\Delta|^2}{E^2  - (E' - i \Omega_m)^2}
        -
        \frac{1}{4 E'}
        \frac{|\Delta|^2}{E^2 - (E' + i \Omega_m)^2}
    \Bigg\rbrace.
\end{multline}
As before, $E' = \sqrt{(\xi - \vb{v} \vdot \vb{q})^2 + |\Delta|^2}$. We again shift $\xi \rightarrow \xi + \vb{v} \vdot \vb{q}$ whereever necessary to ensure $E'$ only shows up as $(E')^2$, and drop terms of $O(\vb{v} \vdot \vb{q} / \Lambda)$, yielding
\begin{multline}
    \lim_{T \rightarrow 0} \sum_{\omega_n}
    \frac{1}{2} \int \frac{\dd^2 \vb{k}}{4 \pi^2}
    \bra{u_{\vb{k}}} \hat{o} \ket{u_{\vb{k}}}^2
    \left(
        G_{k} G_{k-q}
        + G_{-k} G_{-k+q}
        - F_{k} \overline{F}_{k-q}
        - \overline{F}_{k} F_{k-q}
    \right)
    \\
    \approx
    - \int_{\text{FS}} 
    \frac{\dd{k_{\|}}}{4 \pi^2 v}
    \bra{u} \hat{o} \ket{u}^2
    \int_{-\Lambda}^{\Lambda} \dd{\xi}
    \frac{1}{2 E}
    \Bigg\lbrace
        \frac{
           \xi (2 \xi - \vb{v} \vdot \vb{q})
           + i \Omega_m E
        }{
            (i \Omega_m)^2 + 2 i \Omega_m E
            + 2 \xi (\vb{v} \vdot \vb{q})
            - (\vb{v} \vdot \vb{q})^2
        }
        + \frac{
            \xi (2 \xi + \vb{v} \vdot \vb{q})
            - i \Omega_m E
        }{
            (i \Omega_m)^2 - 2 i \Omega_m E
            - 2 \xi (\vb{v} \vdot \vb{q})
            - (\vb{v} \vdot \vb{q})^2
        }
    \Bigg\rbrace
    \\
    =
    \int_{\text{FS}} 
    \frac{\dd{k_{\|}}}{4 \pi^2 v}
    \bra{u} \hat{o} \ket{u}^2
    \int_{-\Lambda}^{\Lambda} \dd{\xi}
    \frac{1}{E}
    \frac{
        4 |\Delta|^2 (i \Omega_m)^2
    }{
        \left[
            (i \Omega_m)^2 + 2 i \Omega_m E
            + 2 \xi (\vb{v} \vdot \vb{q})
            - (\vb{v} \vdot \vb{q})^2
        \right] \left[
            (i \Omega_m)^2 - 2 i \Omega_m E
            - 2 \xi (\vb{v} \vdot \vb{q})
            - (\vb{v} \vdot \vb{q})^2
        \right]
    }
\end{multline}
We handle this integral using the same variable transformation as for the polarization components, $\xi/|\Delta| = \sinh (x)$, $E/|\Delta| = \cosh (x)$, $\vb{v} \vdot \vb{q} / |\Delta| = i a \sinh (i b)$, and $i \Omega_m / |\Delta| = i a \cosh(i b)$. Taking $\Lambda \rightarrow \infty$,
\begin{multline}
    \lim_{T \rightarrow 0} \sum_{\omega_n}
    \frac{1}{2} \int \frac{\dd^2 \vb{k}}{4 \pi^2}
    \bra{u_{\vb{k}}} \hat{o} \ket{u_{\vb{k}}}^2
    \big(
        G_{k} G_{k-q}
        + G_{-k} G_{-k+q}
        - F_{k} \overline{F}_{k-q}
        - \overline{F}_{k} F_{k-q}
    \big)
    \\
    \approx
    \int_{\text{FS}} 
    \frac{\dd{k_{\|}}}{4 \pi^2 v}
    \bra{u} \hat{o} \ket{u}^2
    \int_{-\infty}^{\infty} \dd{x}
    \frac{4 \cosh^2 (i b)}{
        \left(
            2 i \cosh(x + i b) - a       
        \right) \left(
            2 i \cosh(x + i b) + a        
        \right)
    }
    \\
    =
    - \int_{\text{FS}} 
    \frac{\dd{k_{\|}}}{4 \pi^2 v}
    \bra{u} \hat{o} \ket{u}^2
    \int_{-\infty}^{\infty} \dd{x}
    \frac{\cosh^2 (i b)}{
        \cosh^2(x + i b) + (a/2)^2
    }
    \\
    =
    - \int_{\text{FS}} 
    \frac{\dd{k_{\|}}}{4 \pi^2 v}
    \bra{u} \hat{o} \ket{u}^2
    \frac{4 \cosh^2 (i b)}{
        a \sqrt{1 + (a/2)^2}
    }
    \arcsinh \left( \frac{a}{2} \right)
\end{multline}

Recall that $a = -2 i \sin z$ and $\sqrt{1 + (a/2)^2} = \cos z$, so
\begin{multline}
    \lim_{T \rightarrow 0} \sum_{\omega_n}
    \frac{1}{2} \int \frac{\dd^2 \vb{k}}{4 \pi^2}
    \bra{u_{\vb{k}}} \hat{o} \ket{u_{\vb{k}}}^2
    \left(
        G_{k} G_{k-q}
        + G_{-k} G_{-k+q}
        - F_{k} \overline{F}_{k-q}
        - \overline{F}_{k} F_{k-q}
    \right)
    \\
    \approx
    - \int_{\text{FS}} 
    \frac{\dd{k_{\|}}}{4 \pi^2 v}
    \bra{u} \hat{o} \ket{u}^2
    \frac{2 z \cosh^2 (i b)}{
        \sin z \cos z
    }.
\end{multline}
Using the definition of $\cosh (i b)$, we further have
\begin{equation}
    \cosh^2 (i b)
    =
    \left(
        \frac{\Omega_m}{|\Delta|}
        \frac{1}{a}
    \right)^2
    =
    \left(
        \frac{\Omega_m}{|\Delta|}
        \frac{1}{-2 i \sin z}
    \right)^2
    =
    \frac{(i \Omega_m)^2}{4 |\Delta|^2}
    \frac{1}{\sin^2 z},
\end{equation}
yielding
\begin{multline}
    \lim_{T \rightarrow 0} \sum_{\omega_n}
    \frac{1}{2} \int \frac{\dd^2 \vb{k}}{4 \pi^2}
    \bra{u_{\vb{k}}} \hat{o} \ket{u_{\vb{k}}}^2
    \left(
        G_{k} G_{k-q}
        + G_{-k} G_{-k+q}
        - F_{k} \overline{F}_{k-q}
        - \overline{F}_{k} F_{k-q}
    \right)
    \\
    \approx
    - \int_{\text{FS}} 
    \frac{\dd{k_{\|}}}{4 \pi^2 v}
    \bra{u} \hat{o} \ket{u}^2
    \frac{(i \Omega_m)^2}{4 |\Delta|^2}
    \frac{2 z}{
        \sin^3 z \cos z
    }.
\end{multline}

Now for the collective modes. We handle the $+$ case (for the $\mathcal{R}$ sources) first:
\begin{multline}
    \lim_{T \rightarrow 0} \sum_{\omega_n}
    \int \frac{\dd^2 \vb{k}}{4 \pi^2}
    \bra{u_{\vb{k}}} \hat{o} \ket{u_{\vb{k}}}
    \chi^{(\ell)}_{\vb{k}}
    \left(
        F_{k} G_{k-q} + G_{-k} F_{k-q}
        + G_{k} \overline{F}_{k-q} + \overline{F}_{k} G_{-k+q}
    \right)
    \\
    =
    \int_{\text{FS}} \frac{\dd{k_{\|}}}{4 \pi^2 v}
    \bra{u} \hat{o} \ket{u}
    \chi^{(\ell)}
    \int_{-\Lambda}^{\Lambda} \dd{\xi}
    \bigg\lbrace
        \frac{1}{2 E}
        \frac{
            \Delta \left(
                - i \Omega_m + 2 \xi - \vb{v} \vdot \vb{q}
            \right)
            + \overline{\Delta} \left(
                i \Omega_m + 2 \xi - \vb{v} \vdot \vb{q}
            \right)
        }{(E + i \Omega_m)^2 - (E')^2}
        \\
        - \frac{1}{2 E'}
        \frac{
            \Delta \left(
                - i \Omega_m + 2 \xi - \vb{v} \vdot \vb{q}
            \right)
            + \overline{\Delta} \left(
                i \Omega_m + 2 \xi - \vb{v} \vdot \vb{q}
            \right)
        }{E^2 - (E' - i \Omega_m)^2}
    \bigg\rbrace
    \\
    =
    \int_{\text{FS}} \frac{\dd{k_{\|}}}{4 \pi^2 v}
    \bra{u} \hat{o} \ket{u}
    \chi^{(\ell)}
    \int_{-\Lambda}^{\Lambda} \dd{\xi}
    \frac{|\Delta|}{2 E}
    \bigg\lbrace
        \frac{
            e^{i \theta} \left(
                - i \Omega_m + 2 \xi - \vb{v} \vdot \vb{q}
            \right)
            + e^{-i \theta} \left(
                i \Omega_m + 2 \xi - \vb{v} \vdot \vb{q}
            \right)
        }{
            (i \Omega_m)^2 + 2 i \Omega_m E 
            + 2 \xi (\vb{v} \vdot \vb{q}) - (\vb{v} \vdot \vb{q})^2
        }
        \\
        + \frac{
            e^{i \theta} \left(
                - i \Omega_m + 2 \xi + \vb{v} \vdot \vb{q}
            \right)
            + e^{-i \theta} \left(
                i \Omega_m + 2 \xi + \vb{v} \vdot \vb{q}
            \right)
        }{
            (i \Omega_m)^2 - 2 i \Omega_m E
            - 2 \xi (\vb{v} \vdot \vb{q}) - (\vb{v} \vdot \vb{q})^2
        }
    \bigg\rbrace
    + O (\vb{v} \vdot \vb{q} / \Lambda)
    \\
    \approx
    -\int_{\text{FS}} \frac{\dd{k_{\|}}}{4 \pi^2 v}
    \bra{u} \hat{o} \ket{u}
    \chi^{(\ell)}
    \int_{-\Lambda}^{\Lambda} \dd{\xi}
    \frac{2 |\Delta|}{E}
    \Bigg\lbrace
        \cos \theta \left(2 \xi + \vb{v} \vdot \vb{q} \right)
        i \Im \left[
            \frac{1}{
                \Omega_m^2 + 2 \xi (\vb{v} \vdot \vb{q})
                + (\vb{v} \vdot \vb{q})^2 + 2 i \Omega_m E
            }
        \right]
        \\
        +
        \sin \theta \Omega_m
        \Re \left[
            \frac{1}{
                \Omega_m^2 + 2 \xi (\vb{v} \vdot \vb{q})
                + (\vb{v} \vdot \vb{q})^2 + 2 i \Omega_m E
            }
        \right]
    \Bigg\rbrace.
\end{multline}

Note that
\begin{equation}
    \frac{1}{
        \Omega_m^2 + 2 \xi (\vb{v} \vdot \vb{q}) 
        + (\vb{v} \vdot \vb{q})^2 + 2 i \Omega_m E
    }
    =
    \frac{
        \Omega_m^2 + 2 \xi (\vb{v} \vdot \vb{q}) 
        + (\vb{v} \vdot \vb{q})^2 - 2 i \Omega_m E 
    }{
        \left(
            \Omega_m^2 + 2 \xi (\vb{v} \vdot \vb{q}) 
            + (\vb{v} \vdot \vb{q})^2
        \right)^2 
        + \left( 2 \Omega_m E \right)^2
    },
\end{equation}
so
\begin{multline}
    \lim_{T \rightarrow 0} \sum_{\omega_n}
    \int \frac{\dd^2 \vb{k}}{4 \pi^2}
    \bra{u_{\vb{k}}} \hat{o} \ket{u_{\vb{k}}}
    \chi^{(\ell)}_{\vb{k}}
    \left(
        F_{k} G_{k-q} + G_{-k} F_{k-q}
        + G_{k} \overline{F}_{k-q} + \overline{F}_{k} G_{-k+q}
    \right)
    \\
    =
    \int_{\text{FS}} \frac{\dd{k_{\|}}}{4 \pi^2 v}
    \bra{u} \hat{o} \ket{u}
    \chi^{(\ell)}
    \int_{-\Lambda}^{\Lambda} \dd{\xi}
    \frac{2 |\Delta|}{E}
    \frac{
        \cos \theta (2 \xi + \vb{v} \vdot \vb{q})
        (2 i \Omega_m E)
        +
        i \sin \theta (i \Omega_m) \left(
            -(i \Omega_m)^2 + 2 \xi (\vb{v} \vdot \vb{q}) 
            + (\vb{v} \vdot \vb{q})^2
        \right)
    }{
        \left(
            -(i \Omega_m)^2 + 2 \xi (\vb{v} \vdot \vb{q}) 
            + (\vb{v} \vdot \vb{q})^2
        \right)^2 
        - \left( 2 i \Omega_m E \right)^2
    }.
\end{multline}

Using again the same variable transformations as before, we obtain
\begin{multline}
    \lim_{T \rightarrow 0} \sum_{\omega_n}
    \int \frac{\dd^2 \vb{k}}{4 \pi^2}
    \bra{u_{\vb{k}}} \hat{o} \ket{u_{\vb{k}}}
    \chi^{(\ell)}_{\vb{k}}
    \left(
        F_{k} G_{k-q} + G_{-k} F_{k-q}
        + G_{k} \overline{F}_{k-q} + \overline{F}_{k} G_{-k+q}
    \right)
    \\
    =
    2 \int_{\text{FS}} \frac{\dd{k_{\|}}}{4 \pi^2 v}
    \bra{u} \hat{o} \ket{u}
    \chi^{(\ell)}
    \\
    \int_{-\tilde{\Lambda}}^{\tilde{\Lambda}} \dd{x}
    \frac{1}{a}
    \frac{
        2 i \cosh(x) \cosh(i b) \left(
            2 \sinh(x) + i a \sinh(i b)
        \right) \cos \theta
        - \cosh(i b)
        \left(
            a^2 + 2 i a \sinh(x) \sinh(i b)
        \right) \sin \theta
    }{
        \left[
            a + 2 i \sinh(x) \sinh(i b)
        \right]^2
        +
        4 \cosh^2 (x) \cosh^2 (i b)
    }.
\end{multline}
First the $\cos \theta$ term:
\begin{multline}
    \int_{-\tilde{\Lambda}}^{\tilde{\Lambda}} \dd{x}
    \frac{
        2 i \cosh(x) \cosh(i b) \left(
            2 \sinh(x) + i a \sinh(i b)
        \right)
    }{
        \left[
            a + 2 i \sinh(x) \sinh(i b)
        \right]^2
        + 4 \cosh^2 (x) \cosh^2 (i b)
    }
    \\
    =
    \frac{i}{2} \cos b
    \Log \left(
        \frac{
            a^2 + 4 \cos^2 b 
            - 4 a (\Lambda / |\Delta|) \sin b 
            + 4 (\Lambda / |\Delta|)^2
        }{
            a^2 + 4 \cos^2 b 
            + 4 a (\Lambda / |\Delta|) \sin b 
            + 4 (\Lambda / |\Delta|)^2
        }
    \right)
    \xrightarrow{\Lambda \rightarrow \infty} 0.
\end{multline}

The $\sin \theta$ term survives the $\Lambda \rightarrow \infty$ limit. We already found the necessary integral when evaluating the polarization components,
\begin{multline}
    \Re \int_{-\infty}^{\infty} \dd{\xi}
    \frac{1}{E}
    \frac{1}{
        (i \Omega_m)^2 + 2 i \Omega_m E
        + 2 \xi (\vb{v} \vdot \vb{q})
        - (\vb{v} \vdot \vb{q})^2
    }
    =
    \frac{1}{|\Delta|^2}
    \Re \int_{-\infty}^{\infty} \dd{x}
    \frac{1}{
        2 i a \cosh(x + i b) - a^2
    }
    \\
    =
    - \frac{1}{|\Delta|^2}
    \frac{\arcsinh(a / 2)}{
        a \sqrt{1 + (a/2)^2}
    }
    =
    - \frac{1}{|\Delta|^2}
    \frac{z}{2 \sin z \cos z}.
\end{multline}

The drive coupling for $\mathcal{R}^{(\ell)}_{q}$ is then
\begin{equation}
    \Pi^{d \mathcal{R}}_{q; \ell}
    =
    i^{\rho_{\ell}}
    \int_{\text{FS}} 
    \frac{\dd{k_{\|}}}{4 \pi^2 v}
    \bra{u} \hat{o} \ket{u}
    \chi^{(\ell)} \sin \theta 
    \frac{i \Omega_m}{|\Delta|}
    \frac{i z}{ \sin z \cos z }.  \label{eq:Source_R_phi}
\end{equation}
For the $\mathcal{I}$ drive terms, we have a similar result:
\begin{multline}
    \lim_{T \rightarrow 0} \sum_{\omega_n}
    i \int \frac{\dd^2 \vb{k}}{4 \pi^2}
    \bra{u_{\vb{k}}} \hat{o} \ket{u_{\vb{k}}}
    \chi^{(\ell)}_{\vb{k}}
    \left(
        - F_{k} G_{k-q} - G_{-k} F_{k-q}
        + G_{k} \overline{F}_{k-q} + \overline{F}_{k} G_{-k+q}
    \right)
    \\
    =
    i \int_{\text{FS}} \frac{\dd{k_{\|}}}{4 \pi^2 v}
    \bra{u} \hat{o} \ket{u}
    \chi^{(\ell)}
    \int_{-\Lambda}^{\Lambda} \dd{\xi}
    \bigg\lbrace
        \frac{1}{2 E}
        \frac{
            \Delta \left(
                i \Omega_m - 2 \xi + \vb{v} \vdot \vb{q}
            \right)
            + \overline{\Delta} \left(
                i \Omega_m + 2 \xi - \vb{v} \vdot \vb{q}
            \right)
        }{(E + i \Omega_m)^2 - (E')^2}
        \\
        - \frac{1}{2 E'}
        \frac{
            \Delta \left(
                i \Omega_m - 2 \xi + \vb{v} \vdot \vb{q}
            \right)
            + \overline{\Delta} \left(
                i \Omega_m + 2 \xi - \vb{v} \vdot \vb{q}
            \right)
        }{E^2 - (E' - i \Omega_m)^2}
    \bigg\rbrace
    \\
    \approx
    i \int_{\text{FS}} \frac{\dd{k_{\|}}}{4 \pi^2 v}
    \bra{u} \hat{o} \ket{u}
    \chi^{(\ell)}
    \int_{-\Lambda}^{\Lambda} \dd{\xi}
    \frac{|\Delta|}{2 E}
    \bigg\lbrace
        \frac{
            e^{i \theta} \left(
                i \Omega_m - 2 \xi + \vb{v} \vdot \vb{q}
            \right)
            + e^{-i \theta} \left(
                i \Omega_m + 2 \xi - \vb{v} \vdot \vb{q}
            \right)
        }{
            (i \Omega_m)^2 + 2 i \Omega_m E 
            + 2 \xi (\vb{v} \vdot \vb{q}) - (\vb{v} \vdot \vb{q})^2
        }
        \\
        + \frac{
            e^{i \theta} \left(
                i \Omega_m - 2 \xi - \vb{v} \vdot \vb{q}
            \right)
            + e^{-i \theta} \left(
                i \Omega_m + 2 \xi + \vb{v} \vdot \vb{q}
            \right)
        }{
            (i \Omega_m)^2 - 2 i \Omega_m E
            - 2 \xi (\vb{v} \vdot \vb{q}) - (\vb{v} \vdot \vb{q})^2
        }
    \bigg\rbrace
    \\
    =
    i \int_{\text{FS}} \frac{\dd{k_{\|}}}{4 \pi^2 v}
    \bra{u} \hat{o} \ket{u}
    \chi^{(\ell)}
    \int_{-\Lambda}^{\Lambda} \dd{\xi}
    \frac{|\Delta|}{E}
    \bigg\lbrace
        \frac{
            \cos \theta i \Omega_m
            - i \sin \theta \left(
                2 \xi - \vb{v} \vdot \vb{q}
            \right)
        }{
            (i \Omega_m)^2 + 2 i \Omega_m E 
            + 2 \xi (\vb{v} \vdot \vb{q}) - (\vb{v} \vdot \vb{q})^2
        }
        \\
        + \frac{
            \cos \theta i \Omega_m
            - i \sin \theta \left(
                2 \xi + \vb{v} \vdot \vb{q}
            \right)
        }{
            (i \Omega_m)^2 - 2 i \Omega_m E
            - 2 \xi (\vb{v} \vdot \vb{q}) - (\vb{v} \vdot \vb{q})^2
        }
    \bigg\rbrace
    \\
    =
    i \int_{\text{FS}} \frac{\dd{k_{\|}}}{4 \pi^2 v}
    \bra{u} \hat{o} \ket{u}
    \chi^{(\ell)}
    \int_{-\Lambda}^{\Lambda} \dd{\xi}
    \frac{2 |\Delta|}{E}
    \bigg\lbrace
        - \cos \theta i \Omega_m \Re \left[
            \frac{1}{
                \Omega_m^2 + 2 i \Omega_m E 
                + 2 \xi (\vb{v} \vdot \vb{q}) + (\vb{v} \vdot \vb{q})^2
            }
        \right]
        \\
        -
        \sin \theta \left( 2 \xi + \vb{v} \vdot \vb{q} \right)
        \Im \left[
            \frac{1}{
                \Omega_m^2 + 2 i \Omega_m E 
                + 2 \xi (\vb{v} \vdot \vb{q}) + (\vb{v} \vdot \vb{q})^2
            }
        \right]
    \bigg\rbrace
    \\
    =
    - i \int_{\text{FS}} \frac{\dd{k_{\|}}}{4 \pi^2 v}
    \bra{u} \hat{o} \ket{u}
    \chi^{(\ell)}
    \cos \theta i \Omega_m 2 |\Delta|
    \int_{-\Lambda}^{\Lambda} \dd{\xi}
    \frac{1}{E}
    \Re \left[
        \frac{1}{
            \Omega_m^2 + 2 i \Omega_m E 
            + 2 \xi (\vb{v} \vdot \vb{q}) + (\vb{v} \vdot \vb{q})^2
        }
    \right]
    \\
    =
    - i \int_{\text{FS}} \frac{\dd{k_{\|}}}{4 \pi^2 v}
    \bra{u} \hat{o} \ket{u}
    \chi^{(\ell)}
    \cos \theta
    \frac{4 i \Omega_m }{|\Delta|}
    \frac{\arcsinh \left( \frac{a}{2} \right)}{ a \sqrt{4 + a^2} }.
\end{multline}    

The drive coupling for $\mathcal{I}^{(\ell)}_{q}$ is then
\begin{equation}
    \Pi^{d \mathcal{I}}_{q; \ell}
    =
    -i^{\rho_{\ell}}
    \int \frac{\dd{k_{\|}}}{4 \pi^2 v}
    \bra{u} \hat{o} \ket{u} \chi^{(\ell)}
    \cos \theta \frac{i \Omega_m}{|\Delta|}
    \frac{i z}{\sin z \cos z}.  \label{eq:Source_I_phi}
\end{equation}

Finally, we integrate out the fluctuations ($\mathcal{R}$ and $\mathcal{I}$):
\begin{multline}    \label{eq:CM_final}
    \int_{\mathcal{R, I}} \exp{- S_{\text{eff}}}
    =
    \int_{\mathcal{R, I}} \exp{
        \sum_{q X Y \ell \ell'}
        X^{(\ell)}_{-q}
        (\mathcal{D}^{-1}_{q})^{X Y}_{\ell \ell'}
        Y^{(\ell')}_{q}
        -
        \sum_{q X \ell}
        d_{-q} \Pi^{d X}_{q; \ell} X^{(\ell)}_{q}
    }
    \\
    =
    \exp{
        -\sum_{q X Y \ell \ell'}
        \frac{1}{4} d_{-q} d_{q}
        \Pi^{d X}_{-q; \ell}
        (\mathcal{D}_{q})_{\ell \ell'}^{X Y} 
        \Pi^{d Y}_{q; \ell'}
    }.
\end{multline}
$\mathcal{D}_{q}$ is a matrix in the space of fluctuation types $X, Y \in \lbrace \mathcal{R, I} \rbrace$ and channels $\ell, \ell'$, obtained by inverting the inverse propagator given in the previous Appendix. To obtain the real-time response function, we analytically continue $i \Omega_m \rightarrow \Omega + i 0^{+}$ in all expressions.

\section{Single particle Hamiltonian of RTG}
\label{appendix:single-particle_RTG}
We model the single-particle band structure of rhombohedral trilayer graphene using a $6\times6$ tight binding model \cite{zhang2010band}. In the continuum limit, the Bloch Hamiltonian matrix is
\begin{equation}
\label{eq:rtg_hamiltonian}
    h_k 
    = 
    \begin{pmatrix} 
        \delta_0 + \delta_1 + \delta_2 & v_0\pi^* & v_4\pi^* 
        & v_3\pi  & 0 & \frac{\gamma_2}{2} 
        \\ 
        v_0\pi & \delta_1+\delta_2 & \gamma_1 
        & v_4\pi^* & 0 & 0 
        \\ v_4\pi & \gamma_1 &  -2\delta_2 
        & v_0\pi^* & v_4\pi^* & v_3\pi 
        \\
        v_3\pi^* & v_4\pi & v_0\pi & 
        -2\delta_2 & \gamma_1 & v_4\pi^* 
        \\ 
        0 & 0 & v_4\pi 
        & \gamma_1 & -\delta_1+\delta_2 & v_0\pi^* 
        \\ 
        \frac{\gamma_2}{2}  & 0 & v_3\pi^* 
        & v_4\pi & v_0\pi &  \delta_0 -\delta_1+\delta_2
    \end{pmatrix},
\end{equation}
where $\pi=\tau k_x+ik_y$ ($\tau$ is the valley index) and we use the basis ($A_1, B_1, A_2, B_2, A_3, B_3$) with
$A_i$ and $B_i$ denoting the two sites per unit cell (i.e., two sublattices) on layer $i$. Here $v_i = \sqrt{3} a \gamma_i/2$, $a = 2.46$ \AA \, is the lattice constant of monolayer graphene, $\delta_1$ is proportional to the perpendicular electric field, and $\gamma_i$, $\delta_0$, $\delta_2$ are fixed material parameters:
$\gamma_0=3.1$~eV, $\gamma_1= 0.38$~eV, $\gamma_2=-0.015$~eV, $\gamma_3 =-0.29$~eV, $\gamma_4= -0.141$~eV, $\delta_0 =-0.0105$~eV, $\delta_2= -0.0023$~eV~\cite{zhou2021half}.
\begin{figure}[!h]
\centering
\includegraphics[width=\textwidth]{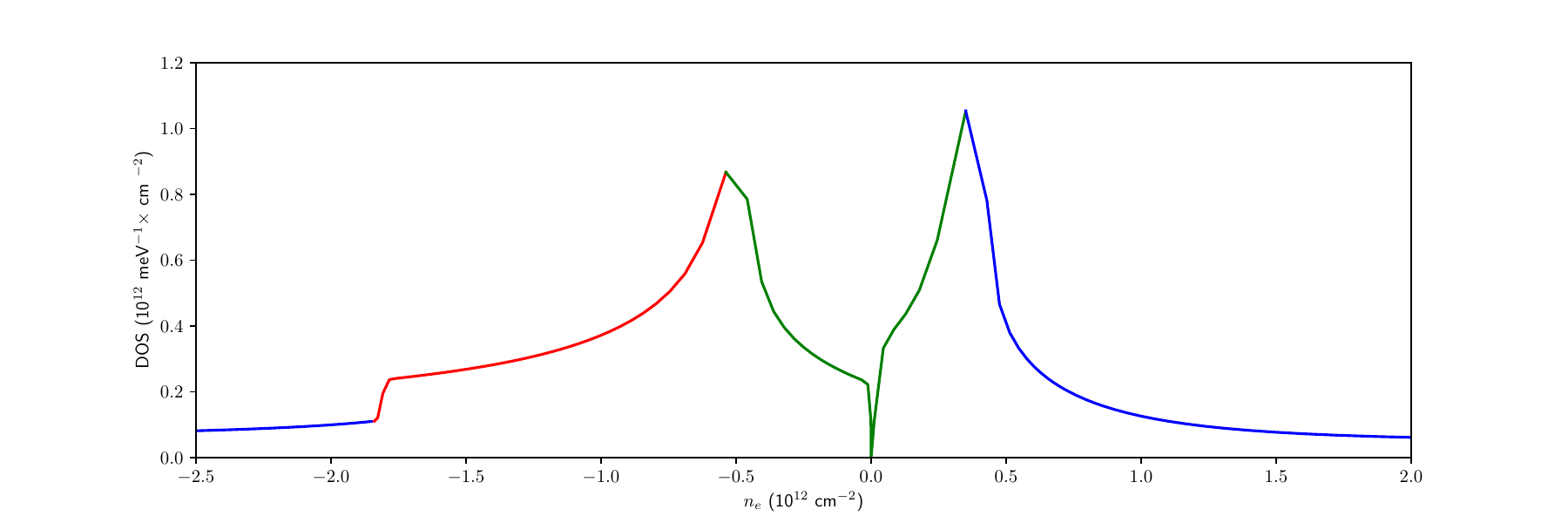}
\caption{\label{fig:DOS} Dependence of density of states on the electron doping in RTG. Perpendicular electric field corresponds to  $\delta_1=20$meV. Colors correspond to different topology of the Fermi contour}
\end{figure}
\begin{figure}[!h]
\centering
\includegraphics[width=\textwidth]{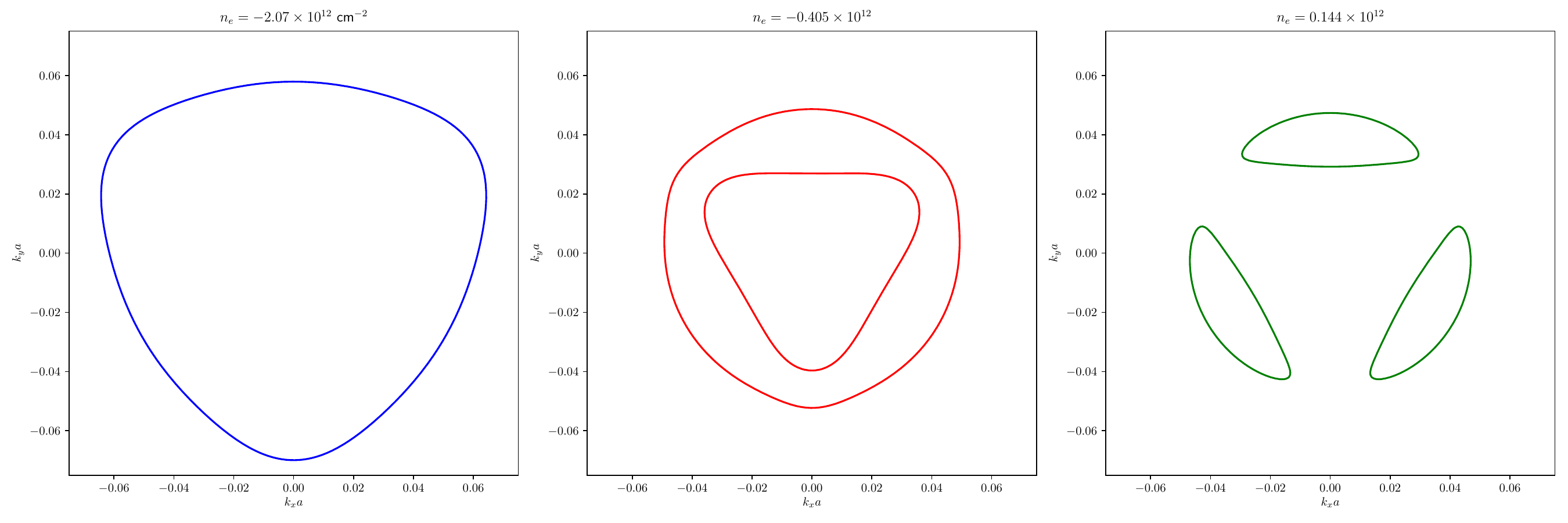}
\caption{\label{fig:FermiContours} Fermi surfaces in the $K$ valley for three different electron concentrations. Colors correspond to the colors in Fig.~\ref{fig:DOS}.}
\end{figure}

In Fig.~\ref{fig:DOS} we show the dependence of the density of states at the Fermi level on the electron doping for $\Delta_1=20$meV. There exist three van Hove singularities (vHS) associated with Lifshitz transitions where the topology of the Fermi surface changes. Three different topologies are shown in Fig.~\ref{fig:FermiContours}. In what follows we will consider the hole-doped region in the vicinity of the vHS where the Fermi surface is annular with two concentric Fermi pockets, as shown in the central subfigure of Fig.~\ref{fig:FermiContours}.
\end{widetext}

\section{Interaction potentials}
\label{appendix:rtg_potentials} \label{eq:phonon_potential}
In the main text we consider two kinds of interaction potentials for RTG. The first is a phonon-mediated interaction, with the potential given by
\begin{align}
\mathcal{V}_{\vb{k,k'}}^{\text{phon}}=-\mathcal{V}_0 |\braket{u_{\vb{k}}}{u_{\vb{k'}}}|^2,
\end{align}
where $|u_{\mathbf{k}}\rangle$ is the Bloch wavevector of the active (valence) band into which we project. The parameter $\mathcal{V}_0$ is chosen such that the critical temperature is the same as for the case of IVC fluctuation-mediated superconductity described below.

The second interaction we consider is mediated by the fluctuations of the intervalley coherent order parameter~\cite{chatterjee2022inter}:
\begin{align}   \label{eq:ivc_potential}
\mathcal{V}_{\vb{k,k'}; \vb{q}}^{\text{IVC}}=\frac{n_e/(\partial n_e / \partial \mu)}{(\vb{k+k'-q})^2+\xi_{IVC}^{-2}} |\tilde{u}_{+,\vb{k}}\tilde{u}_{+,\vb{k'}}|^2,
\end{align}
where $\tilde{u}_{+}$ is the component of the Bloch wavefunction near the $K$ valley projected to the layer and sublattice with maximal electron density. Note that this potential depends on $\vb{k} + \vb{k'}$ as opposed to $\vb{k-k'}$. This is because IVC fluctuations are peaked at the wavevector $\vb{K - K'}$ separating the two valleys (and they hence scatter particles between valleys), and since $\vb{k}$ and $\vb{k'}$ label the outgoing and incoming momenta near the $K$ valley, measured relative to that valley. In this language, IVC scattering in the Cooper channel takes one electron to $\vb{K+k}$ from $\vb{K'-k'+q}$ and the other to $\vb{K'-k+q}$ from $\vb{K+k'}$, i.e.~the total momentum transfer is $\vb{K - K' + k + k' - q}$. Assuming that $|\vb{q}| \xi_{\text{IVC}} \ll 1$, in practice we will evaluate $\mathcal{V}^{\text{IVC}}$ at $\vb{q} = 0$. Since we will be interested in $|\vb{q}|$ up to the scale of the inverse BCS coherence length, this means that we require $\xi_{\text{IVC}} \ll \xi_{\text{BCS}}$.

\section{Self-consistent electromagnetic response}
\label{appendix:anderson-higgs}
In this Appendix, we address the fact that the system responds to the total electric field (including the field induced by polarization), not only the incident field. In the superconducting state, this implements the Anderson-Higgs mechanism, which substantially modifies the dispersion of the global phase (ABG) mode.

For this we will need both the response to the driving electric field ($d$) calculated previously, and the response to the vector potential. The latter can be obtained by the minimal coupling $\vb{k} \rightarrow \vb{k - A}$, where $j=x,y$ and we set the electron charge $e = 1$ as throughout. Neglecting the diamagnetic term (which vanishes for linear dispersion, which we will assume around the Fermi level), the resulting vertices are
\begin{equation}
    \mathds{V}^{(j)}_{k, k-q}
    \approx
     A^{j}_{q} v^{j}_{\vb{k}} \begin{pmatrix}
        1 & 0 \\
        0 & 1
    \end{pmatrix}
\end{equation}
where the matrix acts in Nambu space.
Here we have neglected the contribution of the quantum metric to the vertex, which becomes substantial in flat band systems such as magic angle twisted bilayer graphene~\cite{torma2022superconductivity, PhysRevLett.132.026002}, where the contribution from the band dispersion vanishes. However, in the RTG case, the contribution due to band dispersion is substantial and we may thus neglect the geometrical contribution. We then obtain the expression for the driving terms corresponding to the vector potential similarly to the expressions in Eqs.~\eqref{eq:Source_R_phi} and \eqref{eq:Source_I_phi}:
\begin{multline}
    \Pi^{j \mathcal{R}}_{q; \ell}
    \\
    =
    -i^{\rho_{\ell}}
    \int_{\text{FS}} 
    \frac{\dd{k_{\|}}}{4 \pi^2 v}
    \chi^{(\ell)} \sin \theta 
    \frac{ v^{j} (\vb{v} \vdot \vb{q})}{|\Delta|}
    \frac{i z}{ \sin z \cos z },
\end{multline}
\begin{multline}
    \Pi^{j \mathcal{I}}_{q; \ell}
    \\
    =
    i^{\rho_{\ell}}
    \int_{\text{FS}}  \frac{\dd{k_{\|}}}{4 \pi^2 v}
     \chi^{(\ell)}
    \cos \theta \frac{v^{j}(\vb{v} \vdot \vb{q})}{|\Delta|}
    \frac{i z}{\sin z \cos z}.
\end{multline}

We now introduce the linear response tensor $\tilde{\Pi}^{(\mu,\nu)}_q$, where $\mu,\nu = x,y,d$. The free energy can be written as
\begin{align}
\mathcal{F} = \sum_q  \tilde{\Pi}^{(\mu,\nu)}_q \tilde{A}_{\mu,q} \tilde{A}_{\nu,-q},
\end{align}
where $\tilde{A}_{\mu}= (A_x,A_y,d)$. We note that the variable $d$ corresponds to the drive field, which is the component of the electric field perpendicular to the RTG plane, and not to the scalar potential $A^0$; this is why we put tildes on $A$, since it combines the (gauge-invariant) electric field and (gauge-dependent) vector potential. In what follows we use the Weyl gauge $A^0=0$, allowing us to express the vector potential in terms of the electric field: $\mathcal{E}_j=i \Omega A_j$ for $j \in \{x,y\}$, and $\mathcal{E}_z=d$. $\Omega$ is the (real) frequency of the applied field. We may decompose the response tensor $\tilde{\Pi}$ as the sum of collective mode contributions $\tilde{\Pi}_{\text{CM}}$ and quasiparticle contributions $\tilde{\Pi}_{\text{QP}}$. The former is obtained from an expression analogous to Eq.~\eqref{eq:CM_final}, and the latter is written as
\begin{widetext}
\begin{multline}
    \tilde{\Pi}_{\text{QP}}
    =
    - \int_{\text{FS}} 
    \frac{\dd{k_{\|}}}{4 \pi^2 v}
    \frac{1}{4 |\Delta|^2}
    \frac{2 z}{\sin^3 z \cos z}
    \\
    \begin{pmatrix} 
        v_x^2 (\vb{v} \vdot \vb{q})^2 
        & v_x v_y (\vb{v} \vdot \vb{q})^2 
        & - \bra{u} \hat{o} \ket{u} v_x (\vb{v} \vdot \vb{q})  (i\Omega_m) 
        \\ 
        v_x v_y (\vb{v} \vdot \vb{q})^2
        & v_y^2 (\vb{v} \vdot \vb{q})^2 
        & - \bra{u} \hat{o} \ket{u} v_y (\vb{v} \vdot \vb{q})  (i\Omega_m) 
        \\
        -\bra{u} \hat{o} \ket{u} v_x (\vb{v} \vdot \vb{q})  (i\Omega_m) 
        & -\bra{u} \hat{o} \ket{u} v_y (\vb{v} \vdot \vb{q})  (i\Omega_m) 
        & \bra{u} \hat{o} \ket{u}^2 (i\Omega_m)^2 
    \end{pmatrix}
\end{multline}
\end{widetext}
before analytically continuing $i \Omega_m \rightarrow \Omega + i 0^{+}$.
Again, we note that in deriving the response functions to the in-plane components of the vector potential, we neglected the contributions from the quantum geometric tensor, which might become dominant in flat band systems~\cite{torma2022superconductivity, PhysRevLett.132.026002}. In the case of RTG however the Fermi velocity is large and we thus assume that the quantum geometric contributions to the current matrix elements can be neglected.

The chosen gauge $A^0=0$ allows us to directly relate the response tensor $\tilde{\Pi}$ to the nonlocal electric polarization tensor $\hat{\alpha} = M^{\dag} \tilde{\Pi} M$, where ${P}_i=\alpha_{ij}\mathcal{E}_j$, and $M = \diag(i \Omega, i \Omega, 1)$. We can now solve the classical problem of an electric field $\boldsymbol{\mathcal{E}}_0$ scattering on a two-dimensional polarizable sheet with polarizability $\hat{\alpha}(\Omega,q_x,q_y)\delta(z)$. The answer is written immediately as
\begin{align}
\boldsymbol{\mathcal{E}}_{\vb{q}}(z) = \hat{G}_{\vb{q}}e^{ik_z|z|}\hat{\alpha}(\hat{I}-\hat{G}_{\vb{q}}\hat{\alpha})^{-1}\boldsymbol{\mathcal{E}}_{0,\vb{q}}  \label{eq:FullE}
\end{align}
where $\hat{G}_{\mathbf{q}}$ is the dyadic Green's function in a mixed representation (in real space along $z$ and taken at $z,z'=0$ and in reciprocal space along $x$ and $y$). This Green's function is given by
\begin{align}
\hat{G}_{\mathbf{q}}= \frac{i}{2q_z}\left[\hat{I} \left( \frac{\Omega}{c} \right)^2- (\mathbf{q},q_z)\otimes(\mathbf{q},q_z)\right],
\end{align}
where $q_z=\sqrt{(\Omega/c)^2-\vb{q}^2}$ and $\hat{I}$ is a identity matrix.

Equation \eqref{eq:FullE} can then be used to find the Anderson-Higgs renormalized response function. We rewrite Eq.~\eqref{eq:FullE} as
\begin{equation}
    \boldsymbol{\mathcal{E}}_{\vb{q}}(z) 
    =
    \hat{G}_{\vb{q}}e^{iq_z|z|} \vb{P}
    = 
    \hat{G}_{\vb{q}}(z) \tilde{\hat{\alpha}} \boldsymbol{\mathcal{E}}_{0,\vb{q}},
\end{equation}
with
\begin{equation}
    \tilde{\hat{\alpha}}=\hat{\alpha}(\hat{I}-\hat{G}_{\vb{q}}\hat{\alpha})^{-1}.
\end{equation}
The $zz$ component of $\tilde{\hat{\alpha}}$ then yields the renormalized response function to the perpendicular electric field.

\begin{figure}[b!]
\centering
\includegraphics[width=\columnwidth]{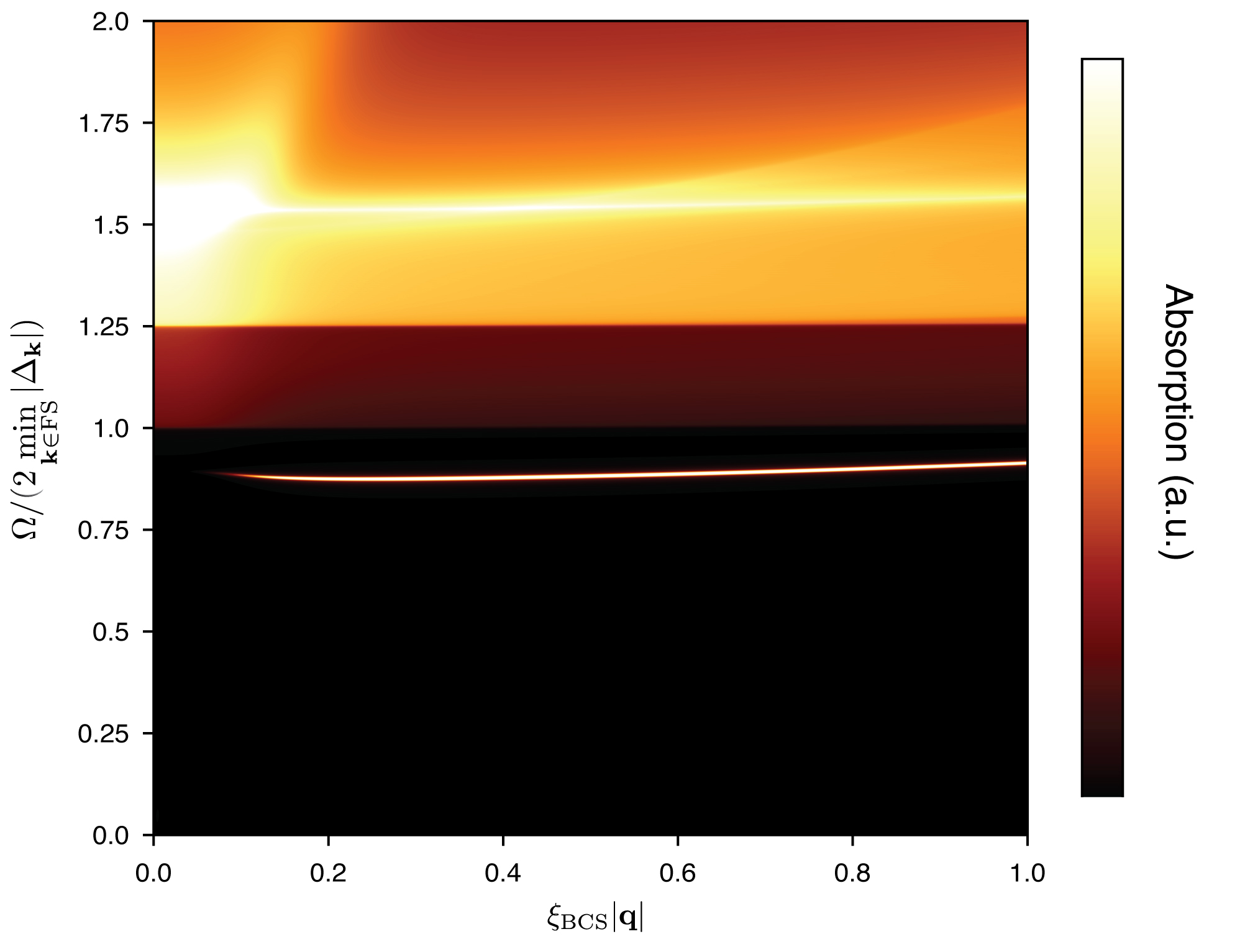}
\caption{\label{fig:RTG_2Dplot} Absorption spectrum of superconducting RTG with IVC-mediated interactions. Parameters are the same as in Fig.~2 of the main text.}
\end{figure}

\section{Clapping mode dispersion for RTG}
\label{appendix:rtg_clapping}
Figure \ref{fig:RTG_2Dplot} shows the absorption spectrum for RTG with IVC-mediated interactions in its $p$-wave superconducting phase, including the Anderson-Higgs mechanism via the self-consistent approach described in Appendix \ref{appendix:anderson-higgs}, as a function of $\vb{q} \| \hat{\vb{x}}$. Different directions of $\vb{q}$ yield only minor differences.

\clearpage

\bibliography{main}

\end{document}